\documentclass[a4paper,fleqn]{cas-sc}
\usepackage[section]{placeins} % For '\input{file}'
\usepackage{amssymb}
\usepackage{amsthm}
\usepackage{amsmath}
\usepackage{upgreek}
\usepackage{pdflscape}
\usepackage{listings}
\usepackage{multirow}
\usepackage[percent]{overpic}
\usepackage{multicol}
\usepackage{color}
\usepackage{mathtools}
\usepackage{stmaryrd}
\usepackage{xfrac}
\usepackage[capitalise]{cleveref}
\usepackage{booktabs}
\usepackage[section]{placeins} % For '\input{file}'
\usepackage[section]{algorithm}
\usepackage[noend]{algpseudocode}
\usepackage{calc}
\usepackage[titletoc]{appendix}
\usepackage[binary-units=true]{siunitx}
\usepackage[sort&compress,square,numbers]{natbib}
\usepackage{dirtytalk}
\usepackage{pifont}
\usepackage{graphicx}
\usepackage{graphics}
\usepackage{inconsolata}
\usepackage{wrapfig}
\usepackage{float}
\usepackage{subfig}
\usepackage[percent]{overpic}
\usepackage{varwidth}
\usepackage{tikz}
\usepackage{pgfplots}
\usepackage{adjustbox}
\usetikzlibrary{arrows,automata,backgrounds,fit,intersections,matrix,mindmap,petri,positioning,shapes,snakes}
\usepackage{tikz-layers}
\usepgfplotslibrary{fillbetween}
\pgfplotsset{compat=newest}
\usepgfplotslibrary{colorbrewer}
\usepgfplotslibrary{patchplots}
\usepgfplotslibrary[colorbrewer]
\usetikzlibrary{pgfplots.colorbrewer}
\usetikzlibrary[pgfplots.colorbrewer]
\usetikzlibrary{patterns}
\usepgfplotslibrary{units}
\usetikzlibrary{spy}
\usepackage{pgfplotstable}
\usepackage{arrayjobx}
\graphicspath{ {./figs/} }
\usetikzlibrary{external}

\newlength\myheight
\newlength\mydepth
\settototalheight\myheight{Xygp}
\newcommand*\inlinegraphics[1]{%
  \settototalheight\myheight{Xygp}%
  \settodepth\mydepth{Xygp}%
  \raisebox{-\mydepth}{\includegraphics[height=\myheight]{#1}}%
}
\newcommand\orcid[1]{\href{https://orcid.org/#1}{\inlinegraphics{orcid_16x16.png}}}

\makeatletter
\def\BState{\State\hskip-\ALG@thistlm}
\makeatother

\newdefinition{definition}{Definition}[section]

\newcommand\ceil[1]{\lceil#1\rceil}

\newcommand\px[2]{\frac{\partial #1}{\partial {#2}}}

\newcommand\dx[2]{\frac{\mathrm{d} #1}{\mathrm{d} #2}}

\newcommand\pxvar[2]{\partial_{#2} #1}

\newcommand{\cmark}{\ding{51}}

\begin{document}

\title[mode=title]{Hyperbolic Diffusion in Flux Reconstruction: Optimisation through Kernel Fusion within Tensor-Product Elements}
\shorttitle{Hyperbolic Diffusion in Flux Reconstruction}
\shortauthors{W. Trojak et al.}

\author[1]{W. Trojak}[orcid=0000-0002-4407-8956]
\cormark[1]
\ead{wt247@tamu.edu}
\cortext[cor1]{Corresponding author}
\address[1]{Department of Ocean Engineering, Texas A\&M University, College Station, TX 77843}

\author[2]{R. Watson}[orcid=0000-0002-8838-9406]
\ead{r.watson@qub.ac.uk}
\address[2]{School of Mechanical and Aerospace Engineering, Queen's University Belfast, Belfast, BT9 5AH}

\author[1]{F. D. Witherden}[orcid=0000-0002-4407-8956]
\ead{fdw@tamu.edu}

\begin{abstract}
    Novel methods are presented in this initial study for the fusion of GPU kernels in the artificial compressibility method (ACM), using tensor product elements with constant Jacobians and flux reconstruction. This is made possible through the hyperbolisation of the diffusion terms, which eliminates the expensive algorithmic steps needed to form the viscous stresses. Two fusion approaches are presented, which offer differing levels of parallelism. This is found to be necessary for the change in workload as the order of accuracy of the elements is increased. Several further optimisations of these approaches are demonstrated, including a generation time memory manager which maximises resource usage. The fused kernels are able to achieve 3-4 times speedup, which compares favourably with a theoretical maximum speedup of 4. In three dimensional test cases, the generated fused kernels are found to reduce total runtime by ${\sim}25\%$, and, when compared to the standard ACM formulation, simulations demonstrate that a speedup of $2.3$ times can be achieved. 
\end{abstract}

\begin{highlights}
\item The hyperbolic diffusion technique is applied to the artificial compressibility method in flux reconstruction.
\item Techniques are developed to fuse large flux evaluation kernels and matrix multiplication on GPUs.
\item Optimisation are explored to fully utilise compute resources, including the development of a generation-time memory manager.
\end{highlights}


%\linenumbers

% ================================================================================

\maketitle

\section{Introduction}\label{sec:intro}
    The application of numerical simulation to fluid flows is a versatile tool both for exploring complex flow physics and for assisting in the engineering design process. Computation has developed this role due the cost reductions it offers in comparison to experimentation, its flexibility to explore new configurations, and its ability to closely examine difficult-to-observe physics. Large eddy simulation (LES) is a term applied to scale resolving simulations in which typically ${\sim}80\%$ of the turbulent kinetic energy of the vortical motions is resolved. The effect of the turbulence which is not resolved is captured by an explicit subgrid scale model description~\citep{Tucker2016} or through implicit LES (ILES), where this modelling is effectively handled by inherent numerical dissipation. The use of scale resolving simulations in fluid dynamics has allowed the detailed exploration of complex flow phenomena. However, even with powerful accelerator hardware such as graphics processing units (GPUs), these calculations remain expensive. In typical flow simulations of engineering interest, a large proportion of the degrees of freedom are concentrated within the boundary layers. The primary cause of this is that, approaching a solid wall, turbulent length scales decrease and develop greater anisotropy. Increased spatial resolution is therefore required to support the structures responsible for the mixing of momentum. Compounding this, viscous effects become increasingly important within the boundary layer, and, close to the wall, dominate the inertial effects. For these viscous effects, the temporal stability limit scales with $h^{-2}$, where $h$ is characteristic of the mesh spacing. This results in rapidly escalating costs for wall resolved LES. \citet{Piomelli2008} argues that the cost of wall resolved LES scales with $Re^{2.4}$, which, given that engineering flows can range from below $Re=10^4$ to $Re=10^8$ and above, poses a significant computational challenge.

    An approach that has been developed in recent years that modifies this $h^{-2}$ dependency to $h^{-1}$ is the method of hyperbolic diffusion, introduced by \citet{Nishikawa2010,Nishikawa2010a}. To illustrate this method, consider a Cauchy problem in one variable and $n$ dimensions:
    \begin{subequations}
        \begin{align}
            \pxvar{u}{t} + \nabla\cdot\mathbb{f} &= \nu\Delta u \quad \mathrm{for}\quad  u(t,\mathbf{x}) : \mathbb{R}_+\times \mathbb{R}^n \rightarrow \mathbb{R}, \\
            u(0,\mathbf{x}) &= u_0,
        \end{align}
    \end{subequations}
    where $\mathbb{f}$ is a flux tensor, and $\Delta$ is some generalised $n$-dimensional Laplacian acting on $u$. First, consider the addition of a pseudo-time derivative, to give:
    \begin{equation}
        \pxvar{u}{t} + \pxvar{u}{\tau} + \nabla\cdot\mathbb{f} = \nu\Delta u.
    \end{equation}
    This extension of the conservation laws is useful when using implicit time integration methods as it can be used as a relaxation mechanism, with a physical time update considered as a steady-state problem with respect to pseudo-time~\citep{Jameson1991}. The hyperbolic diffusion methodology also uses pseudo-time to reframe the second order terms as a set of coupled partial differential equations (PDEs):
    \begin{subequations}\label{eq:hd}
        \begin{align}
            \pxvar{u}{\tau} + \pxvar{u}{t} + \nabla\cdot\left[\mathbb{f} - \nu\,\mathrm{diag}(p_1,p_2\dots,p_n)\right] &= 0, \\
            \pxvar{p_1}{\tau} + \frac{1}{T_r}\pxvar{u}{1} &= \frac{p_1}{T_r}, \\
            &\vdotswithin{=} \\
            \pxvar{p_n}{\tau} + \frac{1}{T_r}\pxvar{u}{n} &= \frac{p_n}{T_r}.
        \end{align}
    \end{subequations}
    To understand what has happened here, consider the limit as $\tau\rightarrow\infty$, where if the pseudo-time derivatives converge on zero, then $p_1\rightarrow\pxvar{u}{1}$. Furthermore, a preconditioning parameter, $T_r$ has been introduced to reduce the stiffness from the auxiliary equations.
    
    Hyperbolic diffusion has been applied to many equation sets that include second derivatives, including the Navier--Stokes equations. Although originally applied to finite volume (FV) methods, several applications have now been made to Discontinuous Galerkin (DG) methods, for example \citet{Lou2017} and \citet{Li2018}. To the authors' knowledge, two previous sets of papers have also explored the application of hyperbolic diffusion to the flux reconstruction (FR) method of \citet{Huynh2007}: the works of \citet{Lou2020}, and those of  \citet{Watson2018} and \citet{McCaughtry2021}. The advantage that hyperbolic diffusion offers is that by hyperbolising the equation set, the temporal stability limit is found to scale with $h^{-1}$, see \citet{Nishikawa2010a,Watson2018}. Given the substantial grid requirements in viscous dominated regions required for wall resolved LES, this offers a route to save considerable computational effort.
    
    The time required for fluid dynamics simulations can also be influenced by the effects of compressibility. Solving low Mach number flows is challenging when using the compressible formulation of the Navier--Stokes equation as the governing equation. This is due to the increased stiffness of the problem as the Mach number is reduced, caused by the acoustic wavespeeds increasing relative to the convective wavespeeds. An alternative is to use one of the many methods similar to SIMPLE~\citep{Patankar1972}. However, this requires the solution of a Poisson equation which requires a separate method that has its own computational challenges~\citep{Fortunato2019}. An alternative is to make use of the artificial compressibility method (ACM) of \citet{Chorin1967} coupled with the dual time stepping of \citet{Jameson1991}. This removes the need to solve a Poisson equation by instead allowing pressure fluctuations in pseudo-time to account for and correct non-zero divergence in the velocity field. The ACM set of equations can be straightforwardly ported to a solver for conservation equations. For example, it has been used together with FR by \citet{Loppi2018}. As this method already uses pseudo-time relaxation to converge the velocity field to a divergence free condition, it is a good candidate for hyperbolic diffusion --- as all that is required is an adjustment to the governing equations.
    
    For approximating the solution to conservation laws such as those governing fluid flows, high-order discontinuous spectral element methods (DSEM) methods exhibit several attractive features. They can make more efficient use of the information in the stencil when compared to lower order approaches such as cell centred finite volumes or even to high-order finite differences~\citep{Trojak2017}. The trade-off made when choosing a high-order method over a method such as second-order vertex centred finite volume is that the number of operations per update is greatly increased. However, in recent years --- for both CPU and GPU hardware --- there has been an abundance of FLOPs and a relative scarcity of memory bandwidth~\citep{Witherden2014}. This limitation is such that the capability of a high-performance PDE solver will typically be bounded by the memory bandwidth. This can be seen in practice when using tools such as PyFR~\citep{Witherden2014}, deal.II~\citep{dealii2019}, or MFEM~\citep{Anderson2021}. For high-order approaches, this means that the high operation count has little impact on runtime and, owing to the latency of memory transactions, can make better use of the available hardware by performing computation while waiting on memory access. A feature of the DSEM and DG methods that makes them particularly attractive is the locally structured computation within elements. This allows many operations to be performed using optimised matrix-matrix multiplication libraries or via optimised pointwise kernels that aim to hide some memory latency.
    
    Graphic processing units have been popular for some years in high performance computing as they offer a high density of FLOPs and a higher memory bandwidth than CPUs. The high flop density originates in huge parallelism, with several thousand cores being commonly found in a single GPU~\citep{Nvidia2017}. As has already been stated, for a number problems in computational physics, calculations are limited by available memory bandwidth. By making better use of the memory hierarchy within GPUs, computations can be effectively optimised. Consider the operation:
    \begin{equation*}
        \mathbf{X} = \mathbf{M}\mathbf{U}.
    \end{equation*}
    Many stages in DSEM simplify to this, and it is typical that the operator matrix, $\mathbf{M}$, is constant and, in some cases, sparse. Using the sparsity and immutability of the operators, the GiMMiK library~\citep{Wozniak2016} produces unique matrix-matrix multiplication kernels by in-lining the operator matrix constants and fully unrolling the operation. In doing so, optimisations, such as removing multiplication by zero, can be performed at generation time. By in-lining the constants, they are streamed through the instruction cache, leaving more bandwidth for the variable matrix. This can lead to highly performant kernels; however, at high-orders or when using double-precision floating-point numbers, performance can suffer through high register pressure. A second approach was explored by \citet{Swirydowicz2019}. For matrix operations in DSEM methods, tensor-product elements, such as maximal-order basis cubes, offer an opportunity for optimisation. By leveraging the tensor product construction, they were able to increase the parallelism and data locality within the memory hierarchy. This was achieved by increasing the number of threads working on a single element and using a user-allocatable L1 cache called shared memory. More recently, \citet{Trojak2021b} presented an in-line compression method for three-dimensional vectors, where a speedup of 50\% could be achieved for memory bound calculations with low compression error. By applying the approaches developed in these works, we hypothesise that we may be able to perform novel optimisation of kernels used within FR when applied to ACM with hyperbolised diffusion (ACM-HD) --- namely, the fusion of two or more kernels which can reduce memory transactions. 
    
    In this paper, we will explore the application of ACM-HD to FR and the opportunities for optimisation that arise. To this end, the paper is structured as follows. FR and ACM-HD are introduced in \cref{sec:prelim},  followed by an in-depth analysis of the possibilities for kernel fusion optimisations in \cref{sec:sparse}.  In \cref{sec:method}, we describe two approaches to kernel fusion and detail their computational performance. In this work we will only focus on elements with constant spatial Jacobians, and further work will be required to develop methods for non-affine elements. These kernels are then applied within the context of PyFR to a three-dimensional test case in \cref{sec:application}, and the results analysed. Finally, in \cref{sec:conclusions}, conclusions are drawn.

\section{Preliminaries}\label{sec:prelim}
    \subsection{Flux Reconstruction}\label{ssec:fr}
        Flux reconstruction (FR)~\citep{Huynh2007} is a high-order DSEM method that has been widely applied to advection-diffusion problems in computational physics. To illustrate the method, take the following first-order PDE in one dimension:
        \begin{equation}
            \px{u}{t} + \px{f(u)}{x} = 0, \quad \mathrm{for} \quad u(x,t): \Omega\times\mathbb{R}_+ \mapsto \mathbb{R}, \quad f(u): \mathbb{R} \rightarrow \mathbb{R}, \quad \mathrm{and} \quad u(x,t=0) = u_0 \;\forall x\in\partial\Omega,
        \end{equation}
        where $\Omega$ is the spatial domain. This domain is partitioned into $N$ compatible sub-domains, $K_i$, such that $\Omega=\bigcup^N_{i=0}K_i$. A reference domain, $\hat{K}$, is specified, which in 1D is taken to be $\hat{K}=[-1,1]$. A mapping $T_i: \hat{K}\mapsto K_i$ can then be defined between the sub-domains and the reference domain. The one-dimensional divergence of the flux at a point $x_j\in K_i$ may then be reconstructed as:
        \begin{equation}
            \px{f(x_j)}{x} = \left(\dx{T_i}{\zeta}\bigg|_{x=x_j}\right)^{-1}\left(\sum^{p}_{k=0}\hat{f}(\hat{x}_k)\dx{l_k(\hat{x}_j)}{\zeta} + (f_l^I - f_l^\delta)\dx{h_l(\hat{x}_j)}{\zeta} + (f_r^I - f_r^\delta)\dx{h_r(\hat{x}_j)}{\zeta}\right).
        \end{equation}
        Here, $l_k$ is the $k$-th Lagrange polynomial of order $p$, and $h_l$ and $h_r$ are the left and right correction functions, respectively. The correction functions have two conditions,  $h_l(-1)=h_r(1) = 1$ and $h_l(1)=h_r(-1) = 0$~\citep{Vincent2010}. In addition, $f_l^I$ and $f_r^I$ are the common interface fluxes at the left and right interfaces, respectively. These are calculated using the solutions interpolated to the interface between adjacent elements. An approximate or exact Riemann solver can then be applied, the aim being to provide physical inter-element communication which also leads to sufficient dissipation to stabilise the method. $f_l^\delta$ is the flux interpolated to the interface.
        
        Once the approximate divergence of the flux has been calculated, it can either be used to perform an explicit update with an appropriate ODE integration method, or as part of an implicit solver. Flux reconstruction has been successfully used in conjunction both with the dual-time stepping implicit method~\citep{Jameson1991,Loppi2018} and pseudo-transient continuation with GMRES~\citep{Wang2020}.
        
        For second-order PDEs, it is necessary to take an extra step and form a continuous approximation of the solution gradient. Take the second-order PDE:
        \begin{equation}
            \px{u}{t} + \px{f(u, \nabla u)}{x} = 0, \quad \mathrm{for} \quad u(x,t): \Omega\times\mathbb{R}_+ \mapsto \mathbb{R}, \quad f(u, \nabla u): \mathbb{R}\times\mathbb{R} \rightarrow \mathbb{R}, \quad \mathrm{and} \quad u(x,t=0) = u_0 \;\forall x\in\partial\Omega,
        \end{equation}
        for which the system, with an additional auxiliary equation, is solved as:
        \begin{subequations}
            \begin{align}
                q_j = &\px{u(x_j)}{x} = \left(\dx{T_i}{\zeta}\bigg|_{x=x_j}\right)^{-1}\left(\sum^{p}_{k=0}\hat{u}(\hat{x}_k)\dx{l_k(\hat{x}_j)}{\zeta} + (u_l^I - u_l^\delta)\dx{h_l(\hat{x}_j)}{\zeta} + (u_r^I - u_r^\delta)\dx{h_r(\hat{x}_j)}{\zeta}\right),\\
                &\px{f(x_j)}{x} = \left(\dx{T_i}{\zeta}\bigg|_{x=x_j}\right)^{-1}\left(\sum^{p}_{k=0}\hat{f}(u(\hat{x}_k),q(\hat{x}_k))\dx{l_k(\hat{x}_j)}{\zeta} + (f_l^I - f_l^\delta)\dx{h_l(\hat{x}_j)}{\zeta} + (f_r^I - f_r^\delta)\dx{h_r(\hat{x}_j)}{\zeta}\right).
            \end{align}
        \end{subequations}
        A disadvantage of this approach to second-order PDEs is that forming this point-wise $H_\mathrm{div}$ approximation to the gradient adds a significant number of algorithmic steps compared to the first-order system.
    
\subsection{The Artificial Compressibility Method}\label{ssec:acm}
    At low Mach numbers, the compressible Euler and Navier--Stokes equations become extremely stiff, with the acoustic waves travelling through the domain far more quickly than the entropy and vorticity waves. This makes them computationally expensive to solve directly. As an alternative, equations which represent the flow of fluids in the incompressible limit can be derived --- the incompressible Euler and Navier--Stokes equations. The artificial compressibility method (ACM) of \citet{Chorin1967} is a technique used to solve these incompressible equations with software originally developed for the compressible ones. It does this by modifying the mass conservation equation to supply the pressure field corresponding to a divergence free velocity field. This is done by adding a relaxation parameter, which is here called $\zeta$. This destroys the intrinsic time accuracy of the equations, but it may be recovered by coupling the approach to the dual-time method of \citet{Jameson1991}, allowing the pressure field to be converged in pseudo-time. The ACM as applied to the incompressible Navier--Stokes equation can be written, in two dimensions, as:
        \begin{equation}\label{eq:acm_2d}
            \px{}{t}\begin{bmatrix}
                0 \\ u \\ v
            \end{bmatrix} 
            + \px{}{\tau}\begin{bmatrix}
                P \\ u \\ v
            \end{bmatrix} 
            + \px{}{x}\begin{bmatrix}
                \zeta u \\ u^2 + P \\ uv
            \end{bmatrix}
            + \px{}{y}\begin{bmatrix}
                \zeta v \\ uv \\ v^2 + P
            \end{bmatrix} = 
            \nu\begin{bmatrix}
                0 \\ \pxvar{u}{xx} \\ \pxvar{v}{yy}
            \end{bmatrix},
        \end{equation}
        where $\tau$ is the pseudo-time variable, and $\nu$ is the kinematic viscosity. Following the method of \citep{Nishikawa2010a}, the diffusion terms may then be hyperbolised by writing the system of equations as:
        \begin{equation}\label{eq:acmhd_2d}
            \px{}{t}\begin{bmatrix}
                0 \\ u \\ v \\ 0 \\ 0 \\ 0 \\ 0
            \end{bmatrix} 
            + \px{}{\tau}\begin{bmatrix}
                P \\ u \\ v \\ q_x \\ q_y \\ r_x \\ r_y
            \end{bmatrix} 
            + \px{}{x}\begin{bmatrix}
                \zeta u \\ u^2 + P - \nu q_x \\ uv - \nu r_x \\ -u/T \\ 0 \\ -v/T \\ 0
            \end{bmatrix}
            + \px{}{y}\begin{bmatrix}
                \zeta v \\ uv - \nu q_y \\ v^2 + P - \nu r_y\\ 0 \\ -u/T \\ 0 \\ -v/T
            \end{bmatrix} = 
            -\frac{1}{T}\begin{bmatrix}
                0 \\ 0 \\ 0 \\ q_x \\ q_y \\ r_x \\ r_y 
            \end{bmatrix},
        \end{equation}
        where $q_x\approx \pxvar{u}{x}$, $r_x \approx \pxvar{v}{x}$, etc. Going further, this may also be defined in three dimensions as:
        \begin{equation}\label{eq:acmhd_nd}
            \px{}{t}\begin{bmatrix}
                0 \\ \mathbf{V} \\ \mathbf{0} \\ \mathbf{0} \\ \mathbf{0} 
            \end{bmatrix} 
            + \px{}{\tau}\begin{bmatrix}
                P \\ \mathbf{V} \\ \mathbf{q} \\ \mathbf{r} \\ \mathbf{s}
            \end{bmatrix} 
            + \nabla\cdot\begin{bmatrix}
                \zeta \mathbf{V}^T \\ \mathbf{V}\otimes\mathbf{V} + P\mathbf{I} - \nu \mathbf{S}^T \\ -\frac{1}{T}\mathbf{V}\otimes\mathbf{I}
            \end{bmatrix} = 
            -\frac{1}{T}\begin{bmatrix}
                0 \\ \mathbf{0} \\ \mathbf{q} \\ \mathbf{r} \\ \mathbf{s} 
            \end{bmatrix},
        \end{equation}
        where $\mathbf{V} = [u,v,w]^T$, $s_x\approx \pxvar{w}{x}$ etc., $\mathbf{s} = [s_x,s_y,s_z]^T$, and $\mathbf{S} = [\mathbf{q},\mathbf{r},\mathbf{s}]$. This system of equations, which makes use of the hyperbolic diffusion method, is herein referred to as the ACM-HD method. A test can be performed on \cref{eq:acmhd_nd} showing that the eigenvalues of the flux Jacobian are all real, and hence new governing equation is indeed hyperbolic.
        
        % raw54 - in equation 9, can we replace q,r,s with S^T?
        % wt247 - in the source term? No I don't think so. S is a matrix the source is a vector. 
        % raw54 - ah, yep, I see.
        
\section{\label{sec:sparse}Problem Setup}
    In this paper, the application of FR to low Mach problems is studied with particular emphasis on the implementation details for GPU accelerators, and although this work specifies the use of the CUDA framework, many of the ideas can be transmitted directly to other vector accelerator type hardware.
    
    Within the last four or more generations of devices, the arithmetic capabilities of GPU devices have greatly outperformed their memory bandwidths~\citep{Witherden2014}, and therefore the optimisation of algorithms is, in large part, concerned with reducing the global memory access. Due to the high number of variables solved for in ACM-HD compared with standard ACM, the global memory bandwidth use in a naive implementation is considerably increased. Full bench-marking data will be presented later --- it suffices at this stage to say that 3D ACM-HD takes twice as long as ACM for a full flux divergence evaluation.

    However, ACM-HD presents several opportunities for optimisation. The first is independent of the numerical scheme: by comparing the flux functions in \cref{eq:acm_2d} to those in \cref{eq:acmhd_2d}, the sparsity of the ACM-HD system can be seen. It can be straightforwardly shown that the sparsity of the ACM-HD flux function is:
    \begin{equation}
        s = 1 - \frac{1 + 2d}{1 + d + d^2}.
    \end{equation}
    Taking advantage of this sparsity would provide a reduction in the bandwidth requirements. For example, with $d=3$, the flux has a sparsity of ${\sim}46\%$. However, to best leverage the opportunity afforded by this sparsity, the second optimisation opportunity is described first. 
    
\begin{figure}[tbhp]
        \centering
        \subfloat[Base scheme.]{\label{fig:dep}\adjustbox{width=0.49\linewidth, valign=b}{\begin{tikzpicture}
    \tikzstyle{data_base}=[draw,rectangle,rounded corners=0.5ex, thick, color={Pastel1-B}, fill={Pastel1-B},
                      text=black,minimum width=10pt]
    \tikzstyle{data_end}=[draw,rectangle,rounded corners=0.5ex, thick, color={Pastel1-C}, fill={Pastel1-C},
                      text=black,minimum width=10pt]
                      
    \tikzstyle{data_int}=[draw,rectangle,rounded corners=0.5ex, thick, color={Pastel1-D}, fill={Pastel1-D},
                      text=black,minimum width=10pt]
                      
    \tikzstyle{data}=[draw,rectangle,rounded corners=0.5ex, thick, color={Pastel1-A}, fill={Pastel1-A},
                      text=black,minimum width=10pt]
                      
    \tikzstyle{dep}=[-latex,black,thick]
                      
    \node[data_base] at (0,1.5) (us) {$\mathsf{U}_s$};
    \node[data_base] at (0,0) (ss) {$\mathsf{S}_s$};
    \node[data_base] at (0,3) (nf) {$\mathsf{n}_f$};
    
    \node[data] at (2,1.5) (uf) {$\mathsf{U}_f$};
    \node[data_int] at (2,3) (fi) {$\mathsf{F}^I_f$};
    \node[data] at (2,0) (fs) {$\mathsf{F}_s$};
    
    \node[data] at (4,0) (dfas) {$\nabla\cdot\mathsf{F}^a_s$};
    \node[data] at (4,1.5) (dfcs) {$\nabla\cdot\mathsf{F}^c_s$};
    \node[data_end] at (6,1.5) (dfs) {$-\nabla\cdot\mathsf{F}_s + \mathsf{s}_s$};
    
    \draw[dep] (us) -- (uf);
    \draw[dep] (uf) -- (fi);
    \draw[dep] (nf) -- (fi);

    \draw[dep] (us) -- (fs);
    \draw[dep] (ss) -- (fs);
    \draw[dep] (fs) -- (dfas);
    \draw[dep] (dfas) -- (dfcs);
    \draw[dep] (fi) -- (dfcs);
    \draw[dep] (dfcs) -- (dfs);
    \path[dep] (us) edge[bend right=-25] node [left] {} (dfs);
    \draw[dep] (ss) -- (dfs);
    
\end{tikzpicture}}}
        \hfil
        \subfloat[Fused scheme.]{\label{fig:dep_red}\adjustbox{width=0.355\linewidth, valign=b}{\begin{tikzpicture}
    \tikzstyle{data_base}=[draw,rectangle,rounded corners=0.5ex, thick, color={Pastel1-B}, fill={Pastel1-B},
                      text=black,minimum width=10pt]
    \tikzstyle{data_end}=[draw,rectangle,rounded corners=0.5ex, thick, color={Pastel1-C}, fill={Pastel1-C},
                      text=black,minimum width=10pt]
                      
    \tikzstyle{data_int}=[draw,rectangle,rounded corners=0.5ex, thick, color={Pastel1-D}, fill={Pastel1-D},
                      text=black,minimum width=10pt]
                      
    \tikzstyle{data}=[draw,rectangle,rounded corners=0.5ex, thick, color={Pastel1-A}, fill={Pastel1-A},
                      text=black,minimum width=10pt]
                      
    \tikzstyle{dep}=[-latex,black,thick]
                      
    \node[data_base] at (0,1.5) (us) {$\mathsf{U}_s$};
    \node[data_base] at (0,0) (ss) {$\mathsf{S}_s$};
    \node[data_base] at (0,3) (nf) {$\mathsf{n}_f$};
    
    \node[data] at (2,1.5) (uf) {$\mathsf{U}_f$};
    \node[data_int] at (2,3) (fi) {$\mathsf{F}^I_f$};
    
    \node[data] at (2,0) (dfas) {$-\nabla\cdot\mathsf{F}^a_s + \mathsf{s}_s$};
    \node[data_end] at (4,1.5) (dfs) {$-\nabla\cdot\mathsf{F}_s + \mathsf{s}_s$};
    
    \draw[dep] (us) -- (uf);
    \draw[dep] (uf) -- (fi);
    \draw[dep] (nf) -- (fi);

    \draw[dep] (us) -- (dfas);
    \draw[dep] (ss) -- (dfas);
    \draw[dep] (dfas) -- (dfs);
    \draw[dep] (fi) -- (dfs);
    
\end{tikzpicture}}}
        \caption{\label{fig:dep_graph}Data dependency graph for purely advective FR.}
    \end{figure}
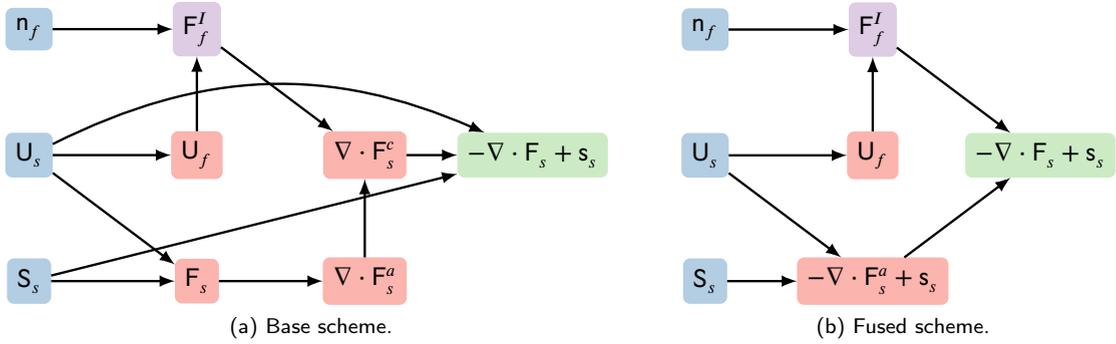

    The ACM-HD approach is purely advective in nature, and so the stages in the FR algorithm concerned with the calculation of diffusion terms may be ignored completely. However, the approach gives rise to possibilities for optimisation beyond merely avoiding performing these stages. Consider the algorithmic stages of FR required for purely advective systems. These are shown in \cref{tab:fr_adv}, for a $d$ dimensional domain of $n_e$ elements, $n_v$ number of variables, $n_s$ solution points, and $n_f$ flux points. Here, we use the notation that $\mathsf{U}$ and $\mathsf{F}$ are the solution and flux respectively, with sub-script $s$ and $f$ notation the variable is at the solution or flux points respectively. There are also two further terms $\mathsf{n}$ and $\mathsf{S}$, which are the outwards facing interface normals and spatial transformation Jacobian. The terms in square brackets of \cref{tab:fr_adv} show the memory space of each variable. It is assumed that the source terms are only dependent on the flow variables, as is the case with the ACM-HD source terms. These stages already contain an algorithmic optimisation that is used in PyFR, the partially corrected flux divergence, where three of the operators are grouped into a single matrix multiplication --- see \citet{Witherden2014} for more details. Regardless, it should be clear that there are several intermediate results that are only required by a single subsequent stage. These data dependencies can be seen more clearly in \cref{fig:dep}. Blocks here represent data that is stored and arrows represent the transformation of data, which has three stages: load; compute; and store.
    
    From considering this graph, there is clearly scope to fuse stages together, thereby reducing both the global memory I/O, and the total memory required to perform the calculation. The proposed reduced graph is shown in \cref{fig:dep_red}, with the component with the greatest potential being the fusion of stages 2 and 3. Stage 6 can also be fused, but this is almost trivial and so will only be considered later as a final stage. The fusion of stages 1 and 4 will not be performed, as stage 4 occurs at the interfaces, and the fusion of M or P kernels with I kernels is not compatible due to the inter-element communication that is required.
    
    \begin{table}[tbhp]
        \centering 
        \caption{\label{tab:fr_adv}Algorithmic stages of the FR method. Here, \emph{I} denotes interface operations, \emph{M} denotes matrix multiplication operations, and \emph{P} denotes point-wise operations.}\begin{tabular}{r c l l l} \toprule
            & Type & Stage & Data in & Data out \\\midrule 
            1 & M & Project solution to $x_f$ & $\mathsf{U}_s[n_e n_v n_s]$ & $\mathsf{U}_f[n_e n_v n_f]$ \\
            2 & P & Calculate flux at $x_s$ & $\mathsf{U}_s[n_e n_v n_s]$, $\mathsf{S}_s[d^2n_en_s]$ & $\mathsf{F}_s[d n_e n_v n_s]$ \\ 
            3 & M & Partially corrected flux div. & $\mathsf{F}_s[d n_e n_v n_s]$ & $\nabla\cdot\mathsf{F}^a_s[n_en_vn_s]$ \\
            4 & I & Common interface flux & $\mathsf{U}_f[n_en_vn_f]$, $\mathsf{n}[dn_en_f]$ & $\mathsf{F}^I_f[n_en_vn_f]$\\
            5 & M & Accumulate correction to flux div. & $\mathsf{F}^I_f[n_en_vn_f]$, $\nabla\cdot\mathsf{F}^a_s[n_en_vn_s]$ & $\nabla\cdot\mathsf{F}^c_s[n_en_vn_s]$ \\
            6 & P & Negative and source term & $\nabla\cdot\mathsf{F}^c_s[n_en_vn_s]$, $\mathsf{U}_s[n_e n_v n_s]$, $\mathsf{S}_s[d^2n_en_s]$ & $\nabla\cdot\mathsf{F}_s[n_en_vn_s]$\\ \bottomrule
        \end{tabular}

    \end{table}
    
    % raw54 - why is stage 3 a partially corrected flux divergence? What's been corrected?
    % wt247 - this is an optimisation that is in pyfr, if you look at equation 4. The bit that does div(f) can be merged with the bit that interpolates the flux to the edges and the does the contribution from the correction function. I.e group df/dx with f_l*dhl/dx and f_r*dhr/dx.
    % wt247 - I have added a clarification
    % raw54 - great, thanks
    
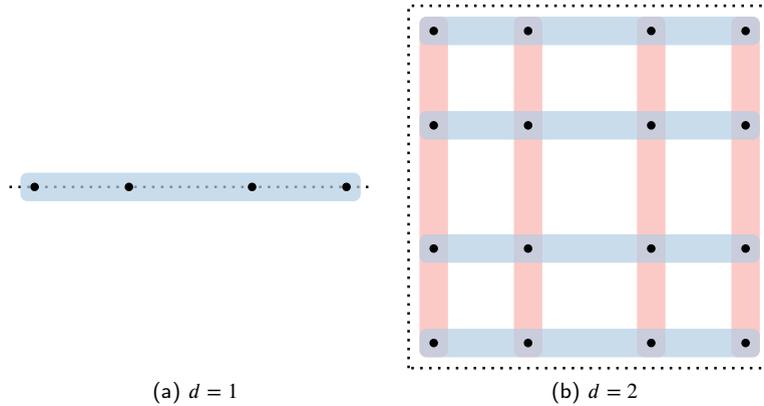
\begin{figure}
        \centering
        \subfloat[$d=1$]{\adjustbox{width=0.3\linewidth,valign=b}{\begin{tikzpicture}[scale=1]
    
    \draw[black!00, dotted, thick] (0,0) rectangle (4,4);
    \draw[black, dotted, thick] (0,2) -- (4,2);
    
    \begin{scope}[on above layer]
        \draw[black, fill=black] (0.27773,2) circle (0.04);
        \draw[black, fill=black] (1.32,2) circle (0.04);
        \draw[black, fill=black] (2.68,2) circle (0.04);
        \draw[black, fill=black] (3.7223,2) circle (0.04);
    \end{scope}
    \draw[Pastel1-B, fill={Pastel1-B}, opacity=0.7, rounded corners=0.5ex] (0.12773,1.85) rectangle (3.8723,2.15);

\end{tikzpicture}}}
        \quad
        \subfloat[$d=2$]{\adjustbox{width=0.3\linewidth,valign=b}{\begin{tikzpicture}[scale=1]

    \draw[black, dotted, thick] (0,0) rectangle (4,4);
    
    \begin{scope}[on above layer]
        \draw[black, fill=black] (0.27773,0.27773) circle (0.04);
        \draw[black, fill=black] (1.32,0.27773) circle (0.04);
        \draw[black, fill=black] (2.68,0.27773) circle (0.04);
        \draw[black, fill=black] (3.7223,0.27773) circle (0.04);
        \draw[black, fill=black] (0.27773,1.32) circle (0.04);
        \draw[black, fill=black] (1.32,1.32) circle (0.04);
        \draw[black, fill=black] (2.68,1.32) circle (0.04);
        \draw[black, fill=black] (3.7223,1.32) circle (0.04);
        \draw[black, fill=black] (0.27773,2.68) circle (0.04);
        \draw[black, fill=black] (1.32,2.68) circle (0.04);
        \draw[black, fill=black] (2.68,2.68) circle (0.04);
        \draw[black, fill=black] (3.7223,2.68) circle (0.04);
        \draw[black, fill=black] (0.27773,3.7223) circle (0.04);
        \draw[black, fill=black] (1.32,3.7223) circle (0.04);
        \draw[black, fill=black] (2.68,3.7223) circle (0.04);
        \draw[black, fill=black] (3.7223,3.7223) circle (0.04);
    \end{scope}
    
    \draw[Pastel1-B, fill={Pastel1-B}, opacity=0.7, rounded corners=0.5ex] (0.12773,0.12773) rectangle (3.8723,0.42773);
    \draw[Pastel1-B, fill={Pastel1-B}, opacity=0.7, rounded corners=0.5ex] (0.12773,1.17) rectangle (3.8723,1.47);
    \draw[Pastel1-B, fill={Pastel1-B}, opacity=0.7, rounded corners=0.5ex] (0.12773,2.53) rectangle (3.8723,2.83);
    \draw[Pastel1-B, fill={Pastel1-B}, opacity=0.7, rounded corners=0.5ex] (0.12773,3.5723) rectangle (3.8723,3.8723);
    
    \begin{scope}[on behind layer]
        \draw[Pastel1-A, fill={Pastel1-A}, opacity=0.7, rounded corners=0.5ex] (0.12773,0.12773) rectangle (0.42773,3.8723);
        \draw[Pastel1-A, fill={Pastel1-A}, opacity=0.7, rounded corners=0.5ex] (1.17,0.12773) rectangle (1.47,3.8723);
        \draw[Pastel1-A, fill={Pastel1-A}, opacity=0.7, rounded corners=0.5ex] (2.53,0.12773) rectangle (2.83,3.8723);
        \draw[Pastel1-A, fill={Pastel1-A}, opacity=0.7, rounded corners=0.5ex] (3.5723,0.12773) rectangle (3.8723,3.8723);
    \end{scope}

    %/draw[Pastel1-B, fill={Pastel1-B}, opacity=0.7, rounded corners=0.5ex] (,-0.08+0.025) rectangle (3.08+0.025,0.08+0.025);

\end{tikzpicture}}}
        \caption{\label{fig:tensor_product}Schematic of $p=4$ tensor-product point layout with Gauss--Legendre points.}
    \end{figure}

    The most complex aspect of the proposed optimisation is the fusion of a pointwise kernel and a matrix multiplication kernel. This is due to the high data reuse and the complexity of the flux functions. Due to this complexity, the simplification of only considering tensor-product elements is made, specifically maximal-order hexahedral elements. It was found in practice that quadrilateral elements were significantly less challenging due to lower memory requirements. \cref{fig:tensor_product} aims to show the benefit of tensor production point sets, when combined with a maximal order basis~\citep{Trefethen2017} many operations simplify to acting along the lines of the points. A further simplification shall be made in assuming that for both the base and fused kernels the Jacobian terms, $\mathsf{S}_s$, will be taken as constant. This will reduce code complexity in an already complex system, and given that both the base and fused kernels will require this data, it is deemed to be a reasonable simplification. Future work will be required to include these terms. 
    
    % raw54 - why are the Jacobian terms S_s in the paragraph above?
    % wt247 - what do you mean?
    % raw54 - I'm not entirely sure, actually, it seems pretty clear now
    
    Based on these assumptions, it is possible to build a simple model for the expected performance gains. If it is a fair assumption that all kernels are memory bandwidth bound, then the resulting speedup can be estimated as:
    \begin{equation}
        \text{speedup} = \frac{\text{Data I/O}_\text{old}}{\text{Data I/O}_\text{new}}
    \end{equation}
    Considering stages 2 and 3, the baseline method will require a minimum of four reads and four writes to global memory. By fusing these steps, this can be reduced to a requirement of a minimum of one read and one write. Hence, if both kernels are bandwidth bound, and the baseline method has no additional global memory access, then the optimal speedup that can be achieved is $4\times$. If the source term is also fused into the kernel, again assuming it is fully memory bandwidth bound and using the source term of \cref{eq:acmhd_nd}, then the optimal theoretical speedup of a single kernel replacing stages 2, 3, and 6 would be $5.346\times$.
    
\subsection{GPU Memory Hierarchy}
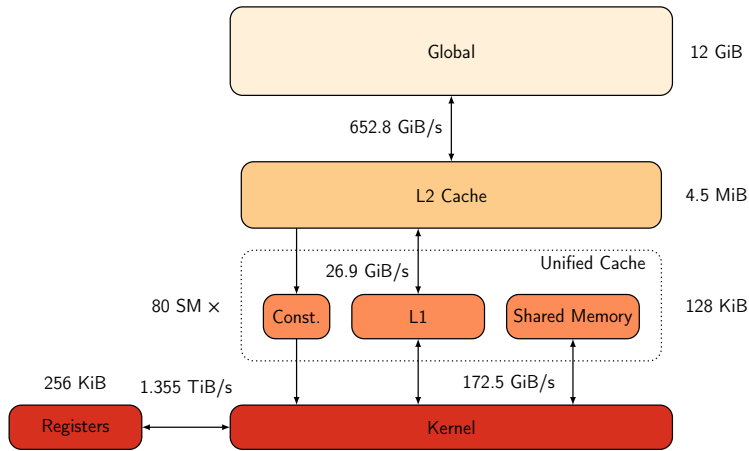
\begin{figure}[tbhp]
        \centering
        \adjustbox{width=0.6\linewidth,valign=b}{\begin{tikzpicture}[scale=1, rotate=90]

   % \draw[black, rounded corners=2ex] (-4,0) rectangle (-2,10) node[pos=.5, rotate=-90]{L2 Cache};
    %\draw[black, rounded corners=2ex] (-4,0) rectangle (-2,10) node[pos=.5, rotate=-90]{L2 Cache};
    \draw[black, fill={OrRd-J}, rounded corners=2ex, thick] (-8,0) rectangle (-7,10) node[pos=.5]{\large{Kernel}};
    \draw[black, fill={OrRd-J}, rounded corners=2ex, thick] (-8,12) rectangle (-7,15) node[pos=.5]{\large{Registers}};
    
    \draw[black, fill=none, rounded corners=2ex, dotted, thick] (-6,0.25) rectangle (-3.5,9.75);
    \node[black] at (-3.775,1.8) {\large{Unified Cache}};
    \draw[black, fill={OrRd-G}, rounded corners=2ex, thick] (-5.5,7.75) rectangle (-4.5,9.25) node[pos=.5]{\large{Const.}};
    \draw[black, fill={OrRd-G}, rounded corners=2ex, thick] (-5.5,4.25) rectangle (-4.5,7.25) node[pos=.5]{\large{L1}};
    \draw[black, fill={OrRd-G}, rounded corners=2ex, thick] (-5.5,0.75) rectangle (-4.5,3.75) node[pos=.5]{\large{Shared Memory}};

    \draw[black, latex-latex, thick] (0,5) -- (-1.5,5); % glb - l2
    \draw[black, latex-latex, thick] (-3,5.75) -- (-4.5,5.75); % l2 - l1
    \draw[black, -latex, thick] (-3,8.5) -- (-4.5,8.5); % l2 - const.
    
    \draw[black, -latex, thick] (-5.5,8.5) -- (-7,8.5); % const. - kernel
    \draw[black, latex-latex, thick] (-5.5,5.75) -- (-7,5.75); % l1 - kernel
    \draw[black, latex-latex, thick] (-5.5,2.25) -- (-7,2.25); % shared - kernel
    \draw[black, latex-latex, thick] (-7.5,12) -- (-7.5,10); % registers - kernel
    
    \draw[black, fill={OrRd-E}, rounded corners=2ex, thick] (-3,0.25) rectangle (-1.5,9.75) node[pos=.5]{\large{L2 Cache}};
    
    \draw[black, fill={OrRd-B}, rounded corners=2ex, thick] (0,0) rectangle (2,10) node[pos=.5]{\large{Global}};

    \node[black] at (-4.75,11) {\textsf{\large{80 SM $\times$}}};

    \node[black] at (-4.75, -1) {\textsf{\large{128 KiB}}};
    \node[black] at (-2.25, -1) {\textsf{\large{4.5 MiB}}};
    \node[black] at (1, -1) {\textsf{\large{12 GiB}}};
    \node[black] at (-6.5, 13.5) {\textsf{\large{256 KiB}}};
    
    \node[black] at (-0.75, 6.25) {\textsf{\large{652.8 GiB/s}}};
    \node[black] at (-6.5, 3.7) {\textsf{\large{172.5 GiB/s}}};
    \node[black] at (-4, 6.9) {\textsf{\large{26.9 GiB/s}}};
    \node[black] at (-6.625, 11) {\textsf{\large{1.355 TiB/s}}};

\end{tikzpicture}}
        
        % \subfloat[Key NVIDIA V100/Titan V GPU stats.]{\label{tab:gpu_stats}\adjustbox{valign=b}{\begin{tabular}{r r}\toprule
        %     Max threads / warp & 32 \\
        %     Max warps / SM & 64 \\
        %     Max FP32 registers / thread & 255 \\
        %     Register file size / SM & \SI{256}{\kibi\byte} \\
        %     Max shared memory / SM & \SI{48}{\kibi\byte} (\SI{96}{\kibi\byte})\\
        %     Max L1 cache / SM & \SI{128}{\kibi\byte}\\\bottomrule 
        % \end{tabular}}}
        \caption{\label{fig:gpu_stats}Simplified NVIDIA TITAN V GPU memory overview and statistics. The bandwidth and memory capacities at the UC level and below are calculated per SM.}
    \end{figure}

    To design an approach for fusing these kernels, it is first useful to understand the memory hierarchy employed by recent NVIDIA GPUs~\citep{Nvidia2017,Nvidia2020}. \cref{fig:gpu_stats} shows a simplified view of the hierarchy of memory and caches. Stream multiprocessors (SM) are the way computation is broken down on a Volta generation GPU. Each SM has 4 schedulers, each capable of issuing instructions to 4 \emph{warps} with a warp being a group of 32 threads. The unified cache (UC) sits at the SM level, and is partitioned into several spaces. The three important partitions for this application are: constant memory (\texttt{cmem}); L1 cache; and shared memory (\texttt{smem}); with texture and surface memory not applicable here. The constant memory partition is read only and has a fixed size --- however, the split between L1 and shared memory is configurable. The maximum configurable sizes for the Volta series are shown in \cref{fig:gpu_stats}\footnote{It should be noted that the constant cache does extend up the cache hierarchy, however, here use is only made of a small quantity, and hence the discussion is restricted to L1.}. To achieve the maximum shared memory allocation of \SI{96}{\kibi\byte} per SM, dynamic shared memory has to be used, with the allocation specified at runtime. To maximise memory throughput, a mixture of the UC memory types should be used, reducing pressure on any single memory and maximising available bandwidth.
    
    % \begin{table}[tbhp]
    %     \caption{\label{tab:baseline}Run times for the flux and partially corrected flux divergence kernels, for various orders and precisions.}
    %     \centering
    %     \begin{tabular}{c c r r | r}
    %         \toprule
    %         $p$ & Precision & Flux $[\SI{}{\micro\second}]$ & Div $[\SI{}{\micro\second}]$ & Total $[\SI{}{\micro\second}]$ \\ \midrule
    %         \multirow{2}{*}{1} & single & 103.296 & 88.928 & 192.224\\
    %          & double & 205.312 & 176.224 & 381.536\\
    %         \multirow{2}{*}{2} & single & 370.176 & 305.216 & 675.392\\
    %          & double & 710.560 & 602.080 & 1312.640\\
    %         \multirow{2}{*}{3} & single & 897.792 & 726.784 & 1624.578\\
    %          & double & 1697.568 & 1426.496 & 3124.064\\
    %         \multirow{2}{*}{4} & single & 1794.560 & 1419.264 & 3213.824\\
    %          & double & 3338.592 & 4369.024 & 7707.616\\
    %         \multirow{2}{*}{5} & single & 3139.360 & 2789.152 & 5928.512\\
    %          & double & 5781.216 & 11694.848 & 17476.064\\
    %         \multirow{2}{*}{6} & single & 5013.472 & 8499.776 & 13513.248\\ 
    %          & double & 9172.544 & 22398.720 & 31571.264\\\bottomrule
    %     \end{tabular}
    % \end{table}
\section{\label{sec:method}Methods}
    Two kernel fusion approaches have been developed, following a similar pattern to that seen in \citet{Swirydowicz2019}. These two methods differ in their degree of parallelism, and, consequently, in their shared memory requirements. The development of the two differing methods was motivated by a key challenge that was encountered: namely, for a fixed problem size as order increases, the cardinality of a points stencil grows. This means that the optimal level of parallelism will be dependent on order. At lower orders, the resources required to evaluate the contribution from a single tensor-product line is lower, and, therefore, to give threads sufficient work, lower parallelism and fully unrolling the kernels can be beneficial. However, as the contribution along the tensor product line becomes more significant, the resources required become higher --- and hence increased parallelism and not unrolling the operation can be beneficial. 
    
\subsection{Planar Method}\label{ssec:planar}
% raw54 - Check that the second sentence below is correct, I may have changed its meaning inadvertently.
    Due to the re-use of data within a flux function evaluation, it is logical that a collection of $N_{et}$ threads will perform all the calculation associated with a single element. This reduces the global loads and keeps the storage of an element to a single block. Furthermore, it would be ideal if all the data could be read into shared memory, which has significantly higher bandwidth and lower latency. This was the approach taken in the related work of \citet{Swirydowicz2019}. For a tensor product element, this would then require the following number of entries:
    \begin{equation*}
        \text{Shared memory} = \frac{N_tn_vn_s}{N_{et}} = \frac{N_t(1 + d + d^2)(m^d)}{N_{et}},
    \end{equation*}
    where $N_t$ is the total number of threads per block and $m=p+1$. As an example of the memory usage required for storing the entire element in shared memory by this method, take $N_t=64$, $m=4$, $N_{et}=m$, and $d=3$. This would require approximately \SI{53}{\kibi\byte} FP-32 or \SI{106}{\kibi\byte} FP-64 of shared memory. It is clear that, due to the high number of variables, there is a considerable demand on the shared memory, and, at higher values of  $m$, it will become unfeasible to read the entire element in to shared. 
    
    \begin{figure}[tbhp]
        \centering
        \subfloat[Arrangement of planes.]{\label{fig:plane_or}\adjustbox{width=0.4\linewidth,valign=b}{\begin{tikzpicture}[scale=1]

    \draw[black!00, fill={Pastel1-A}, fill opacity=0.5, thick] (0,0) -- (0.866,0.500) -- (0.866, 2.5) -- (0, 2) -- (0,0);
    \draw[black!00, fill={Pastel1-A}, fill opacity=0.5, thick] (1,0) -- (1.866,0.500) -- (1.866, 2.5) -- (1, 2) -- (1,0);
    \draw[black!00, fill={Pastel1-A}, fill opacity=0.5, thick] (2,0) -- (2.866,0.500) -- (2.866, 2.5) -- (2, 2) -- (2,0);
    \draw[black!00, fill={Pastel1-A}, fill opacity=0.5, thick] (3,0) -- (3.866,0.500) -- (3.866, 2.5) -- (3, 2) -- (3,0);
    
    \begin{scope}[on above layer]
        \draw[black, fill=black] (0.025,0.025) circle (0.05);
        \draw[black, fill=black] (0.286,0.165) circle (0.05);
        \draw[black, fill=black] (0.572,0.33) circle (0.05);
        \draw[black, fill=black] (-0.025+0.866,-0.025+0.50) circle (0.05);
        
        \draw[black, fill=black] (0.025,0.025+0.666) circle (0.05);
        \draw[black, fill=black] (0.286,0.165+0.666) circle (0.05);
        \draw[black, fill=black] (0.572,0.33+0.666) circle (0.05);
        \draw[black, fill=black] (-0.025+0.866,-0.025+0.50+0.666) circle (0.05);
        
        \draw[black, fill=black] (0.025,0.025+1.333) circle (0.05);
        \draw[black, fill=black] (0.286,0.165+1.333) circle (0.05);
        \draw[black, fill=black] (0.572,0.33+1.333) circle (0.05);
        \draw[black, fill=black] (-0.025+0.866,-0.025+0.50+1.333) circle (0.05);
        
        \draw[black, fill=black] (0.025,2.025) circle (0.05);
        \draw[black, fill=black] (0.286,2.165) circle (0.05);
        \draw[black, fill=black] (0.572,2.33) circle (0.05);
        \draw[black, fill=black] (-0.025+0.866,-0.025+2.50) circle (0.05);
    \end{scope}
    
    \draw[black, -latex, thick] (0,-0.3) -- (0.8,-0.3) node[pos=1.1, below] {i};
    \draw[black, -latex, thick] (-0.3,0) -- (-0.3, 0.8) node[pos=1.1, below, xshift=-0.75em] {k};
    \draw[black, -latex, thick] (3.5,0) -- (4.106, 0.35) node[pos=1.1, below, xshift=0.2em] {j};

\end{tikzpicture}}}
        ~
        \subfloat[Memory stencil for first point: purple - register, green - shared, blue- global.]{\label{fig:plane_mem}\adjustbox{width=0.4\linewidth,valign=b}{\begin{tikzpicture}[scale=1]

    \begin{scope}[on behind layer]
        \draw[Pastel1-A, fill=Pastel1-A, opacity=0.5, thick] (0,0) -- (0.866,0.500) -- (0.866, 2.5) -- (0, 2) -- (0,0);
    \end{scope}
    
    \begin{scope}[on above layer]
        \draw[black, fill=black] (0.025,0.025) circle (0.05);
        \draw[black, fill=black] (0.286,0.165) circle (0.05);
        \draw[black, fill=black] (0.572,0.33) circle (0.05);
        \draw[black, fill=black] (-0.025+0.866,-0.025+0.50) circle (0.05);
        
        \draw[black, fill=black] (0.025,0.025+0.666) circle (0.05);
        \draw[black, fill=black] (0.025,2.025) circle (0.05);
        \draw[black, fill=black] (0.025,0.025+1.333) circle (0.05);
        \draw[black, fill=black] (1.025,0.025) circle (0.05);
        \draw[black, fill=black] (2.025,0.025) circle (0.05);
        \draw[black, fill=black] (3.025,0.025) circle (0.05);
        
        \draw[black, fill=black, opacity=0.07] (0.286,0.165+0.666) circle (0.05);
        \draw[black, fill=black, opacity=0.07] (0.572,0.33+0.666) circle (0.05);
        \draw[black, fill=black, opacity=0.07] (-0.025+0.866,-0.025+0.50+0.666) circle (0.05);
        
        \draw[black, fill=black, opacity=0.07] (0.286,0.165+1.333) circle (0.05);
        \draw[black, fill=black, opacity=0.07] (0.572,0.33+1.333) circle (0.05);
        \draw[black, fill=black, opacity=0.07] (-0.025+0.866,-0.025+0.50+1.333) circle (0.05);
        
        \draw[black, opacity=0.07, fill=black] (0.286,2.165) circle (0.05);
        \draw[black, opacity=0.07, fill=black] (0.572,2.33) circle (0.05);
        \draw[black, opacity=0.07, fill=black] (-0.025+0.866,-0.025+2.50) circle (0.05);

    \end{scope}
    
    \begin{scope}
        
        \draw[Pastel1-B, fill={Pastel1-B}, opacity=0.7, rounded corners=0.5ex] (-0.08+0.025,-0.08+0.025) rectangle (3.08+0.025,0.08+0.025);
        
        \draw[Pastel1-C, fill={Pastel1-C}, opacity=0.7, rounded corners=0.5ex] (-0.08+0.025,-0.08+0.025) rectangle (0.08+0.025,2.08+0.025);
        
        \draw[Pastel1-D, fill={Pastel1-D}, opacity=0.7, rounded corners=0.5ex, rotate=30] (-0.08+0.025,-0.1+0.025) rectangle (1.08+0.025,0.07+0.025);
    \end{scope}
\end{tikzpicture}}}
        \caption{\label{fig:planar_method}Schematic for planar method.}
    \end{figure}
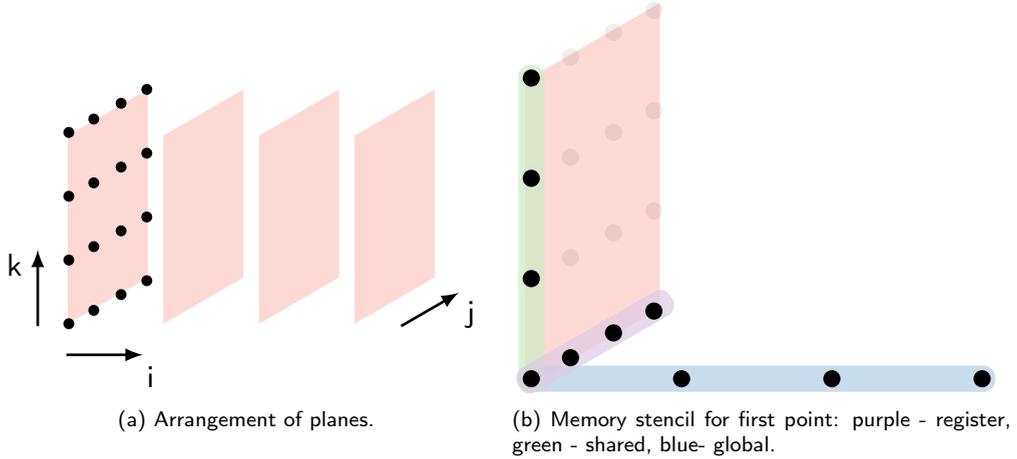
    
    Instead, the first approach proposed here is a planar method, where $N_{et} = m$. So, for $d=3$, each thread computes an x-y plane of the tensor product element. In this method, instead of an entire element, a single plane will be loaded into shared at a time. \cref{fig:plane_or} shows the cardinal axes and how a tensor product element is broken into planes. Sequentially, $y$-$z$ planes are loaded into shared memory, and each thread calculates the flux divergence for all points along a y-line. To reduce the pressure on the shared memory, some variables can be explicitly stored in registers, and as the threads are working along a $y$-line these are a sensible candidate, the y-lines are also how threads load variables into shared memory. Finally, to calculate the flux divergence contribution in $x$, values are needed that are not resident in either the register file or shared memory, and therefore global loads are used. The memory hierarchy for a single thread is shown in \cref{fig:plane_mem} for the first point on the $y$-line. The complete algorithm used here is shown in \cref{alg:plane}. As set out in the introduction, the GiMMiK framework allows for entire matrix multiplication kernels to be fully unrolled, these unrolled kernels gave significantly increased performance for this method compared to without explicit unrolling. Therefore, all planar method kernels will be fully unrolled.
    
    \begin{algorithm}[tbhp]
        \caption{\label{alg:plane}Fusion of flux evaluation and divergence calculation with one thread per x-y plane.}
        \begin{algorithmic}[1]
        \Procedure{FusedPlane}{$N,\mathsf{U}_s,-\nabla\cdot\mathsf{F}_s^a$}
            \State Set global element number, $e_g$.
            \State $k := \texttt{((threadIdx.x) \% warp\_size) \% (p+1)}$
            \If{$e_g < N$}
                \For{$i\in\{0,\dots,p\}$} \Comment{Loop over $y-z$ planes}
                    \For{$j\in\{0,\dots,p\}$} \Comment{Loop over points on $y$-line}
                        \State Store $k$-th $y$-$z$ plane of $\mathsf{U}_s$ in \texttt{smem}.
                        \State Accumulate $x$-line contribution to $\nabla\cdot\mathsf{F}_s^a(i,j,k)$ from \texttt{gmem}
                        \State Accumulate $y$-line contribution to $\nabla\cdot\mathsf{F}_s^a(i,j,k)$ from \texttt{rmem}
                        \State Barrier, arrive, and wait
                        \State Accumulate $z$-line contribution to $\nabla\cdot\mathsf{F}_s^a(i,j,k)$ from \texttt{smem}
                    \EndFor
                \EndFor
            \EndIf
        \EndProcedure
        
        \end{algorithmic}
    \end{algorithm}
    
    \subsubsection{\label{ssec:opt_static}Static Optimisations}
    There are several optimisations that can be applied to \cref{alg:plane} in an attempt to further reduce run-time. At this stage there are four which we will consider. These are:
    
    \begin{enumerate}
        \item As \cref{fig:planar_method} makes clear, there is some overlap in which a register can be used instead of performing a global load. One optimisation is to consider using the register when possible. 
        \item Access to shared memory at the warp level is banked, where the bank number is calculated for single-precision using the following formula:
        \begin{equation}
            \mathrm{bank} = \mathrm{mod}\left(\frac{\mathrm{addr.}}{4}, 32 \right).
        \end{equation}
        It is therefore possible for all 32 threads in a warp to access shared memory in parallel; however, if the access of two or more threads requires the same bank, the access will be serialised, reducing bandwidth. To overcome this, a constant offset can be added to each element in a warp, such that bank conflicts can be reduced to theoretically zero. This is only theoretical, as with a fully unrolled kernel some conflicts can occur due to the compiler reordering operations.
        \item The Volta series of GPUs have \SI{64}{\kibi\byte} of constant memory (\texttt{cmem}), which uses a separate read only cache, spread throughout the cache hierarchy. In the case of a cache hit, this can reduce the need to stream constants through the instruction cache. Therefore, one strategy is to explicitly load the divergence matrix constants into \texttt{cmem}.
        \item When directly loading a global variable into shared, it will be implicitly stored in a register within the kernel by the compiler. This is to facilitate several optimisation steps, namely if a variable in shared memory is only used by a single thread, then it may be possible to optimise away the shared load. A further and more common occurrence is that a variable is only guaranteed to be in shared after a thread sync --- hence loading a variable into a register first will enable the compiler to better optimise the shared access. However, in the planar algorithm, we know \emph{a priori} that $y$-lines will be stored in registers. Therefore, global variables can first be explicitly stored in registers then in shared, removing the use of implicit registers. 
    \end{enumerate}
    
    To test the effectiveness of these optimisations, we use a standardised test case at $p=3$ with $1024\times32$ elements and a block size of $128$ threads. As previously reported by \citet{Trojak2021b}, the performance of test kernels can be significantly different if the problem fits in its entirety into L2 cache. The number of elements used here is thus made sufficiently large so that, for all orders $p\geqslant1$, the L2 cache size of \SI{4.5}{\mebi\byte} is exceeded. Tests were performed on a NVIDIA Titan V using the Nsight Compute profiling suite, and CUDA V11.2.
    
    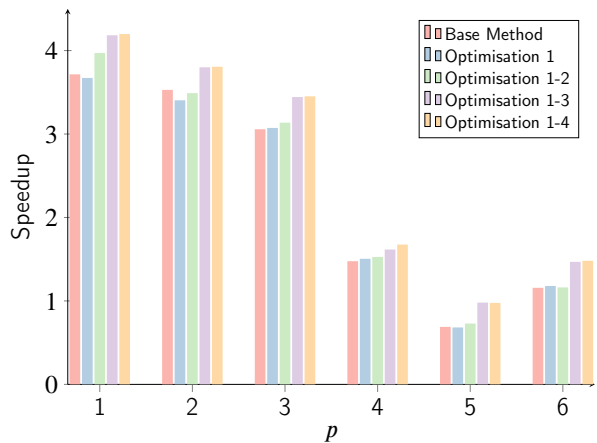
\begin{figure}[thbp]
        \centering
        \adjustbox{width=0.477\linewidth,valign=b}{\begin{tikzpicture}[spy using outlines={rectangle, height=3cm,width=2.5cm, magnification=3, connect spies}]
    \pgfplotsset{scaled x ticks=false}
    \begin{axis}
    [
        axis line style={latex-latex},
        axis y line=left,
        axis x line=left,
        width=12cm,height=9cm,
        ybar=1pt,
        bar width=6pt,
        xmode=linear, % not log
        ymode=linear, % not log
        xlabel = {$p$},
        ylabel = {Speedup},
        ymin = 0, ymax = 4.5,
        symbolic x coords={1,2,3,4,5,6},
        xtick={1,2,3,4,5,6},
        enlarge x limits={abs=3*\pgfplotbarwidth},
        legend cell align={left},
        legend style={font=\normalsize, at={(0.97, 0.97)},anchor=north east},
        %nodes near coords,
        %nodes near coords align={vertical},
        x tick label style={/pgf/number format/.cd, fixed, fixed zerofill, precision=0, /tikz/.cd},
        y tick label style={/pgf/number format/.cd, fixed, fixed zerofill, precision=0, /tikz/.cd},
        label style={font=\Large},
        tick label style={font=\Large},
    ]
        
        \addplot[fill={Pastel1-A}, draw=none] coordinates {(1,3.715) (2,3.529) (3,3.057) (4,1.476) (5,0.689) (6,1.157)};
        \addplot[fill={Pastel1-B}, draw=none] coordinates {(1,3.672) (2,3.404) (3,3.073) (4,1.505) (5,0.682) (6,1.179)};
        \addplot[fill={Pastel1-C}, draw=none] coordinates {(1,3.970) (2,3.490) (3,3.136) (4,1.527) (5,0.729) (6,1.162)};
        \addplot[fill={Pastel1-D}, draw=none] coordinates {(1,4.183) (2,3.799) (3,3.443) (4,1.616) (5,0.980) (6,1.468)};
        \addplot[fill={Pastel1-E}, draw=none] coordinates {(1,4.198) (2,3.806) (3,3.452) (4,1.675) (5,0.976) (6,1.481)};
        \legend{Base Method,Optimisation 1, Optimisation 1-2,Optimisation 1-3,Optimisation 1-4};
    \end{axis}
\end{tikzpicture}}
        \caption{\label{fig:um_opt}Effect of sequentially applying optimisation strategies 1-4 to the planar method with $128$ threads per block at FP-32.}
    \end{figure}
    
    \cref{fig:um_opt} shows the cumulative effect of applying these optimisation strategies on performance. It is clear that the largest improvement is made by strategy 3 --- moving the matrix multiplication constants into constant memory. In the GiMMiK package, constant memory is not generally explicitly used, as it often negatively impacts performance. However, the operator matrix used to calculate the divergence on a tensor product element has only $(p+1)^2$ or fewer unique values. This is reduced to $\ceil{(p+1)^2/2}$ if the sign is removed. Hence, due to the small number of constants compared to more general sparse matrix multiplication kernels, this optimisation is effective here. 
    
    The other optimisation strategies were found to have only limited effects, although they generally led to small improvements. The exception to this was the use of optimisation strategy 1 at low $p$, where the cache pressure is lower. At these low values of $p$, loads are reduced in favour of registers, thereby increasing the register pressure, and overall reducing the performance. 
    
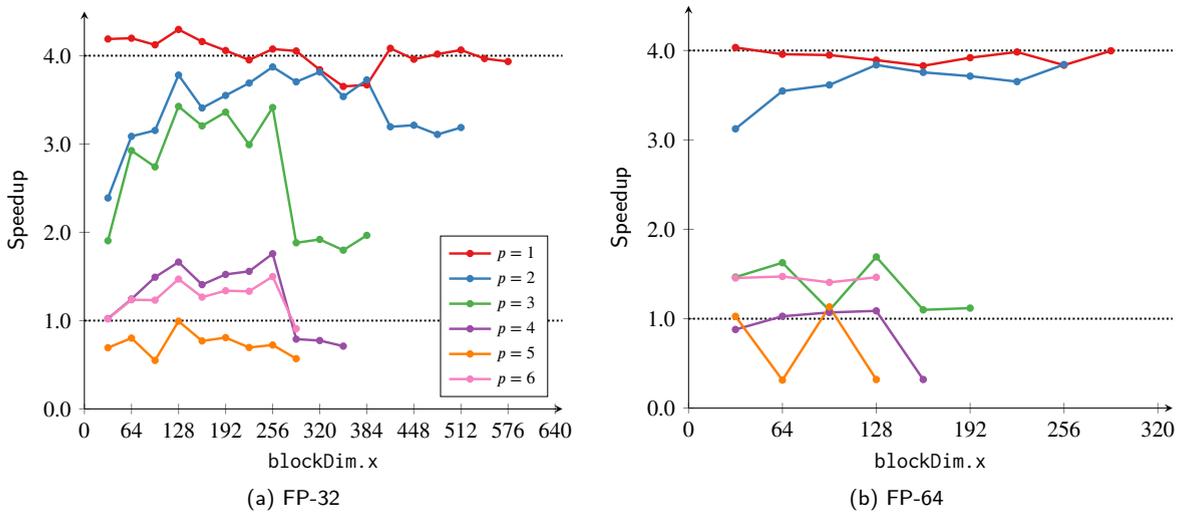
\begin{figure}[tbhp]
        \centering
        \subfloat[FP-32]{\adjustbox{width=0.47\linewidth,valign=b}{\begin{tikzpicture}[spy using outlines={rectangle, height=3cm,width=2.5cm, magnification=3, connect spies}]
\definecolor{agrey}{rgb}{0.6,0.6,0.6}
    \pgfplotsset{scaled x ticks=false}
    \begin{axis}
    [
        axis line style={latex-latex},
        axis y line=left,
        axis x line=left,
        xmode=linear, % not log
        ymode=linear, % not log
        xlabel = {$\texttt{blockDim.x}$},
        ylabel = {Speedup},
        xmin = 0, xmax = 650,
        ymin = 0, ymax = 4.5,
        xtick = {0, 64, 128, 192, 256, 320, 384, 448, 512, 576, 640},
        legend cell align={left},
        legend style={font=\scriptsize, at={(0.97, 0.03)},anchor=south east},
        %axis line style={draw=none},
        %tick style={draw=none},
        x tick label style={/pgf/number format/.cd, fixed, fixed zerofill, precision=0, /tikz/.cd},
        y tick label style={/pgf/number format/.cd, fixed, fixed zerofill, precision=1, /tikz/.cd},
    ]
        \addplot[thick, black, densely dotted, forget plot] coordinates{(0,4) (650,4)};
        \addplot[thick, black, densely dotted, forget plot] coordinates{(0,1) (650,1)};
    
        \addplot[line width=1pt,color={Set1-A}, mark=*, mark options={solid},mark size=1pt] table[x expr=(\thisrowno{6}), y expr=(192224/\thisrowno{17}), 
                                                      col sep=comma]{./figs/data/unmanaged/bdc/gimmik_data_p1_fp32.csv};
        \addlegendentry{$p=1$}
                                                      
        \addplot[line width=1pt,color={Set1-B}, mark=*, mark options={solid},mark size=1pt] table[x expr=(\thisrowno{6}), y expr=(675392/\thisrowno{17}), 
                                                      col sep=comma]{./figs/data/unmanaged/bdc/gimmik_data_p2_fp32.csv};
        \addlegendentry{$p=2$}
                                                      
        \addplot[line width=1pt,color={Set1-C}, mark=*, mark options={solid},mark size=1pt] table[x expr=(\thisrowno{6}), y expr=(1624578/\thisrowno{17}), 
                                                      col sep=comma]{./figs/data/unmanaged/bdc/gimmik_data_p3_fp32.csv};
        \addlegendentry{$p=3$}
                                                      
        \addplot[line width=1pt,color={Set1-D}, mark=*, mark options={solid},mark size=1pt] table[x expr=(\thisrowno{6}), y expr=(3213824/\thisrowno{17}), 
                                                      col sep=comma]{./figs/data/unmanaged/bdc/gimmik_data_p4_fp32.csv};
        \addlegendentry{$p=4$}
                                                      
        \addplot[line width=1pt,color={Set1-E}, mark=*, mark options={solid},mark size=1pt] table[x expr=(\thisrowno{6}), y expr=(5928512/\thisrowno{17}), 
                                                      col sep=comma]{./figs/data/unmanaged/bdc/gimmik_data_p5_fp32.csv};
        \addlegendentry{$p=5$}
                                                      
        \addplot[line width=1pt,color={Set1-H}, mark=*, mark options={solid},mark size=1pt] table[x expr=(\thisrowno{6}), y expr=(13513248/\thisrowno{17}), 
                                                      col sep=comma]{./figs/data/unmanaged/bdc/gimmik_data_p6_fp32.csv};
        \addlegendentry{$p=6$}
    \end{axis}
\end{tikzpicture}}}
        ~
        \subfloat[FP-64]{\adjustbox{width=0.47\linewidth,valign=b}{\begin{tikzpicture}[spy using outlines={rectangle, height=3cm,width=2.5cm, magnification=3, connect spies}]
\definecolor{agrey}{rgb}{0.6,0.6,0.6}
    \pgfplotsset{scaled x ticks=false}
    \begin{axis}
    [
        axis line style={latex-latex},
        axis y line=left,
        axis x line=left,
        xmode=linear, % not log
        ymode=linear, % not log
        xlabel = {$\texttt{blockDim.x}$},
        ylabel = {Speedup},
        xmin = 0, xmax = 330,
        ymin = 0, ymax = 4.5,
        xtick = {0, 64, 128, 192, 256, 320, 384, 448, 512, 576, 640},
        legend cell align={left},
        legend style={font=\scriptsize, at={(0.97, 0.03)},anchor=south east},
        %axis line style={draw=none},
        %tick style={draw=none},
        x tick label style={/pgf/number format/.cd, fixed, fixed zerofill, precision=0, /tikz/.cd},
        y tick label style={/pgf/number format/.cd, fixed, fixed zerofill, precision=1, /tikz/.cd},
    ]
        \addplot[thick, black, densely dotted, forget plot] coordinates{(0,4) (650,4)};
        \addplot[thick, black, densely dotted, forget plot] coordinates{(0,1) (650,1)};
    
        \addplot[line width=1pt,color={Set1-A}, mark=*, mark options={solid},mark size=1pt] table[x expr=(\thisrowno{6}), y expr=(381536/\thisrowno{17}), 
                                                      col sep=comma]{./figs/data/unmanaged/bdc/gimmik_data_p1_fp64.csv};
        %\addlegendentry{$p=1$}
                                                      
        \addplot[line width=1pt,color={Set1-B}, mark=*, mark options={solid},mark size=1pt] table[x expr=(\thisrowno{6}), y expr=(1312640/\thisrowno{17}), 
                                                      col sep=comma]{./figs/data/unmanaged/bdc/gimmik_data_p2_fp64.csv};
        %\addlegendentry{$p=2$}
                                                      
        \addplot[line width=1pt,color={Set1-C}, mark=*, mark options={solid},mark size=1pt] table[x expr=(\thisrowno{6}), y expr=(3124064/\thisrowno{17}), 
                                                      col sep=comma]{./figs/data/unmanaged/bdc/gimmik_data_p3_fp64.csv};
        %\addlegendentry{$p=3$}
                                                      
        \addplot[line width=1pt,color={Set1-D}, mark=*, mark options={solid},mark size=1pt] table[x expr=(\thisrowno{6}), y expr=(7707616/\thisrowno{17}), 
                                                      col sep=comma]{./figs/data/unmanaged/bdc/gimmik_data_p4_fp64.csv};
        %\addlegendentry{$p=4$}
                                                      
        \addplot[line width=1pt,color={Set1-E}, mark=*, mark options={solid},mark size=1pt] table[x expr=(\thisrowno{6}), y expr=(17476064/\thisrowno{17}), 
                                                      col sep=comma]{./figs/data/unmanaged/bdc/gimmik_data_p5_fp64.csv};
        %\addlegendentry{$p=5$}
                                                      
        \addplot[line width=1pt,color={Set1-H}, mark=*, mark options={solid},mark size=1pt] table[x expr=(\thisrowno{6}), y expr=(31571264/\thisrowno{17}), 
                                                      col sep=comma]{./figs/data/unmanaged/bdc/gimmik_data_p6_fp64.csv};
        %\addlegendentry{$p=6$}
    \end{axis}
\end{tikzpicture}}}
        \caption{\label{fig:umanaged_rt}Unmanaged kernel runtime vs. block size.}
    \end{figure}

    For this previous series of tests, a block size of 128 threads was used, as, in our experience with PyFR, this value typically gives optimal performance. For the planar method, however, changing the block size will vary the number of active warps per SM due to the shared memory constraints. \cref{fig:umanaged_rt} shows the effect on the speedup as the block size is varied, for the planar method with all optimisation applied. Firstly, for single-precision, 128 threads per block is near optimal for all orders. Secondly, the double-precision results with this approach are poor for $p>2$. This is due to the high register usage of unrolled methods coupled to the larger 64-bit variable, requiring double the space in the register file. If register usage demands more registers than are available, then a process called spillage occur. In this process, space in the register file is freed by writing some of its contents to local memory in the L1 cache. This uses the same memory pathways as global memory access and so can reduce the bandwidth available for global memory transactions. Once the register limit is reached, spillage can quickly become significant. The consequence, due to the high local and global memory usage, is that the primary warp stall reason was given as waiting on scoreboard dependency (\say{Stall Long Scoreboard}). This means that threads were primarily stalled waiting for memory for long latency memory access, such as global or local memory.

\subsubsection{Greedy Memory Manager}
    The benchmarks displayed in \cref{fig:umanaged_rt} show that the planar approach taken in three dimensions can achieve close to peak performance in some cases. It does not, however, make full use of the shared memory available. Consequently, the number of global reads per variable is fixed at $p+2$ or $p+1$, depending on the use of optimisation strategy 1. The resulting latency was hidden somewhat through caching and compute interleaving performed either by the programmer or compiler; however, as detailed in the previous section, the primary stall mechanism was reported as long scoreboard stalls. Therefore, performance may be further improved by increasing shared memory usage, reducing the pressure on the cache. 
    
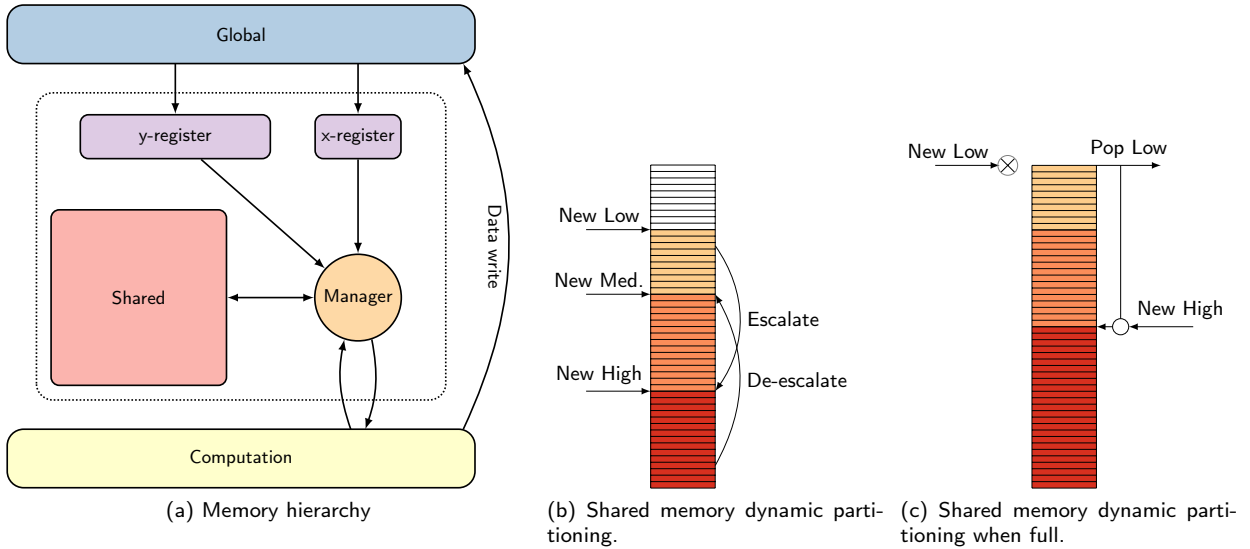
\begin{figure}[tbhp]
        \centering
        \subfloat[Memory hierarchy]{\label{fig:mem_arc}\adjustbox{width=0.42\linewidth,valign=b}{\begin{tikzpicture}
    
    \tikzstyle{memory}=[draw,rectangle,rounded corners=1ex, thick, color={black},
                      text=black,minimum width=1cm,minimum height=0.75cm]
                      
    \tikzstyle{manager}=[draw,circle, thick, color={black},
                      text=black,minimum width=0.75cm,minimum height=0.75cm]
                      
    \tikzstyle{data_r}=[-latex,black,thick]
    \tikzstyle{data_rw}=[latex-latex,black,thick]

    \draw[black, fill=Pastel1-B, rounded corners=2ex, thick] (0,0) rectangle (8,1) node[pos=.5]{Global};
    \node[memory,rounded corners=2ex,fill=Pastel1-F,minimum width=8cm,minimum height=1cm] at (4,-6.75) (comp) {Computation};
    
    \begin{scope}[on behind layer]
        \draw[black, densely dotted, rounded corners=2ex, thick] (0.5,-5.75) rectangle (7.5,-0.5);
    \end{scope}
    
    \begin{scope}
        %\draw[black, rounded corners=1ex, thick] (1.25,-1.5) rectangle (4,-0.75) node[pos=.5]{y register};
        %\draw[black, rounded corners=1ex, thick] (5.25,-1.5) rectangle (6.75,-0.75) node[pos=.5]{x register};
        \node[memory,fill=Pastel1-D,minimum width=3.25cm] at (2.875,-1.25) (ry) {y-register};
        \node[memory,fill=Pastel1-D] at (6,-1.25) (rx) {x-register};
        \node[memory,fill=Pastel1-A,minimum width=3cm,minimum height=3.00cm] at (2.25,-4) (s1) {Shared};
        
        \node[manager,fill=Pastel1-E] at (6,-4) (mm) {Manager};
        
        \draw[data_r] (ry) -- (mm);
        \draw[data_r] (rx) -- (mm);
        \draw[data_rw] (s1) -- (mm);
        \draw[data_r] (2.875,0) -- (ry);
        \draw[data_r] (6,0) -- (rx);
        \path[data_r] (7.8,-6.25) edge[bend right=25] node [left,rotate=-90,midway,below] {Data write} (7.8,0);
        
        \path[data_r] (mm) edge[bend right=-15] node [right] {} (6.125,-6.25);
        \path[data_r] (5.875,-6.25) edge[bend right=-15] node [right] {} (mm);
    \end{scope}

    %\draw[black, rounded corners=2ex, thick] (0,-7.25) rectangle (8,-6.25) node[pos=.5]{Computation};

\end{tikzpicture}}}
        ~
        \subfloat[Shared memory dynamic partitioning.]{\label{fig:shr_part}\adjustbox{width=0.27\linewidth,valign=b}{\begin{tikzpicture}[scale=1]
    \draw[black!00] (-1.5,5.25)--(3.5,5.25);

    \begin{scope}
	    \draw[black] (0,0) grid[xstep=1,ystep=0.1] (1,1.5);
	    \draw[black] (0,1.5) grid[xstep=1,ystep=0.1] (1,3);
	    \draw[black] (0,3) grid[xstep=1,ystep=0.1] (1,4);
	    \draw[black] (0,4) grid[xstep=1,ystep=0.1] (1,5);
	\end{scope}
	
	\begin{scope}[on behind layer]
        \fill[{OrRd-J}] (0,0) rectangle (1,1.5);
        \fill[{OrRd-G}] (0,1.5) rectangle (1,3);
        \fill[{OrRd-E}] (0,3) rectangle (1,4);
    \end{scope}

    \draw[black,-latex] (-1,4) -- (0,4) node [pos=0.2, above] {New Low};
    \draw[black,-latex] (-1,3) -- (0,3) node [pos=0.2, above] {New Med.};
    \draw[black,-latex] (-1,1.5) -- (0,1.5) node [pos=0.2, above] {New High};
    \path[black,-latex] (1,3.75) edge[bend right=-35] node [right] {Escalate} (1,1.5);
    \path[black,-latex] (1,0.35) edge[bend right=29] node [right] {De-escalate} (1,3);

\end{tikzpicture}}}
        ~
        \subfloat[Shared memory dynamic partitioning when full.]{\label{fig:shr_part_full}\adjustbox{width=0.27\linewidth,valign=b}{\begin{tikzpicture}[scale=1, cross/.style={path picture={ 
    \draw[black]
        (path picture bounding box.south east) -- (path picture bounding box.north west) (path picture bounding box.south west) -- (path picture bounding box.north east);
    }}]

    \begin{scope}
	    \draw[black] (0,0) grid[xstep=1,ystep=0.1] (1,2.5);
	    \draw[black] (0,2.5) grid[xstep=1,ystep=0.1] (1,4);
	    \draw[black] (0,4) grid[xstep=1,ystep=0.1] (1,5);
	\end{scope}
	
	\begin{scope}[on behind layer]
        \fill[{OrRd-J}] (0,0) rectangle (1,2.5);
        \fill[{OrRd-G}] (0,2.5) rectangle (1,4);
        \fill[{OrRd-E}] (0,4) rectangle (1,5);
    \end{scope}

    \draw[black,-latex] (-1.5,5) -- (-0.5,5) node [pos=0.2, above] {New Low};
    \node [draw,circle,cross,minimum width=0.25cm] (L) at (-0.375,5){};
    
    \draw[black,-latex] (2.5,2.5) -- (1.5,2.5) node [pos=0.2, above] {New High};
    \draw[black] (1.375,2.625) -- (1.375,5);
    \draw[black,-latex] (1,5) -- (2,5) node [midway, above] {Pop Low};
    \draw[black] (1.375,2.5) circle (0.125cm);
    \draw[black,-latex] (1.25,2.5) -- (1,2.5);

\end{tikzpicture}}}
        \caption{\label{fig:memory_manger}Memory manger architecture.}
    \end{figure}

    To allow a variable amount of shared memory to be used, a more monolithic approach has to be taken. This monolithic approach will permit several novel optimisation strategies to be explored later. To enable this, we introduce the idea of a generation time memory manager --- the architecture of the manager for this problem is shown in \cref{fig:mem_arc}. This was facilitated by producing a flexible framework of sub-managers to handle the different memory types explicitly addressable by users, namely: register; shared; and global. Due to the algorithmic requirements of our application, all global data passes through the $x$ and $y$ registers, which from \cref{ssec:planar} are intended to store a $y$-line and $x$-point. 
    
    At generation time, the computation will request the location of data within the memory hierarchy from the manager, which will pass back the variable name and index within the hierarchy. A greedy approach is used to achieve this, so the manager will give preference in the order: registers over shared over global. In responding to a request, the manager can assess the state of the shared memory. If a variable is only resident in global memory, the manager can cache it into shared memory. To facilitate this, the shared memory is dynamically partitioned by the manager into four partitions: free, low, medium, and high priority, with each partition treated like a stack, as shown in \cref{fig:shr_part}. The priority of a given variable is set by the computation when making a request. An example of the behaviour when no free space is available is shown in \cref{fig:shr_part_full}; for higher priority variables, space is freed from lower priority variables. If no lower priority space is available, the data is not cached in shared. Lastly, data can have their priority escalated or de-escalated --- in the planar algorithm this is important to ensure that all variables in the working plane remain in shared, and it also prevents re-loading of variables that have already been cached.
    
    This shared partitioning has the benefit of being able to scale easily to any amount of shared memory, which, for this planar algorithm, is assumed to be greater than the minimum required to store a single plane in shared. The memory manager does account for the number of threads accessing memory --- to avoid branch divergence in this algorithm, this means that the data required at a given point is stored in shared memory sequentially for each thread, permitting coalesced access. This does impose the constraint that a variable can only be cached in shared if there is sufficient space for all threads to cache it. In practice, however, this turns out to be of little consequence. The first-in last-out nature of a stack does mean that memory is not ordered in a structured way. However, by requiring that an element is handled by a single block, and furthermore that an element is handled by a single warp, it is possible to coalesce shared access with a low number of bank conflicts.
    
    % \begin{algorithm}[tbhp]
    %     \caption{\label{alg:manager}Architecture of memory manager.}
    %     \begin{algorithmic}[1]
    %     \Procedure{MemoryManager}{$\textsf{v}, \textsf{x}, \textsf{ld}$}
            
    %     \EndProcedure
    %     \end{algorithmic}
    % \end{algorithm}
    
    As the memory manager works at generation time, there is no recurring overhead, and the overhead at generation time is low, typically on the order of a second.
    
\subsubsection{Generation Time Optimisations}
    At generation time, there are several optimisation approaches that may be applied line-by-line, or as post-generation actions by analysing the generated source, facilitated by the monolithic approach. We now set out some of these optimisation strategies before comparing their performance. Due to the clear increase in performance seen earlier by storing the matrix constant in \texttt{cmem}, this optimisation is assumed throughout. The other optimisations are as follows:
    \begin{description}
        \item[Deconfliction] The first optimisation is the shared deconfliction that was used in the unmanaged kernel --- it is expected to have a larger impact here due to the increased use of shared memory.
        
        \item[GSR/GRS] The second is the ordering of the explicit shared loads and opportunistic shared load operations. Although not a requirement, it is common in the planar algorithm that when global memory is loaded for storage in shared memory it is then also required in a register. Therefore, the operation can either be ordered as global to shared to registers (GSR), or global to register to shared (GRS). The reason to differentiate this is that, when loading a global variable and storing it to shared on Volta and earlier devices, the compiler will first implicitly load the global to a register, then store the register to shared. However, by explicitly loading a variable into register and then storing that to shared, it is theorised that this is likely to alter the behaviour of the compiler.
        
        \item[Load/Store] A third category of optimisations affects when loading or storing from global memory, CUDA has respectively the functions \texttt{\_\_ldX(...)} and \texttt{\_\_stX(...)}, where \texttt{X} is a modifier that can influence the compiler's choice of caching strategy. For the load instructions the choices are: \texttt{ca}, \texttt{cs}, \texttt{cg}, \texttt{lu}, and \texttt{cv}. More information is available from the PTX programming guide but, briefly, these are respectively: cache in all levels, bypass L1, evict first, last use, and do not cache/fetch again. For the store instructions there are: \texttt{wb}, \texttt{cg}, \texttt{cs}, and \texttt{wt}, which are respectively write back in all levels, bypass L1, used once, and write through. As the output from the kernel is only written to once, both the \texttt{cs} and \texttt{wt} were deemed to be most applicable --- with the latter found to give the best performance. For the load operations, the source can be analysed and when a variable is loaded multiple times the base \texttt{g} modifier may be used. However, if a variable is loaded once, or if a particular load is the last use, then the \texttt{lu} modifier is used. A more complex method could be devised where the memory manager attempts to track the L1 usage in the unified cache, and the modifier could then be then set based on the predicted usage --- however, this method was found to perform poorly in practice. As these modifiers only act as hints to the compiler, it is probable that the manager's model of L1 is inaccurate, and hence changing the caching strategy reduces performance. As reported by \citet{Jia2018}, the actual L1 cache size can vary from the theoretical expectation depending on shared memory usage, with up to \SI{7}{\kibi\byte} unaccounted for. Given the variable usage of \texttt{smem}, this is also a contributing factor to the inaccuracy of the approximate L1 cache model.
        
        \item[Compute Interleaving] It was demonstrated by \citet{Swirydowicz2019} that by interleaving computation with memory transactions, memory latency could be reduced or hidden by keeping threads busy with computation. A simple strategy was to give the compiler a good chance of interleaving computation in the generated assembly by applying an ASAP strategy. This was performed as a post generation optimisation by first categorising source lines as either memory transactions or computation. A dependency graph is then produced, and compute lines moved to be performed ASAP. There are several nuances to this. Firstly, two lines that both accumulate to the same register were deemed to independent. Secondly, thread synchronisation barriers clearly formed a hard dependency. Without a substantial ratio of computation to memory transaction, interleaving can only have a moderate impact. This can be seen from the memory latencies. For example, from \citet{Jia2018}, the shared latency without conflicts is approximately 19 SM warp cycles and the average L2 hit latency is 193 SM warp cycles, with a float fused multiple-add (FFMA) being 4 SM warp cycles. A differentiation is made here between clock cycles and SM warp cycles, with each SM capable of handling 4 schedulers at a time, and a SM warp cycle is defined as one cycle issuing instructions to each of the active warps in the SM.
    
        \item[Compute Agglomeration] A second interleaving algorithm was produced which we will term agglomeration. This method was designed with a view to future hardware features, where shared loads can bypass register and cache levels. Rather than an ASAP interleaving strategy, we look to group the source lines into blocks of memory and computation that can then interleaved. To do this we again categorise lines as compute or memory transactions and form a preliminary dependency graph. This graph is used to move shared loads as early as possible. Then the source is divided into blocks of compute and memory transaction. It is common to have blocks of a range of sizes, with several blocks of just a single line occurring later in the algorithm due to more data cached in shared. These smaller blocks can be agglomerated into larger ones by more closely studying the local dependencies, moving shared loads later, with this method a minimum block size of 13 lines was used for this application. Once blocked, the larger blocks can then be broken up to a user specified length. Memory blocks can then be interleaved with compute blocks by further analysing the inter-block dependencies.
    \end{description}
     
    The combination of optimisation options tested is shown in \cref{tab:managed_opt_cases}, together with their speedup compared to the un-fused case. In all cases, the minimum amount of shared memory was used. Reviewing the results, the basic interleaving strategy did not give a speedup; however, comparing case 2 with case 14, agglomeration interleaving could increase performance. Yet, when agglomeration is coupled to load or store optimisations, a degradation was observed, although it is not clear why. In a comparison between the GRS and GSR store strategies, GSR more often outperforms GRS; however, in the best performing case, GRS outperforms GSR by the largest margin. It should be noted that the best performing managed kernel, case 13, was outperformed by the similar unmanaged kernel, a speedup of $3.452\times$ compared to the $3.250\times$ obtained here. The distinction seems to stem from a single difference: the unmanaged kernel performs the shared store of plane in a block. So, all the global to register loads are in one block, followed by a block performing the register to shared stores, whereas the managed method interleaves the global to register to shared process, and as seen from case 14, this is the origin of the speedup offered by the agglomeration approach. 
    
    \begin{table}[tbhp]
        \centering
        \caption{\label{tab:managed_opt_cases}Effect of optimisation approaches at $p=3$, single precision, with the minimum required shared memory, and 128 threads per block.}
        \begin{tabular}{r c c c c c c |c}
            \toprule 
            Case & Shared & Deconflict & Interleave & Agglomerate & Load & Store & Speedup \\ \midrule
            0 & \textit{GSR} & & & & & & 3.058\\
            \rowcolor{Pastel1-A}1 & GRS & & & & & & 3.049\\
            2 & \textit{GSR} & \cmark & & & & & 3.110\\
            3 & GRS & \cmark & & & & & 3.140\\
            4 & \textit{GSR} & \cmark & \cmark & & & & 3.070\\
            5 & GRS & \cmark & \cmark & & & & 3.045\\
            6 & \textit{GSR} & \cmark & \cmark & & & \cmark & 3.104\\
            7 & GRS & \cmark & \cmark & &  & \cmark & 3.091\\
            8 & \textit{GSR} & \cmark & \cmark & & \cmark & & 3.052\\
            9 & GRS & \cmark & \cmark & & \cmark & & 3.050\\
            \rowcolor{Pastel1-A}10 & \textit{GSR} & \cmark & \cmark & &\cmark & \cmark & 3.049\\
            11 & GRS & \cmark & \cmark & & \cmark & \cmark & 3.104\\
            12 & \textit{GSR} & \cmark &  & & \cmark & & 3.177\\
            \rowcolor{Pastel1-C} 13 & GRS & \cmark &  & & \cmark & & 3.250\\
            14 & \textit{GSR} & \cmark &  & \cmark & & & 3.164\\
            15 & \textit{GSR} & \cmark &  & \cmark & \cmark & & 3.108\\
            16 & \textit{GSR} & \cmark &  & \cmark &  & \cmark & 3.153\\
            17 & \textit{GSR} & \cmark &  & \cmark & \cmark & \cmark & 3.089\\ \bottomrule
        \end{tabular}
    \end{table}
    
\begin{figure}[tbhp]
        \centering
        \subfloat[FP-32]{\adjustbox{width=0.477\linewidth,valign=b}{\begin{tikzpicture}[spy using outlines={rectangle, height=3cm,width=2.5cm, magnification=3, connect spies}]
    \pgfplotsset{scaled x ticks=false}
    \begin{axis}
    [
        axis line style={latex-latex},
        axis y line=left,
        axis x line=left,
        width=12cm,height=9cm,
        ybar=2pt,
        bar width=12pt,
        xmode=linear, % not log
        ymode=linear, % not log
        xlabel = {$p$},
        ylabel = {Speedup},
        ymin = 0, ymax = 4.5,
        symbolic x coords={1,2,3,4,5,6},
        xtick={1,2,3,4,5,6},
        enlarge x limits={abs=1.5*\pgfplotbarwidth},
        legend cell align={left},
        legend style={font=\normalsize, at={(0.97, 0.97)},anchor=north east},
        %nodes near coords,
        %nodes near coords align={vertical},
        x tick label style={/pgf/number format/.cd, fixed, fixed zerofill, precision=0, /tikz/.cd},
        y tick label style={/pgf/number format/.cd, fixed, fixed zerofill, precision=0, /tikz/.cd},
        label style={font=\Large},
        tick label style={font=\Large},
    ]
        
        \addplot[fill={Pastel1-A}, draw=none] coordinates {(1,4.297) (2,3.780) (3,3.426) (4,1.663) (5,0.993) (6,1.470)};
        \addplot[fill={Pastel1-B}, draw=none] coordinates {(1,4.248) (2,3.723) (3,3.713) (4,1.858) (5,1.039) (6,1.562)};
        \legend{Unmanaged, Managed}
    \end{axis}
\end{tikzpicture}}}
        ~
        \subfloat[FP-64]{\adjustbox{width=0.477\linewidth,valign=b}{\begin{tikzpicture}[spy using outlines={rectangle, height=3cm,width=2.5cm, magnification=3, connect spies}]
    \pgfplotsset{scaled x ticks=false}
    \begin{axis}
    [
        axis line style={latex-latex},
        axis y line=left,
        axis x line=left,
        width=12cm,height=9cm,
        ybar=2pt,
        bar width=12pt,
        xmode=linear, % not log
        ymode=linear, % not log
        xlabel = {$p$},
        ylabel = {Speedup},
        ymin = 0, ymax = 4.5,
        symbolic x coords={1,2,3,4,5,6},
        xtick={1,2,3,4,5,6},
        enlarge x limits={abs=1.5*\pgfplotbarwidth},
        legend cell align={left},
        legend style={font=\normalsize, at={(0.97, 0.97)},anchor=north east},
        %nodes near coords,
        %nodes near coords align={vertical},
        x tick label style={/pgf/number format/.cd, fixed, fixed zerofill, precision=0, /tikz/.cd},
        y tick label style={/pgf/number format/.cd, fixed, fixed zerofill, precision=0, /tikz/.cd},
        label style={font=\Large},
        tick label style={font=\Large},
    ]
        
        \addplot[fill={Pastel1-A}, draw=none] coordinates {(1,3.89259) (2,3.838667) (3,1.69139) (4,1.08779) (5,0.31998) (6,1.463414)};
        \addplot[fill={Pastel1-B}, draw=none] coordinates {(1,3.96508) (2,4.01448) (3,2.26908) (4,1.29950) (5,1.39205) (6,1.67991)};
    \end{axis}
\end{tikzpicture}}}
        \caption{\label{fig:planar_su}Unmanaged kernels compared to optimal managed kernels.}
    \end{figure}
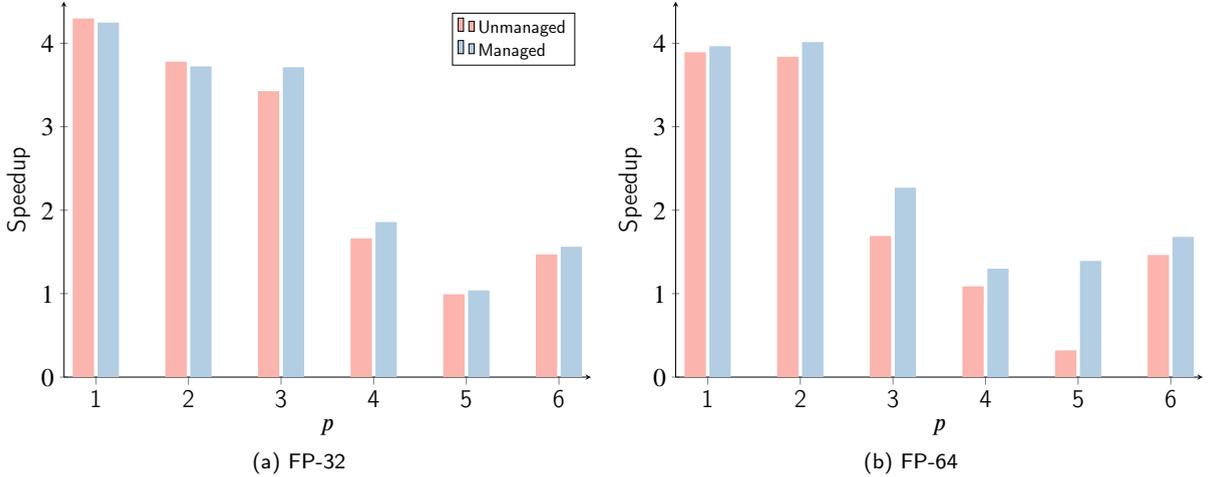

    Regardless, the benefit of the managed approach is the ability to use more of the shared memory --- and, using the optimisations of case 13, a sweep over the available shared memory was performed. The results of this are shown in \cref{fig:man_smem_rt}, with the best performing kernels compared to the unmanaged approach in \cref{fig:planar_su}. This highlights that the planar approach can produce near optimal kernels at low orders. However, performance suffers at higher orders. Furthermore, the memory manager can increase performance at high orders, with the most stark improvement in performance occurring with the FP-64 kernels. As order increases, so too does the workload per thread and register usage. For FP-32, the register limit is reached at $p=4$, and at $p=3$ for FP-64. This is the cause of the steep degradation in the performance with order. Therefore, an alternative method should be sought to improve the performance at higher orders. 
    
    \begin{figure}[tbhp]
        \centering
        \subfloat[FP-32]{\adjustbox{width=0.447\linewidth,valign=b}{\begin{tikzpicture}[spy using outlines={rectangle, height=3cm,width=2.5cm, magnification=3, connect spies}]
\definecolor{agrey}{rgb}{0.6,0.6,0.6}
    \pgfplotsset{scaled x ticks=false}
    \begin{axis}
    [
        axis line style={latex-latex},
        axis y line=left,
        axis x line=left,
        xmode=linear, % not log
        ymode=linear, % not log
        xlabel = {Shared memory / bytes},
        ylabel = {Speedup},
        xmin = 16384, xmax = 110000,
        ymin = 0, ymax = 4.5,
        xtick = {0, 2^14, 2^15, 3*2^14, 4*2^14, 5*2^14, 6*2^14},
        xticklabels = {$0$, $1$, $2$, $3$, $4$, $5$, $6$},
        legend cell align={left},
        legend style={font=\scriptsize, at={(0.03, 0.03)},anchor=south west},
        %axis line style={draw=none},
        %tick style={draw=none},
        x tick label style={/pgf/number format/.cd, fixed, fixed zerofill, precision=1, /tikz/.cd},
        y tick label style={/pgf/number format/.cd, fixed, fixed zerofill, precision=1, /tikz/.cd},
    ]
        \addplot[thick, black, densely dotted, forget plot] coordinates{(0,4) (110000,4)};
        \addplot[thick, black, densely dotted, forget plot] coordinates{(0,1) (110000,1)};
        \addplot[thick, black, densely dashed, forget plot] coordinates{(3*2^14,0) (3*2^14,4.5)};
    
        \addplot[line width=1pt,color={Set1-A}] table[x expr=(\thisrowno{16}), y expr=(192224/\thisrowno{17}), 
                                                      col sep=comma]{./figs/data/managed/128/gimmik_data_b128_p1_fp32.csv};
        %\addlegendentry{$p=1$}
                                                      
        \addplot[line width=1pt,color={Set1-B}] table[x expr=(\thisrowno{16}), y expr=(675392/\thisrowno{17}), 
                                                      col sep=comma]{./figs/data/managed/128/gimmik_data_b128_p2_fp32.csv};
        %\addlegendentry{$p=2$}
                                                      
        \addplot[line width=1pt,color={Set1-C}] table[x expr=(\thisrowno{16}), y expr=(1624578/\thisrowno{17}), 
                                                      col sep=comma]{./figs/data/managed/128/gimmik_data_b128_p3_fp32.csv};
        %\addlegendentry{$p=3$}
                                                      
        \addplot[line width=1pt,color={Set1-D}] table[x expr=(\thisrowno{16}), y expr=(3213824/\thisrowno{17}), 
                                                      col sep=comma]{./figs/data/managed/128/gimmik_data_b128_p4_fp32.csv};
        %\addlegendentry{$p=4$}
                                                      
        \addplot[line width=1pt,color={Set1-E}] table[x expr=(\thisrowno{16}), y expr=(5928512/\thisrowno{17}), 
                                                      col sep=comma]{./figs/data/managed/128/gimmik_data_b128_p5_fp32.csv};
        %\addlegendentry{$p=5$}
                                                      
        \addplot[line width=1pt,color={Set1-H}] table[x expr=(\thisrowno{16}), y expr=(13513248/\thisrowno{17}), 
                                                      col sep=comma]{./figs/data/managed/128/gimmik_data_b128_p6_fp32.csv};
        %\addlegendentry{$p=6$}
    \end{axis}
    \node[black] at (6.65,-0.5) {$\cdot2^{14}$};
\end{tikzpicture}}}
        ~
        \subfloat[FP-64]{\adjustbox{width=0.447\linewidth,valign=b}{\begin{tikzpicture}[spy using outlines={rectangle, height=3cm,width=2.5cm, magnification=3, connect spies}]
\definecolor{agrey}{rgb}{0.6,0.6,0.6}
    \pgfplotsset{scaled x ticks=false}
    \begin{axis}
    [
        axis line style={latex-latex},
        axis y line=left,
        axis x line=left,
        xmode=linear, % not log
        ymode=linear, % not log
        xlabel = {Shared memory / bytes},
        ylabel = {Speedup},
        xmin = 40960, xmax = 105000,
        ymin = 0, ymax = 4.5,
        xtick = {2.5*2^14, 3*2^14, 3.5*2^14, 4*2^14, 4.5*2^14, 5*2^14, 5.5*2^14,  6*2^14},
        xticklabels = { ,$3$, ,$4$, ,$5$, ,$6$},
        legend cell align={left},
        legend style={font=\scriptsize, at={(0.03, 0.03)},anchor=south west},
        %axis line style={draw=none},
        %tick style={draw=none},
        x tick label style={/pgf/number format/.cd, fixed, fixed zerofill, precision=1, /tikz/.cd},
        y tick label style={/pgf/number format/.cd, fixed, fixed zerofill, precision=1, /tikz/.cd},
    ]
        \addplot[thick, black, densely dotted, forget plot] coordinates{(0,4) (110000,4)};
        \addplot[thick, black, densely dotted, forget plot] coordinates{(0,1) (110000,1)};
        \addplot[thick, black, densely dashed, forget plot] coordinates{(3*2^14,0) (3*2^14,4.5)};
    
        \addplot[line width=1pt,color={Set1-A}] table[x expr=(\thisrowno{16}), y expr=(381536/\thisrowno{17}), col sep=comma]{./figs/data/managed/128/gimmik_data_b128_p1_fp64.csv};
                                                      
        \addplot[line width=1pt,color={Set1-B}] table[x expr=(\thisrowno{16}), y expr=(1312640/\thisrowno{17}), col sep=comma]{./figs/data/managed/128/gimmik_data_b128_p2_fp64.csv};
                                                      
        \addplot[line width=1pt,color={Set1-C}] table[x expr=(\thisrowno{16}), y expr=(3124064/\thisrowno{17}), col sep=comma]{./figs/data/managed/128/gimmik_data_b128_p3_fp64.csv};
                                                      
        \addplot[line width=1pt,color={Set1-D}] table[x expr=(\thisrowno{16}), y expr=(7707616/\thisrowno{17}), col sep=comma]{./figs/data/managed/128/gimmik_data_b128_p4_fp64.csv};
                                                      
        \addplot[line width=1pt,color={Set1-E}] table[x expr=(\thisrowno{16}), y expr=(17478064/\thisrowno{17}), col sep=comma]{./figs/data/managed/128/gimmik_data_b128_p5_fp64.csv};
                                                      
        \addplot[line width=1pt,color={Set1-H}] table[x expr=(\thisrowno{16}), y expr=(31571264/\thisrowno{17}), col sep=comma]{./figs/data/managed/128/gimmik_data_b128_p6_fp64.csv};
        \addlegendentry{$p=1$}
        \addlegendentry{$p=2$}
        \addlegendentry{$p=3$}
        \addlegendentry{$p=4$}
        \addlegendentry{$p=5$}
        \addlegendentry{$p=6$}
    \end{axis}
    \node[black] at (6.65,-0.5) {$\cdot2^{14}$};
\end{tikzpicture}

% wt247 - rerunning this after I messed up the data
}}
        \caption{\label{fig:man_smem_rt}Effect of shared memory usage on speedup in managed kernels.}
    \end{figure}
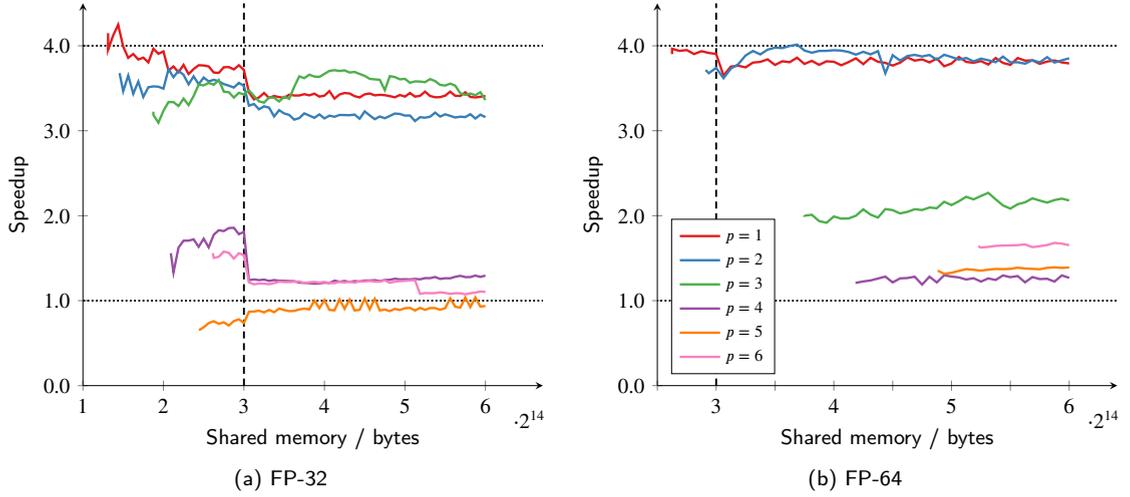
    
    In general the memory manager approach was able to give small performance when the register limit is not reached, and larger improvement once spillage occurs. The memory manager also introduces several parameters, in practical applications software developers have three potential options. The first is to run a sweep over the parameter space and find the best performing kernel in each case \emph{a priori}, the second is to repeat this at run time, however this will add significant overhead. The third option is to set the block size to 128 threads and the shared memory based on the underlying hardware, in this case $3\times2^{14}$ bits. This configuration was generally performant and will translate well to other architectures, such as AMD GPUs. Finally, the benefit of memory manager are small given the time required to develop the memory manager. However, this approach is useful in a many runtime code generation application where shared memory can be leveraged, and we foresee that the agglomeration approaches will be increasingly useful as asynchronous shared memory acceleration gains adoption.
    
\subsection{Lines Method}
    Although the planar method was able to produce performant fused kernels, it is clear that, due to the deficiencies already discussed, at higher values of $p$, there is a significant drop off. To remedy this, an alternative approach may be used where a thread is allocated a line in $z$ to calculate, increasing the number of threads per element from $p+1$ to $(p+1)^2$. Alternative approaches in which a thread works on a single flux variable were attempted, however the significant branch divergence, coupled to a constant number of threads per element, limited performance. 
    
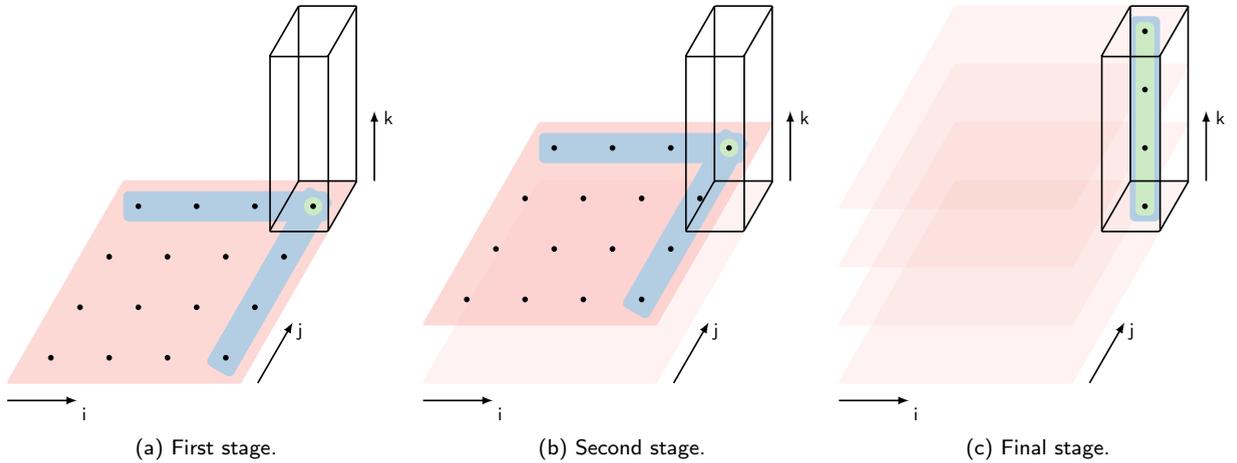
\begin{figure}[tbhp]
        \centering
        \subfloat[First stage.]{\label{fig:lines_z1}\adjustbox{width=0.32\linewidth, valign=b}{\begin{tikzpicture}[scale=1]

    %\begin{scope}[on behind layer]
    %    \draw[Pastel1-A, fill=Pastel1-A, opacity=0.5, thick] (0,0) -- (0.866,0.500) -- (0.866, 2.5) -- (0, 2) -- (0,0);
    %\end{scope}
    
    \begin{scope}[on behind layer]
        \draw[Pastel1-A, fill=Pastel1-A, opacity=0.5, thick,rounded corners=0.2ex] (0,0) -- (4,0) -- (6,3.4641) -- (2,3.4641) -- cycle;
    \end{scope}

    \draw[Pastel1-B,name path = B1,rounded corners=0.6ex] (2.0,2.78) rectangle (5.5,3.28);
    \draw[Pastel1-B,name path = A1,rounded corners=0.6ex,rotate around={60:(3.75,0.43301)}] (3.5,0.183) rectangle (7.0,0.683);
    \tikzfillbetween[of=A1 and B1]{Pastel1-B};
    
    \draw[Pastel1-C, fill=Pastel1-C] (5.25,3.0311) circle (0.15);

    \begin{scope}[on above layer]
        \draw[black, fill=black] (0.75,0.43301) circle (0.04);
        \draw[black, fill=black] (1.75,0.43301) circle (0.04);
        \draw[black, fill=black] (2.75,0.43301) circle (0.04);
        \draw[black, fill=black] (3.75,0.43301) circle (0.04);
        \draw[black, fill=black] (1.25,1.299) circle (0.04);
        \draw[black, fill=black] (2.25,1.299) circle (0.04);
        \draw[black, fill=black] (3.25,1.299) circle (0.04);
        \draw[black, fill=black] (4.25,1.299) circle (0.04);
        \draw[black, fill=black] (1.75,2.1651) circle (0.04);
        \draw[black, fill=black] (2.75,2.1651) circle (0.04);
        \draw[black, fill=black] (3.75,2.1651) circle (0.04);
        \draw[black, fill=black] (4.75,2.1651) circle (0.04);
        \draw[black, fill=black] (2.25,3.0311) circle (0.04);
        \draw[black, fill=black] (3.25,3.0311) circle (0.04);
        \draw[black, fill=black] (4.25,3.0311) circle (0.04);
        \draw[black, fill=black] (5.25,3.0311) circle (0.04);
    \end{scope}

    \begin{scope}[on above layer]
        \draw[black,rounded corners=0.2ex, thick] (4.5,2.598) -- (5.5,2.598) -- (6,3.4641) -- (5,3.4641) -- cycle;
        \draw[black,rounded corners=0.2ex, thick] (4.5,5.598) -- (5.5,5.598) -- (6,6.4641) -- (5,6.4641) -- cycle;
        \draw[black, thick] (4.51,2.62) -- (4.51,5.62);
        \draw[black, thick] (5.51,2.62) -- (5.51,5.62);
        \draw[black, thick] (5.99,3.45) -- (5.99,6.46);
        \draw[black, thick] (5,3.456) -- (5,6.456);
    \end{scope}

    \draw[black, -latex, thick] (0,-0.3) -- (1.2,-0.3) node[pos=1.1, below] {i};
    \draw[black, -latex, thick] (6.3,3.4641) -- (6.3, 4.6641) node[pos=1.1, below, xshift=0.75em] {k};
    \draw[black, -latex, thick] (4.3,0) -- (4.9, 1.039) node[pos=1.1, below, xshift=0.2em] {j};
    
\end{tikzpicture}}}
        ~
        \subfloat[Second stage.]{\label{fig:lines_z2}\adjustbox{width=0.32\linewidth, valign=b}{\begin{tikzpicture}[scale=1]

    %\begin{scope}[on behind layer]
    %    \draw[Pastel1-A, fill=Pastel1-A, opacity=0.5, thick] (0,0) -- (0.866,0.500) -- (0.866, 2.5) -- (0, 2) -- (0,0);
    %\end{scope}
    
    \begin{scope}[on behind layer]
        \draw[Pastel1-A, fill=Pastel1-A, opacity=0.15, thick,rounded corners=0.2ex] (0,0) -- (4,0) -- (6,3.4641) -- (2,3.4641) -- cycle;
        \draw[Pastel1-A, fill=Pastel1-A, opacity=0.5, thick,rounded corners=0.2ex] (0,1) -- (4,1) -- (6,4.4641) -- (2,4.4641) -- cycle;
    \end{scope}

    \draw[Pastel1-B,name path = B1,rounded corners=0.6ex] (2.0,3.78) rectangle (5.5,4.28);
    \draw[Pastel1-B,name path = A1,rounded corners=0.6ex,rotate around={60:(3.75,1.43301)}] (3.5,1.183) rectangle (7.0,1.683);
    \tikzfillbetween[of=A1 and B1]{Pastel1-B};
    \draw[Pastel1-C, fill=Pastel1-C] (5.25,4.0311) circle (0.15);

    \begin{scope}[on above layer]
        \draw[black, fill=black] (0.75,1.43301) circle (0.04);
        \draw[black, fill=black] (1.75,1.43301) circle (0.04);
        \draw[black, fill=black] (2.75,1.43301) circle (0.04);
        \draw[black, fill=black] (3.75,1.43301) circle (0.04);
        \draw[black, fill=black] (1.25,2.299) circle (0.04);
        \draw[black, fill=black] (2.25,2.299) circle (0.04);
        \draw[black, fill=black] (3.25,2.299) circle (0.04);
        \draw[black, fill=black] (4.25,2.299) circle (0.04);
        \draw[black, fill=black] (1.75,3.1651) circle (0.04);
        \draw[black, fill=black] (2.75,3.1651) circle (0.04);
        \draw[black, fill=black] (3.75,3.1651) circle (0.04);
        \draw[black, fill=black] (4.75,3.1651) circle (0.04);
        \draw[black, fill=black] (2.25,4.0311) circle (0.04);
        \draw[black, fill=black] (3.25,4.0311) circle (0.04);
        \draw[black, fill=black] (4.25,4.0311) circle (0.04);
        \draw[black, fill=black] (5.25,4.0311) circle (0.04);
    \end{scope}

    \begin{scope}[on above layer]
        \draw[black,rounded corners=0.2ex, thick] (4.5,2.598) -- (5.5,2.598) -- (6,3.4641) -- (5,3.4641) -- cycle;
        \draw[black,rounded corners=0.2ex, thick] (4.5,5.598) -- (5.5,5.598) -- (6,6.4641) -- (5,6.4641) -- cycle;
        \draw[black, thick] (4.51,2.62) -- (4.51,5.62);
        \draw[black, thick] (5.51,2.62) -- (5.51,5.62);
        \draw[black, thick] (5.99,3.45) -- (5.99,6.46);
        \draw[black, thick] (5,3.456) -- (5,6.456);
    \end{scope}

    \draw[black, -latex, thick] (0,-0.3) -- (1.2,-0.3) node[pos=1.1, below] {i};
    \draw[black, -latex, thick] (6.3,3.4641) -- (6.3, 4.6641) node[pos=1.1, below, xshift=0.75em] {k};
    \draw[black, -latex, thick] (4.3,0) -- (4.9, 1.039) node[pos=1.1, below, xshift=0.2em] {j};
    
\end{tikzpicture}}}
        ~
        \subfloat[Final stage.]{\label{fig:lines_z}\adjustbox{width=0.32\linewidth, valign=b}{\begin{tikzpicture}[scale=1]

    %\begin{scope}[on behind layer]
    %    \draw[Pastel1-A, fill=Pastel1-A, opacity=0.5, thick] (0,0) -- (0.866,0.500) -- (0.866, 2.5) -- (0, 2) -- (0,0);
    %\end{scope}
    
    \begin{scope}[on behind layer]
        \draw[Pastel1-A, fill=Pastel1-A, opacity=0.15, thick,rounded corners=0.2ex] (0,0) -- (4,0) -- (6,3.4641) -- (2,3.4641) -- cycle;
        \draw[Pastel1-A, fill=Pastel1-A, opacity=0.15, thick,rounded corners=0.2ex] (0,1) -- (4,1) -- (6,4.4641) -- (2,4.4641) -- cycle;
        \draw[Pastel1-A, fill=Pastel1-A, opacity=0.15, thick,rounded corners=0.2ex] (0,2) -- (4,2) -- (6,5.4641) -- (2,5.4641) -- cycle;
        \draw[Pastel1-A, fill=Pastel1-A, opacity=0.15, thick,rounded corners=0.2ex] (0,3) -- (4,3) -- (6,6.4641) -- (2,6.4641) -- cycle;
    \end{scope}

    \draw[Pastel1-B,fill={Pastel1-B},name path = B1,rounded corners=0.6ex] (5.0,2.78) rectangle (5.5,6.28);
    \draw[Pastel1-C,fill={Pastel1-C},name path = B1,rounded corners=0.6ex] (5.1,2.88) rectangle (5.4,6.18);
    
    \begin{scope}[on above layer]
        \draw[black, fill=black] (5.25,3.0311) circle (0.04);
        \draw[black, fill=black] (5.25,4.0311) circle (0.04);
        \draw[black, fill=black] (5.25,5.0311) circle (0.04);
        \draw[black, fill=black] (5.25,6.0311) circle (0.04);
    \end{scope}

    \begin{scope}[on above layer]
        \draw[black,rounded corners=0.2ex, thick] (4.5,2.598) -- (5.5,2.598) -- (6,3.4641) -- (5,3.4641) -- cycle;
        \draw[black,rounded corners=0.2ex, thick] (4.5,5.598) -- (5.5,5.598) -- (6,6.4641) -- (5,6.4641) -- cycle;
        \draw[black, thick] (4.51,2.62) -- (4.51,5.62);
        \draw[black, thick] (5.51,2.62) -- (5.51,5.62);
        \draw[black, thick] (5.99,3.45) -- (5.99,6.46);
        \draw[black, thick] (5,3.456) -- (5,6.456);
    \end{scope}

    \draw[black, -latex, thick] (0,-0.3) -- (1.2,-0.3) node[pos=1.1, below] {i};
    \draw[black, -latex, thick] (6.3,3.4641) -- (6.3, 4.6641) node[pos=1.1, below, xshift=0.75em] {k};
    \draw[black, -latex, thick] (4.3,0) -- (4.9, 1.039) node[pos=1.1, below, xshift=0.2em] {j};
    
\end{tikzpicture}}}
        \caption{\label{fig:lines}Progression of the method of lines for one thread. Red indicates data that is present in shared memory, green indicated the point being worked on by the current thread, and blue shows the data requirements for the current point.}
    \end{figure}

    In the procedure for this lines based method, the $x$-$y$ planes are first sequentially stored into shared memory, with each thread loading a single point from the plane. An optimisation that will be described in more detail later is the changing of the number of variables stored in shared. Once the variables have been stored, to ensure there visibility to other threads a synchronisation is required, after which each thread can then calculate and accumulate the $x$ and $y$ contributions to the flux divergence. For a single thread, this is shown diagrammatically for the first and second plane in \cref{fig:lines_z1,fig:lines_z2}. After all planes have been stored in shared, the $z$ line contribution can be calculated and stored in global, shown diagrammatically in \cref{fig:lines_z}. The stages of the calculation are fully outlined in \cref{alg:lines}.
    
    The generation procedure for this method was straightforward as an unrolled strategy was not used in this case. This was done as it was found to give significantly reduced performance, and did not solve the stated problem of high register usage compared to the planar method, which did benefit from full unrolling. 
    
    \begin{algorithm}[tbhp]
        \caption{\label{alg:lines}Fusion of flux evaluation and divergence calculation with one thread per $z$-line.}
        \begin{algorithmic}[1]
        
        \Procedure{FusedLines}{$N,\mathsf{U}_s,-\nabla\cdot\mathsf{F}_s^a$}
            \State Set global element number, $e_g$.
            \State $i := \mathrm{mod}\left(\texttt{threadIdx.x}/n,\; p+1\right)$
            \State $j := \mathrm{mod}\left(\texttt{threadIdx.x}/(n(p+1)),\; p+1\right)$
            \If{$e_g < N$}
                \For{$k\in\{0,\dots,p\}$}
                    \State Store k-th x-y plane of $\mathsf{U}_s$ in \texttt{smem}.
                    \State Barrier, arrive and wait
                    \State Accumulate $x$ line contribution to $\nabla\cdot\mathsf{F}_s^a(i,j,k)$
                    \State Accumulate $y$ line contribution to $\nabla\cdot\mathsf{F}_s^a(i,j,k)$
                \EndFor
                \State Barrier, arrive and wait
                \For{$k\in\{0,\dots,p\}$}
                    \State Accumulate $z$ line contribution to $\nabla\cdot\mathsf{F}_s^a(i,j,k)$.
                    \State Store $\nabla\cdot\mathsf{F}_s^a(i,j,k)$ in global.
                \EndFor
            \EndIf
        \EndProcedure
        \end{algorithmic}
    \end{algorithm}
    
    As was discussed for the unmanaged planar kernel, the storage pattern used for the shared memory can have a significant impact on performance through bank conflicts, and the differences between the planar and lines based algorithm required a different deconfliction approach. A naive AoSoA shared layout will lead to a situation that almost guarantees banks conflicts. To design a memory layout, two things have to be considered: first, how is the global data arranged, and secondly, how will shared memory be used. In response to the former, the global data, inherited from PyFR, uses an AoSoA structure where the same variable for multiple elements are stored sequentially. Therefore, to achieve coalesced global loads, sequential threads should work on sequential elements. Therefore, if a block works on $n$ elements with $n(p+1)^2$ threads, the global element number for a thread is calculated as:
    \begin{equation}
        e_g = \mathrm{mod}(\texttt{threadIdx.x},\; n) + n(\texttt{blockIdx.x}).
    \end{equation}
    The $i$ and $j$ indices may then be subsequently calculated as shown in \cref{alg:lines}. When shared memory is used, it will be loaded one variable at a time by all threads in a warp. Therefore, after elements, $i$ and $j$ are the next major indices, followed by $k$, and with variable numbering last. Combining this we can produce the following indexing:
    \begin{equation}
        \mathsf{U}_s(e_l,i,j,k,v) \rightarrow e_l + n\left(i + j(p+1) + k(p+1)^2 + v(p+1)^3\right),
    \end{equation}
    where $e_l$ is the block local element number, $e=\mathrm{mod}(\texttt{threadIdx.x},\; n)$. For single-precision numbers, the effect of this layout is that, when loading or storing either a variable or accumulator to shared, each thread uses a different shared memory bank. There is one exception to this: for some values of $k$ aliasing can lead to a small number of bank conflicts. However, this number is generally very small and in practice the number of bank conflicts was in the range  $0\sim 5\%$.
    
\subsubsection{Static Optimisations}
    The base method shown in \cref{alg:lines} results in all accumulation stages being performed using variables from shared. This change, together with not fully unrolling the kernel, makes several new optimisations possible. Firstly, it is not necessary to store pressure, as it could be combined with several of the gradient terms. Furthermore, in the flux kernel the gradients are always multiplied by the viscosity, therefore rather than repeatedly multiplying by $\nu$ the following are actually stored in shared memory:
    \begin{equation*}
        \begin{bmatrix}
            u& v& w\\ -\nu q_x + P& -\nu q_y& -\nu q_z\\ -\nu r_x& -\nu r_y + P& -\nu r_z\\ -\nu s_x&-\nu s_y&-\nu s_z + P
        \end{bmatrix}
    \end{equation*}
    Consequently, together with the accumulator the base method requires 25 variables per point to be allocated in shared memory. 
    
    A further optimisation can be made by recognising that the first term in the flux divergence can be straightforwardly reconstructed from 3 of the other flux divergence terms, thereby reducing the total number of variables to 24 per point. The final optimisation is to vary the number of variables stored in shared memory. As there is one point per line per element, decreasing the shared memory usage can increase the block size, which may help to increase occupancy at higher $p$. A second effect of reducing the number of variables in shared memory is that it will increase the maximum theoretical bandwidth. Consider the $z$-line accumulation: once all variables are in shared, the global bandwidth is then under utilised. To determine the validity of this hypothesis, the several options for which variables are stored were tested, see \cref{tab:lines_mem} for these configurations.
    
    \begin{table}[tbhp]
        \caption{\label{tab:lines_mem}Variable storage in the lines based method.}
        \centering
        \begin{tabular}{c c c c}
            \toprule
            No. variables & No. accumulators & Total & Variables stored\\\midrule
            12 & 13 & 25 & Velocity and pressure augmented gradients \\
            12 & 12 & 24 & Velocity and pressure augmented gradients\\
            6 & 12 & 18 & Velocity, $(-\nu q_x+P)$, $(-\nu r_y+P)$, and $(-\nu s_z+P)$\\
            3 & 12 & 15 & Velocity\\
            0 & 12 & 12 & None\\ \bottomrule
        \end{tabular}
    \end{table}
    
    Several unsuccessful optimisation approaches were also attempted, and we would like to briefly mention one of note. Several vector types are defined in CUDA, such as \texttt{float4}, and, when preforming a global load or store operation, it is possible to increase the sectors per request by ensuring memory alignment and then utilising these types. However, due to the global AoSoA data layout of PyFR~\citep{Witherden2014}, this necessitates that the elements per block is divisible by the vector width and, furthermore, that only some threads participate in the load. As a result, a more optimal sectors per request was produced --- but at a lower bandwidth, due to the overhead of the synchronisation and data shuffle steps required.
    
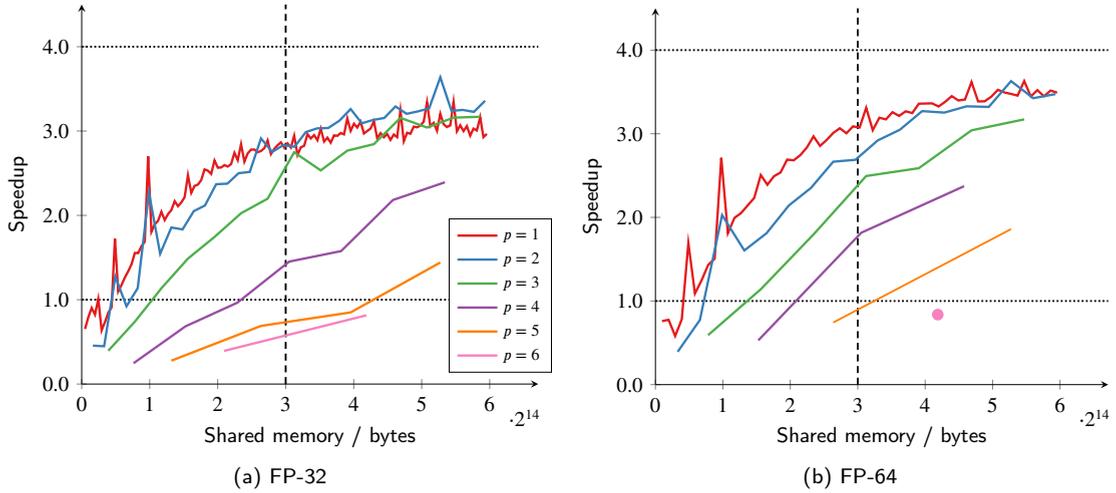
\begin{figure}[tbhp]
        \centering
        \subfloat[FP-32]{\label{fig:l25_smem_rt_32}\adjustbox{width=0.447\linewidth, valign=b}{\begin{tikzpicture}[spy using outlines={rectangle, height=3cm,width=2.5cm, magnification=3, connect spies}]
\definecolor{agrey}{rgb}{0.6,0.6,0.6}
    \pgfplotsset{scaled x ticks=false}
    \begin{axis}
    [
        axis line style={latex-latex},
        axis y line=left,
        axis x line=left,
        xmode=linear, % not log
        ymode=linear, % not log
        xlabel = {Shared memory / bytes},
        ylabel = {Speedup},
        xmin = 0, xmax = 110000,
        ymin = 0, ymax = 4.5,
        xtick = {0, 2^14, 2^15, 3*2^14, 4*2^14, 5*2^14, 6*2^14},
        xticklabels = {$0$, $1$, $2$, $3$, $4$, $5$, $6$},
        legend cell align={left},
        legend style={font=\scriptsize, at={(1.03, 0.03)},anchor=south east},
        %axis line style={draw=none},
        %tick style={draw=none},
        x tick label style={/pgf/number format/.cd, fixed, fixed zerofill, precision=1, /tikz/.cd},
        y tick label style={/pgf/number format/.cd, fixed, fixed zerofill, precision=1, /tikz/.cd},
    ]
        \addplot[thick, black, densely dotted, forget plot] coordinates{(0,4) (110000,4)};
        \addplot[thick, black, densely dotted, forget plot] coordinates{(0,1) (110000,1)};
        \addplot[thick, black, densely dashed, forget plot] coordinates{(3*2^14,0) (3*2^14,4.5)};
    
        \addplot[line width=1pt,color={Set1-A}] table[x expr=(\thisrowno{6}), y expr=(192224/\thisrowno{7}), 
                                                      col sep=comma]{./figs/data/lines/25/gimmik_data_lines_p1_fp32.csv};
        \addlegendentry{$p=1$}
                                                      
        \addplot[line width=1pt,color={Set1-B}] table[x expr=(\thisrowno{6}), y expr=(675392/\thisrowno{7}), 
                                                      col sep=comma]{./figs/data/lines/25/gimmik_data_lines_p2_fp32.csv};
        \addlegendentry{$p=2$}
                                                      
        \addplot[line width=1pt,color={Set1-C}] table[x expr=(\thisrowno{6}), y expr=(1624578/\thisrowno{7}), 
                                                      col sep=comma]{./figs/data/lines/25/gimmik_data_lines_p3_fp32.csv};
        \addlegendentry{$p=3$}
                                                      
        \addplot[line width=1pt,color={Set1-D}] table[x expr=(\thisrowno{6}), y expr=(3213824/\thisrowno{7}), 
                                                      col sep=comma]{./figs/data/lines/25/gimmik_data_lines_p4_fp32.csv};
        \addlegendentry{$p=4$}
                                                      
        \addplot[line width=1pt,color={Set1-E}] table[x expr=(\thisrowno{6}), y expr=(5928512/\thisrowno{7}), 
                                                      col sep=comma]{./figs/data/lines/25/gimmik_data_lines_p5_fp32.csv};
        \addlegendentry{$p=5$}
                                                      
        \addplot[line width=1pt,color={Set1-H}] table[x expr=(\thisrowno{6}), y expr=(13513248/\thisrowno{7}), 
                                                      col sep=comma]{./figs/data/lines/25/gimmik_data_lines_p6_fp32.csv};
        \addlegendentry{$p=6$}
    \end{axis}
    \node[black] at (6.65,-0.5) {$\cdot2^{14}$};
\end{tikzpicture}}}
        ~
        \subfloat[FP-64]{\label{fig:l25_smem_rt_64}\adjustbox{width=0.44\linewidth, valign=b}{\begin{tikzpicture}[spy using outlines={rectangle, height=3cm,width=2.5cm, magnification=3, connect spies}]
\definecolor{agrey}{rgb}{0.6,0.6,0.6}
    \pgfplotsset{scaled x ticks=false}
    \begin{axis}
    [
        axis line style={latex-latex},
        axis y line=left,
        axis x line=left,
        xmode=linear, % not log
        ymode=linear, % not log
        xlabel = {Shared memory / bytes},
        ylabel = {Speedup},
        xmin = 0, xmax = 110000,
        ymin = 0, ymax = 4.5,
        xtick = {0, 2^14, 2^15, 3*2^14, 4*2^14, 5*2^14, 6*2^14},
        xticklabels = {$0$, $1$, $2$, $3$, $4$, $5$, $6$},
        legend cell align={left},
        legend style={font=\scriptsize, at={(1.1, 0.03)},anchor=south east},
        %axis line style={draw=none},
        %tick style={draw=none},
        x tick label style={/pgf/number format/.cd, fixed, fixed zerofill, precision=1, /tikz/.cd},
        y tick label style={/pgf/number format/.cd, fixed, fixed zerofill, precision=1, /tikz/.cd},
    ]
        \addplot[thick, black, densely dotted, forget plot] coordinates{(0,4) (110000,4)};
        \addplot[thick, black, densely dotted, forget plot] coordinates{(0,1) (110000,1)};
        \addplot[thick, black, densely dashed, forget plot] coordinates{(3*2^14,0) (3*2^14,4.5)};
    
        \addplot[line width=1pt,color={Set1-A}] table[x expr=(\thisrowno{6}), y expr=(381536/\thisrowno{7}), col sep=comma]{./figs/data/lines/25/gimmik_data_lines_p1_fp64.csv};
                                                      
        \addplot[line width=1pt,color={Set1-B}] table[x expr=(\thisrowno{6}), y expr=(1312640/\thisrowno{7}), col sep=comma]{./figs/data/lines/25/gimmik_data_lines_p2_fp64.csv};
                                                      
        \addplot[line width=1pt,color={Set1-C}] table[x expr=(\thisrowno{6}), y expr=(3124064/\thisrowno{7}), col sep=comma]{./figs/data/lines/25/gimmik_data_lines_p3_fp64.csv};
                                                      
        \addplot[line width=1pt,color={Set1-D}] table[x expr=(\thisrowno{6}), y expr=(7707616/\thisrowno{7}), col sep=comma]{./figs/data/lines/25/gimmik_data_lines_p4_fp64.csv};
                                                      
        \addplot[color={Set1-E}, thick] table[x expr=(\thisrowno{6}), y expr=(17476064/\thisrowno{7}), col sep=comma]{./figs/data/lines/25/gimmik_data_lines_p5_fp64.csv};
                                                      
        \addplot[line width=1pt,color={Set1-H}, mark=*, mark options={solid}, smooth] table[x expr=(\thisrowno{6}), y expr=(31571264/\thisrowno{7}), col sep=comma]{./figs/data/lines/25/gimmik_data_lines_p6_fp64.csv};
        
    \end{axis}
    \node[black] at (6.65,-0.5) {$\cdot2^{14}$};
\end{tikzpicture}}}
        \caption{\label{fig:lines25_smem_rt}Method of lines performance with 25 variables and varying shared memory usage.}
    \end{figure}

    As the block size is set by the number of elements per block, we will investigate the performance with order and precision, from one element up to the maximum number of elements per block based on shared memory size. The results for the base method with 25 values per point is shown in \cref{fig:lines25_smem_rt}. Here the vertical dashed line indicates \SI{48}{\kibi\byte}. The results for 12 values per point, i.e. all values read from global, are presented in \cref{fig:lines12_smem_rt}, and additionally the other variable configurations can be found in Appendix~\ref{app:lines}.
    
    Firstly, it is worth noting the change that occurs for $\texttt{smem}>\SI{48}{\kibi\byte}$, which is half the maximum size of shared memory. Around this value, the performance curve flattens, and in some cases, there is a dip in performance. This is caused by a reduction in the occupancy, as once $\texttt{smem}>\SI{48}{\kibi\byte}$, only a single block per SM can be operated on at a time. For example, take $p=3$ with 18 values per point and 160 threads per block, so the shared memory usage is just below half. Therefore, two blocks can be worked on at once, with an achieved warp occupancy of 9.5 warps per SM. Compare that with 176 threads per block, where only a single block can be worked on due to shared memory exceeding half of the max, this achieved 5.85 warps per SM. This also links to a shift in performance between lower $p$ and higher $p$. At lower $p$, higher thread counts are possible with lower shared memory usage, therefore leading to high occupancy, whereas at higher $p$ performance becomes more monotonic with shared memory.
    
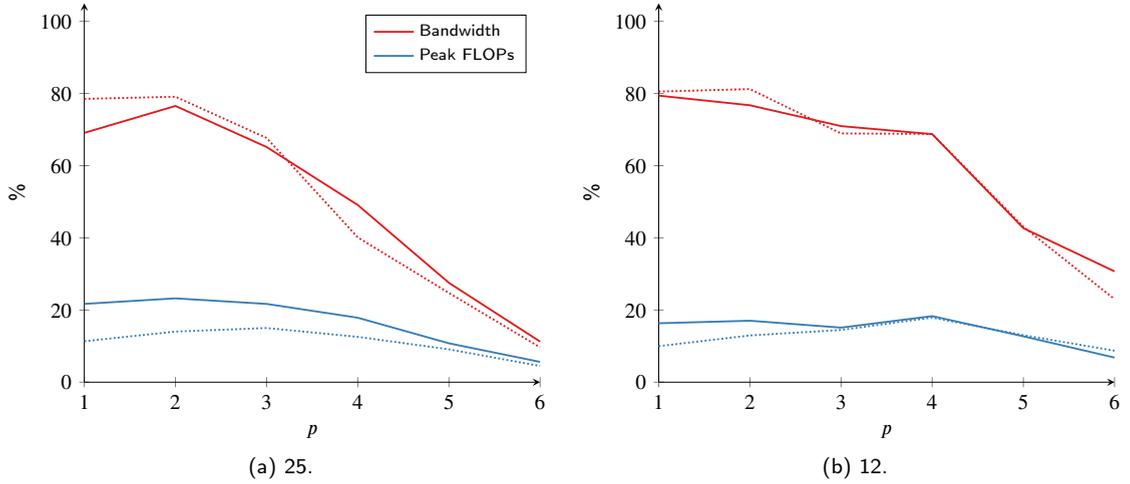
\begin{figure}[tbhp]
        \centering
        \subfloat[25.]{\label{fig:l25_opt_ratios}\adjustbox{width=0.447\linewidth, valign=b}{\begin{tikzpicture}[spy using outlines={rectangle, height=3cm,width=2.5cm, magnification=3, connect spies}]
\definecolor{agrey}{rgb}{0.6,0.6,0.6}
    \pgfplotsset{scaled x ticks=false}
    \begin{axis}
    [
        axis line style={latex-latex},
        axis y line=left,
        axis x line=left,
        xmode=linear, % not log
        ymode=linear, % not log
        xlabel = {$p$},
        ylabel = {\%},
        xmin = 1, xmax = 6,
        ymin = 0, ymax = 105,
        legend cell align={left},
        legend style={font=\scriptsize, at={(0.97, 0.97)},anchor=north east},
        %axis line style={draw=none},
        %tick style={draw=none},
        x tick label style={/pgf/number format/.cd, fixed, fixed zerofill, precision=0, /tikz/.cd},
        y tick label style={/pgf/number format/.cd, fixed, fixed zerofill, precision=0, /tikz/.cd},
    ]
               
        \addplot[thick,color={Set1-A}] table[x expr=\thisrowno{0}, y expr=\thisrowno{22}), col sep=comma]{./figs/data/lines/25/gimmik_data_lines_opt_fp32.csv};
        \addplot[thick,color={Set1-A}, densely dotted, forget plot] table[x expr=\thisrowno{0}, y expr=\thisrowno{22}), col sep=comma]{./figs/data/lines/25/gimmik_data_lines_opt_fp64.csv};
        \addlegendentry{Bandwidth}
        
        \addplot[thick,color={Set1-B}] table[x expr=\thisrowno{0}, y expr=\thisrowno{8}), col sep=comma]{./figs/data/lines/25/gimmik_data_lines_opt_fp32.csv};
        \addplot[thick,color={Set1-B}, densely dotted, forget plot] table[x expr=\thisrowno{0}, y expr=\thisrowno{8}), col sep=comma]{./figs/data/lines/25/gimmik_data_lines_opt_fp64.csv};
        \addlegendentry{Peak FLOPs}
        
        %\addplot[thick,color={Set1-C}] table[x expr=\thisrowno{0}, y expr=((\thisrowno{20} + \thisrowno{21})/(\thisrowno{11} + \thisrowno{12})), 
        %                                              col sep=comma]{./figs/data/lines/gimmik_data_lines_opt_fp32.csv};
        %\addplot[thick,color={Set1-C}, densely dotted, forget plot] table[x expr=\thisrowno{0},  y expr=((\thisrowno{20} + \thisrowno{21})/(\thisrowno{11} + \thisrowno{12})), 
        %                                              col sep=comma]{./figs/data/lines/gimmik_data_lines_opt_fp64.csv};
        %\addlegendentry{Bank Conflicts}
        % \addplot[thick,color={Set1-C}] table[x expr=\thisrowno{0}, y expr=\thisrowno{9}), col sep=comma]{./figs/data/lines/25/gimmik_data_lines_opt_fp32.csv};
        
        % \addplot[thick,color={Set1-C}, densely dotted, forget plot] table[x expr=\thisrowno{0}, y expr=\thisrowno{9}), col sep=comma]{./figs/data/lines/25/gimmik_data_lines_opt_fp64.csv};
        % \addlegendentry{L2}
    \end{axis}
\end{tikzpicture}}}
        ~
        \subfloat[12.]{\label{fig:l12_opt_ratios}\adjustbox{width=0.447\linewidth, valign=b}{\begin{tikzpicture}[spy using outlines={rectangle, height=3cm,width=2.5cm, magnification=3, connect spies}]
\definecolor{agrey}{rgb}{0.6,0.6,0.6}
    \pgfplotsset{scaled x ticks=false}
    \begin{axis}
    [
        axis line style={latex-latex},
        axis y line=left,
        axis x line=left,
        xmode=linear, % not log
        ymode=linear, % not log
        xlabel = {$p$},
        ylabel = {\%},
        xmin = 1, xmax = 6,
        ymin = 0, ymax = 105,
        legend cell align={left},
        legend style={font=\scriptsize, at={(0.97, 0.97)},anchor=north east},
        %axis line style={draw=none},
        %tick style={draw=none},
        x tick label style={/pgf/number format/.cd, fixed, fixed zerofill, precision=0, /tikz/.cd},
        y tick label style={/pgf/number format/.cd, fixed, fixed zerofill, precision=0, /tikz/.cd},
    ]
               
        \addplot[thick,color={Set1-A}] table[x expr=\thisrowno{0}, y expr=\thisrowno{22}), col sep=comma]{./figs/data/lines/12/gimmik_data_lines_opt_fp32.csv};
        \addplot[thick,color={Set1-A}, densely dotted, forget plot] table[x expr=\thisrowno{0}, y expr=\thisrowno{22}), col sep=comma]{./figs/data/lines/12/gimmik_data_lines_opt_fp64.csv};
        %\addlegendentry{Bandwidth}
        
        \addplot[thick,color={Set1-B}] table[x expr=\thisrowno{0}, y expr=\thisrowno{8}), col sep=comma]{./figs/data/lines/12/gimmik_data_lines_opt_fp32.csv};
        \addplot[thick,color={Set1-B}, densely dotted, forget plot] table[x expr=\thisrowno{0}, y expr=\thisrowno{8}), col sep=comma]{./figs/data/lines/12/gimmik_data_lines_opt_fp64.csv};
        %\addlegendentry{Peak FLOPs}
        
        %\addplot[thick,color={Set1-C}] table[x expr=\thisrowno{0}, y expr=((\thisrowno{20} + \thisrowno{21})/(\thisrowno{11} + \thisrowno{12})), 
        %                                              col sep=comma]{./figs/data/lines/gimmik_data_lines_opt_fp32.csv};
        %\addplot[thick,color={Set1-C}, densely dotted, forget plot] table[x expr=\thisrowno{0},  y expr=((\thisrowno{20} + \thisrowno{21})/(\thisrowno{11} + \thisrowno{12})), 
        %                                              col sep=comma]{./figs/data/lines/gimmik_data_lines_opt_fp64.csv};
        %\addlegendentry{Bank Conflicts}
        % \addplot[thick,color={Set1-C}] table[x expr=\thisrowno{0}, y expr=\thisrowno{9}), 
        %                                               col sep=comma]{./figs/data/lines/12/gimmik_data_lines_opt_fp32.csv};
        % \addplot[thick,color={Set1-C}, densely dotted, forget plot] table[x expr=\thisrowno{0}, y expr=\thisrowno{9}), 
        %                                               col sep=comma]{./figs/data/lines/12/gimmik_data_lines_opt_fp64.csv};
        % \addlegendentry{L2}
    \end{axis}
\end{tikzpicture}}}
        \caption{\label{fig:lines_opt}Global memory bandwidth and FLOPs trends for optimally configured kernels with 25 and 12 variables in shared memory per point. Solid lines are FP-32, whereas dotted lines are FP-64.}
    \end{figure}

    Studying \cref{fig:lines_opt}, it can be seen that reducing the usage of shared memory in favour of greater diversity of memory access can increase performance at higher order, most notably for FP-64. The advantage comes from increasing the global memory bandwidth. \cref{fig:l25_opt_ratios} shows that even for these highly optimised kernels, at higher order it is difficult to achieve high global memory bandwidth. By reducing the amount of data stored in shared, this defers some global access. The effect is reduced pressure on shared memory, but by adding a second memory location to some parts of the calculation, the effective bandwidth is increased. The latency of shared and global memory are significantly different, 19 SM warp cycles compared to 193 SM warp cycles for L2 on Volta, and given that parts of the flux function involve multiple terms, there is not a clear relationship to inform how much of shared should be deferred to global. But the general trend of \cref{fig:l12_opt_ratios} is a flattening of the bandwidth curve with order. 
    
    \begin{figure}[tbhp]
        \centering
        \subfloat[FP-32]{\label{fig:l12_smem_rt_32}\adjustbox{width=0.447\linewidth, valign=b}{\begin{tikzpicture}[spy using outlines={rectangle, height=3cm,width=2.5cm, magnification=3, connect spies}]
\definecolor{agrey}{rgb}{0.6,0.6,0.6}
    \pgfplotsset{scaled x ticks=false}
    \begin{axis}
    [
        axis line style={latex-latex},
        axis y line=left,
        axis x line=left,
        xmode=linear, % not log
        ymode=linear, % not log
        xlabel = {Shared memory / bytes},
        ylabel = {Speedup},
        xmin = 0, xmax = 110000,
        ymin = 0, ymax = 4.5,
        xtick = {0, 2^14, 2^15, 3*2^14, 4*2^14, 5*2^14, 6*2^14},
        xticklabels = {$0$, $1$, $2$, $3$, $4$, $5$, $6$},
        legend cell align={left},
        legend style={font=\scriptsize, at={(1.03, 0.03)},anchor=south east},
        %axis line style={draw=none},
        %tick style={draw=none},
        x tick label style={/pgf/number format/.cd, fixed, fixed zerofill, precision=1, /tikz/.cd},
        y tick label style={/pgf/number format/.cd, fixed, fixed zerofill, precision=1, /tikz/.cd},
    ]
        \addplot[thick, black, densely dotted, forget plot] coordinates{(0,4) (110000,4)};
        \addplot[thick, black, densely dotted, forget plot] coordinates{(0,1) (110000,1)};
        \addplot[thick, black, densely dashed, forget plot] coordinates{(3*2^14,0) (3*2^14,4.5)};
    
        \addplot[line width=1pt,color={Set1-A}] table[x expr=(\thisrowno{6}), y expr=(192224/\thisrowno{7}), 
                                                      col sep=comma]{./figs/data/lines/12/gimmik_data_lines_p1_fp32.csv};
        \addlegendentry{$p=1$}
                                                      
        \addplot[line width=1pt,color={Set1-B}] table[x expr=(\thisrowno{6}), y expr=(675392/\thisrowno{7}), 
                                                      col sep=comma]{./figs/data/lines/12/gimmik_data_lines_p2_fp32.csv};
        \addlegendentry{$p=2$}
                                                      
        \addplot[line width=1pt,color={Set1-C}] table[x expr=(\thisrowno{6}), y expr=(1624578/\thisrowno{7}), 
                                                      col sep=comma]{./figs/data/lines/12/gimmik_data_lines_p3_fp32.csv};
        \addlegendentry{$p=3$}
                                                      
        \addplot[line width=1pt,color={Set1-D}] table[x expr=(\thisrowno{6}), y expr=(3213824/\thisrowno{7}), 
                                                      col sep=comma]{./figs/data/lines/12/gimmik_data_lines_p4_fp32.csv};
        \addlegendentry{$p=4$}
                                                      
        \addplot[line width=1pt,color={Set1-E}] table[x expr=(\thisrowno{6}), y expr=(5928512/\thisrowno{7}), 
                                                      col sep=comma]{./figs/data/lines/12/gimmik_data_lines_p5_fp32.csv};
        \addlegendentry{$p=5$}
                                                      
        \addplot[line width=1pt,color={Set1-H}] table[x expr=(\thisrowno{6}), y expr=(13513248/\thisrowno{7}), 
                                                      col sep=comma]{./figs/data/lines/12/gimmik_data_lines_p6_fp32.csv};
        \addlegendentry{$p=6$}
    \end{axis}
    
    \node[black] at (6.65,-0.5) {$\cdot2^{14}$};
\end{tikzpicture}}}
        ~
        \subfloat[FP-64]{\label{fig:l12_smem_rt_64}\adjustbox{width=0.44\linewidth, valign=b}{\begin{tikzpicture}[spy using outlines={rectangle, height=3cm,width=2.5cm, magnification=3, connect spies}]
\definecolor{agrey}{rgb}{0.6,0.6,0.6}
    \pgfplotsset{scaled x ticks=false}
    \begin{axis}
    [
        axis line style={latex-latex},
        axis y line=left,
        axis x line=left,
        xmode=linear, % not log
        ymode=linear, % not log
        xlabel = {Shared memory / bytes},
        ylabel = {Speedup},
        xmin = 0, xmax = 110000,
        ymin = 0, ymax = 4.5,
        xtick = {0, 2^14, 2^15, 3*2^14, 4*2^14, 5*2^14, 6*2^14},
        xticklabels = {$0$, $1$, $2$, $3$, $4$, $5$, $6$},
        legend cell align={left},
        legend style={font=\scriptsize, at={(1.1, 0.03)},anchor=south east},
        %axis line style={draw=none},
        %tick style={draw=none},
        x tick label style={/pgf/number format/.cd, fixed, fixed zerofill, precision=1, /tikz/.cd},
        y tick label style={/pgf/number format/.cd, fixed, fixed zerofill, precision=1, /tikz/.cd},
    ]
        \addplot[thick, black, densely dotted, forget plot] coordinates{(0,4) (110000,4)};
        \addplot[thick, black, densely dotted, forget plot] coordinates{(0,1) (110000,1)};
        \addplot[thick, black, densely dashed, forget plot] coordinates{(3*2^14,0) (3*2^14,4.5)};
    
        \addplot[line width=1pt,color={Set1-A}] table[x expr=(\thisrowno{6}), y expr=(381536/\thisrowno{7}), 
                                                      col sep=comma]{./figs/data/lines/12/gimmik_data_lines_p1_fp64.csv};
                                                      
        \addplot[line width=1pt,color={Set1-B}] table[x expr=(\thisrowno{6}), y expr=(1312640/\thisrowno{7}), 
                                                      col sep=comma]{./figs/data/lines/12/gimmik_data_lines_p2_fp64.csv};
                                                      
        \addplot[line width=1pt,color={Set1-C}] table[x expr=(\thisrowno{6}), y expr=(3124064/\thisrowno{7}), 
                                                      col sep=comma]{./figs/data/lines/12/gimmik_data_lines_p3_fp64.csv};
                                                      
        \addplot[line width=1pt,color={Set1-D}] table[x expr=(\thisrowno{6}), y expr=(7707616/\thisrowno{7}), 
                                                      col sep=comma]{./figs/data/lines/12/gimmik_data_lines_p4_fp64.csv};
                                                      
        \addplot[color={Set1-E}, thick] table[x expr=(\thisrowno{6}), y expr=(17476064/\thisrowno{7}), 
                                                      col sep=comma]{./figs/data/lines/12/gimmik_data_lines_p5_fp64.csv};
                                                      
        \addplot[line width=1pt,color={Set1-H}] table[x expr=(\thisrowno{6}), y expr=(31571264/\thisrowno{7}), 
                                                      col sep=comma]{./figs/data/lines/12/gimmik_data_lines_p6_fp64.csv};
        
    \end{axis}
    \node[black] at (6.65,-0.5) {$\cdot2^{14}$};
\end{tikzpicture}}}
        \caption{\label{fig:lines12_smem_rt}Method of lines performance with 12 variables and varying shared memory usage.}
    \end{figure}
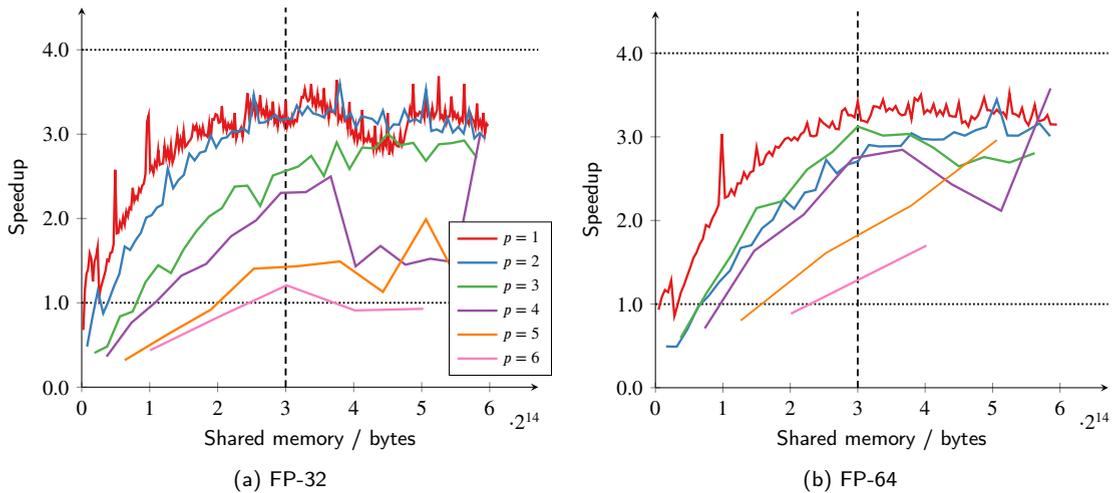
    
    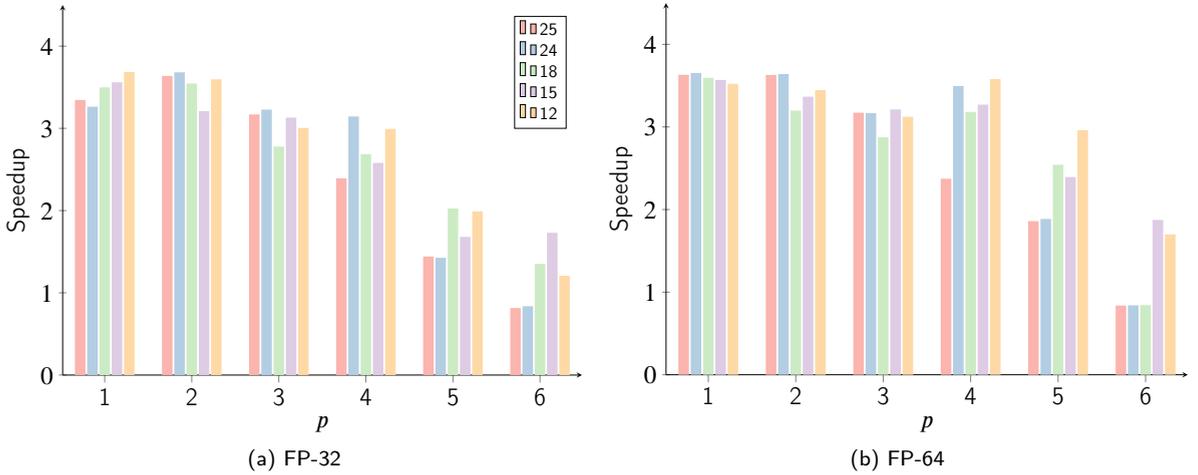
\begin{figure}[tbhp]
        \centering
        \subfloat[FP-32]{\adjustbox{width=0.47\linewidth, valign=b}{\begin{tikzpicture}[spy using outlines={rectangle, height=3cm,width=2.5cm, magnification=3, connect spies}]
    \pgfplotsset{scaled x ticks=false}
    \begin{axis}
    [
        axis line style={latex-latex},
        axis y line=left,
        axis x line=left,
        width=12cm,height=9cm,
        ybar=1pt,
        bar width=6pt,
        xmode=linear, % not log
        ymode=linear, % not log
        xlabel = {$p$},
        ylabel = {Speedup},
        ymin = 0, ymax = 4.5,
        symbolic x coords={1,2,3,4,5,6},
        xtick={1,2,3,4,5,6},
        enlarge x limits={abs=4*\pgfplotbarwidth},
        legend cell align={left},
        legend style={font=\normalsize, at={(0.97, 0.97)},anchor=north east},
        %nodes near coords,
        %nodes near coords align={vertical},
        x tick label style={/pgf/number format/.cd, fixed, fixed zerofill, precision=0, /tikz/.cd},
        y tick label style={/pgf/number format/.cd, fixed, fixed zerofill, precision=0, /tikz/.cd},
        label style={font=\Large},
        tick label style={font=\Large},
    ]
        
        \addplot[fill={Pastel1-A}, draw=none] coordinates {(1,3.34465) (2,3.637084) (3,3.16904) (4,2.392114) (5,1.44122) (6,0.815096)};
        \addplot[fill={Pastel1-B}, draw=none] coordinates {(1,3.262900) (2,3.68085) (3,3.22767) (4,3.1443957) (5,1.42577) (6,0.8370496)};
        \addplot[fill={Pastel1-C}, draw=none] coordinates {(1,3.498543) (2,3.545439) (3,2.778158) (4,2.686784377) (5,2.0258278) (6,1.3523937)};
        \addplot[fill={Pastel1-D}, draw=none] coordinates {(1,3.56) (2,3.21) (3,3.13) (4,2.58) (5,1.68) (6,1.73)};
        \addplot[fill={Pastel1-E}, draw=none] coordinates {(1,3.685276) (2,3.5961833) (3,3.0047385) (4,2.9929670) (5,1.989946) (6,1.2070263)};
        \legend{25,24,18,15,12}
    \end{axis}
\end{tikzpicture}}}
        ~
        \subfloat[FP-64]{\adjustbox{width=0.47\linewidth, valign=b}{\begin{tikzpicture}[spy using outlines={rectangle, height=3cm,width=2.5cm, magnification=3, connect spies},scale=2]
    \pgfplotsset{scaled x ticks=false}
    \begin{axis}
    [
        axis line style={latex-latex},
        axis y line=left,
        axis x line=left,
        width=12cm,height=9cm,
        ybar=1pt,
        bar width=6pt,
        xmode=linear, % not log
        ymode=linear, % not log
        xlabel = {$p$},
        ylabel = {Speedup},
        ymin = 0, ymax = 4.5,
        symbolic x coords={1,2,3,4,5,6},
        xtick={1,2,3,4,5,6},
        enlarge x limits={abs=4*\pgfplotbarwidth},
        legend cell align={left},
        legend style={font=\scriptsize, at={(0.97, 0.97)},anchor=north east},
        %nodes near coords,
        %nodes near coords align={vertical},
        x tick label style={/pgf/number format/.cd, fixed, fixed zerofill, precision=0, /tikz/.cd},
        y tick label style={/pgf/number format/.cd, fixed, fixed zerofill, precision=0, /tikz/.cd},
        label style={font=\Large},
        tick label style={font=\Large},
    ]
        
        \addplot[fill={Pastel1-A}, draw=none] coordinates {(1,3.63063) (2,3.629125) (3,3.17187) (4,2.37292) (5,1.859622) (6,0.837559)};
        \addplot[fill={Pastel1-B}, draw=none] coordinates {(1,3.65288) (2,3.640398) (3,3.16806) (4,3.495175) (5,1.88602) (6,0.840110)};
        \addplot[fill={Pastel1-C}, draw=none] coordinates {(1,3.59451311) (2,3.197443) (3,2.873749) (4,3.1818518) (5,2.5415747) (6,0.8435336)};
        \addplot[fill={Pastel1-D}, draw=none] coordinates {(1,3.568692) (2,3.3658817) (3,3.2114144) (4,3.269796) (5,2.393332) (6,1.874032)};
        \addplot[fill={Pastel1-E}, draw=none] coordinates {(1,3.5223043) (2,3.4461900) (3,3.1232644) (4,3.5795301) (5,2.96081000) (6,1.6997455)};
        %\legend{25,24,18,15,12}
    \end{axis}
\end{tikzpicture}}}
        \caption{\label{fig:lines_sppedup}Speedup comparison for optimal configuration for differing numbers of variables per point.}
    \end{figure}
    
    Although these lines based approaches were developed for the Volta architecture, we do not foresee any limitations in porting them to other GPUs. However, as is evident from the performance curves, the general trend is that more shared memory leads to better performance. Given that this method aims to avoid global memory bottlenecks, this is to be expected, but consequently some architectures that have limited shared memory may not perform as well. However, it does appear that trend in new generations of GPU devices is for the shared memory per block to increase, suggesting that the methods outlined here will perform well on forthcoming devices.  
\section{Application of Methods}\label{sec:application}
    In the previous section, two approaches were outlined, and it was shown that it was possible to fuse together two ACM-HD kernels, which provided a speedup over the un-fused method. These kernels have been implemented in the open-source solver PyFR~\citep{Witherden2014}. Currently, two simplifying restrictions have been made: the mesh must have a constant Jacobian, and it must be composed solely of hexahedral elements. With this in mind, the Taylor--Green vortex (TGV)~\citep{Taylor1937,Brachet1983} test case was used to benchmark the methods. The domain for the TGV is a three-dimensional periodic box, $\Omega=[0,2\pi]^3$, with the initial condition defined as:
    \begin{subequations}
        \begin{align}
            u &=  \sin(x)\cos(y)\cos(z), \\
            v &= -\cos(x)\sin(y)\cos(z), \\
            w &= 0, \\
            P &= \frac{1}{\gamma M^2} + \frac{1}{16}\cos(2z + 2)\left[\cos(2x) + \cos(2y)\right],
        \end{align}
    \end{subequations}
    where $\gamma$ is the ratio of specific heats and $M$ is the Mach number, which we take to be $0.08$. To initialise the gradient terms, the exact derivative of the initial condition may be straightforwardly found. The Reynolds number was set at $1600$ and the artificial compressibility parameter to $\zeta=2.5$, with a time period of $T=5\times10^{-3}$. From the prior testing, the optimal kernels for single-precision floating-point were used, the configurations of which are shown in \cref{tab:pmg_config_fp32,tab:pmg_config_fp64}. 
    
    \begin{table}[tbhp]
        \centering
        \caption{\label{tab:pmg_config_fp32}Optimal kernel configuration with polynomial order for FP-32 working precision.}
        \begin{tabular}{l r r}
            \toprule
            $p$ & Method & Configuration\\\midrule
            1 & Planar, un-managed & $\texttt{blockDim.x} = 128$ \\
            2 & Planar, un-managed & $\texttt{blockDim.x} = 128$ \\
            3 & Planar, managed & $\texttt{smem} = \SI{66}{\kibi\byte}$, $\texttt{blockDim.x} = 128$\\
            4 & Lines, 24 & $\texttt{smem} = \SI{93.75}{\kibi\byte}$, $\texttt{blockDim.x} = 200$ \\
            \bottomrule
        \end{tabular}
    \end{table}
    
    \begin{table}[tbhp]
        \centering
        \caption{\label{tab:pmg_config_fp64}Optimal kernel configuration with polynomial order for FP-64 working precision.}
        \begin{tabular}{l r r}
            \toprule
            $p$ & Method & Configuration\\\midrule
            1 & Planar, managed & $\texttt{smem} = \SI{42}{\kibi\byte}$, $\texttt{blockDim.x} = 128$ \\
            2 & Planar, managed & $\texttt{smem} = \SI{59}{\kibi\byte}$, $\texttt{blockDim.x} = 128$ \\
            3 & Lines, 15 & $\texttt{smem} = \SI{90}{\kibi\byte}$, $\texttt{blockDim.x} = 192$\\
            4 & Lines, 12 & $\texttt{smem} = \SI{93.75}{\kibi\byte}$, $\texttt{blockDim.x} = 200$ \\
            \bottomrule
        \end{tabular}
    \end{table}
    
\begin{figure}[tbhp]
        \centering
        \subfloat[Single step runtime comparison of ACM-HD fusions, $p=4$.]{\label{fig:tgv_p4_fp32}\adjustbox{width=0.56\linewidth,valign=b}{\begin{tikzpicture}[spy using outlines={rectangle, height=3cm,width=2.5cm, magnification=3, connect spies}]
    \pgfplotsset{scaled y ticks=false}
    \begin{axis}
    [
        axis line style={latex-latex},
        axis y line=left,
        axis x line=left,
        ybar stacked,
        bar width=35pt,
        xmode=linear, % not log
        ymode=linear, % not log
        ylabel = {Runtime [\SI{}{\micro\second}]},
        ymin = 0, ymax = 10000,
        symbolic x coords={1,2,3},
        xtick={1,2,3},
        xticklabels={ACM-HD,ACM-HD (F),ACM-HD (FS)},
        ytick={0,2000,4000,6000,8000,10000},
        yticklabels={0,2000,4000,6000,8000,10000},
        enlarge x limits={abs=1*\pgfplotbarwidth},
        legend cell align={left},
        legend style={font=\footnotesize, at={(0.97, 0.97)},anchor=north east},
        %nodes near coords,
        %nodes near coords align={vertical},
        x tick label style={/pgf/number format/.cd, fixed, fixed zerofill, precision=0, /tikz/.cd,rotate=90},
        y tick label style={/pgf/number format/.cd, fixed, fixed zerofill, precision=0, /tikz/.cd},
        label style={font=\normalsize},
        tick label style={font=\normalsize},
    ]
        
        \addplot+[ybar,fill={Pastel1-A}, draw=none] coordinates {(1, 626) (2, 628) (3, 630)};
        
        \addplot+[ybar,fill={Set1-B!25},  draw=none] coordinates {(1, 1270) (2,0) (3,0)};
        \addplot+[ybar,fill={Set1-B!50}, draw=none] coordinates {(1, 1100) (2, 963) (3, 1100)};
        \addplot+[ybar,fill={Set1-B!80}, draw=none] coordinates {(1, 742) (2, 740) (3,0)};
        
        \addplot+[ybar,fill={Pastel1-C}, draw=none] coordinates {(1, 1010) (2, 1020) (3,1020)};
        \addplot+[ybar,fill={Pastel1-D}, draw=none] coordinates {(1, 1110) (2, 1100) (3,1100)};
        \addplot+[ybar,fill={Pastel1-E}, draw=none] coordinates {(1, 252) (2, 253) (3, 252)};
        \addplot+[ybar,fill={Pastel1-H}, draw=none] coordinates {(1, 2880) (2, 2880) (3, 2890)};

    \end{axis}
    
    \begin{scope}[on above layer]
        \draw[fill=Pastel1-A,Pastel1-A] (7,1.0) rectangle (7.3,1.3) node[black,right,above=-0.15,anchor=west]{Interface projection};
        \draw[fill=Set1-B!25,Set1-B!25] (7,1.5) rectangle (7.3,1.8) node[black,right,above=-0.15,anchor=west]{Flux};
        \draw[fill=Set1-B!50,Set1-B!50] (7,2.0) rectangle (7.3,2.3) node[black,right,above=-0.15,anchor=west]{Flux div.};
        \draw[fill=Set1-B!50,Set1-B!80] (7,2.5) rectangle (7.3,2.8) node[black,right,above=-0.15,anchor=west]{Source term};
        \draw[fill=Pastel1-C,Pastel1-C] (7,3.0) rectangle (7.3,3.3) node[black,right,above=-0.15,anchor=west]{Interface flux};
        \draw[fill=Pastel1-D,Pastel1-D] (7,3.5) rectangle (7.3,3.8) node[black,right,above=-0.15,anchor=west]{Interface div. projection};
        \draw[fill=Pastel1-E,Pastel1-E] (7,4.0) rectangle (7.3,4.3) node[black,right,above=-0.15,anchor=west]{Real-time derivative};
        \draw[fill=Pastel1-H,Pastel1-H] (7,4.5) rectangle (7.3,4.8) node[black,right,above=-0.15,anchor=west]{Pseudo-time update};
    \end{scope}
    \draw[black] (6.8,0.8) rectangle (10.85,5.0);
    
\end{tikzpicture}}}
        ~
        \subfloat[Runtime comparison with $P$-multigrid, $p=4$ for $t_\text{end}=10$.]{\label{fig:tgv_pmg_fp32}\adjustbox{width=0.38\linewidth,valign=b}{\begin{tikzpicture}[spy using outlines={rectangle, height=3cm,width=2.5cm, magnification=3, connect spies}]
    \pgfplotsset{scaled y ticks=false}
    \begin{axis}
    [
        axis line style={latex-latex},
        axis y line=left,
        axis x line=left,
        ybar,
        bar width=32pt,
        xmode=linear, % not log
        ymode=linear, % not log
        ylabel = {Runtime [\SI{}{\second}]},
        ymin = 0, ymax = 4200,
        symbolic x coords={1,2,3,4},
        xtick={1,2,3,4},
        xticklabels={ACM,ACM-HD,ACM-HD (F),ACM-HD (FS)},
        ytick={0,500,1000,1500,2000,2500,3000,3500,4000},
        yticklabels={0, ,1000, ,2000, ,3000, ,4000},
        enlarge x limits={abs=1*\pgfplotbarwidth},
        legend cell align={left},
        legend style={font=\footnotesize, at={(0.97, 0.97)},anchor=north east},
        %nodes near coords,
        %nodes near coords align={vertical},
        x tick label style={/pgf/number format/.cd, fixed, fixed zerofill, precision=0, /tikz/.cd,rotate=90},
        y tick label style={/pgf/number format/.cd, fixed, fixed zerofill, precision=0, /tikz/.cd},
        label style={font=\normalsize},
        tick label style={font=\normalsize},
    ]
        
        \addplot[ybar,fill={Pastel1-A}, draw=none] coordinates {(1, 3230.778) (2, 1929.334) (3, 1625.957) (4, 1404.686)};
        %\addplot[ybar,fill={Pastel1-A}, draw=none] coordinates {(1, 3854.228331991) (2,2372.094805) (3, 1914.703765) (4,1673.34878039360)};
        
    \end{axis}
\end{tikzpicture}}}
        \caption{\label{fig:tgv_runtime_fp32}TGV runtime comparison, FP-32, on a NVIDIA V100.}
    \end{figure}
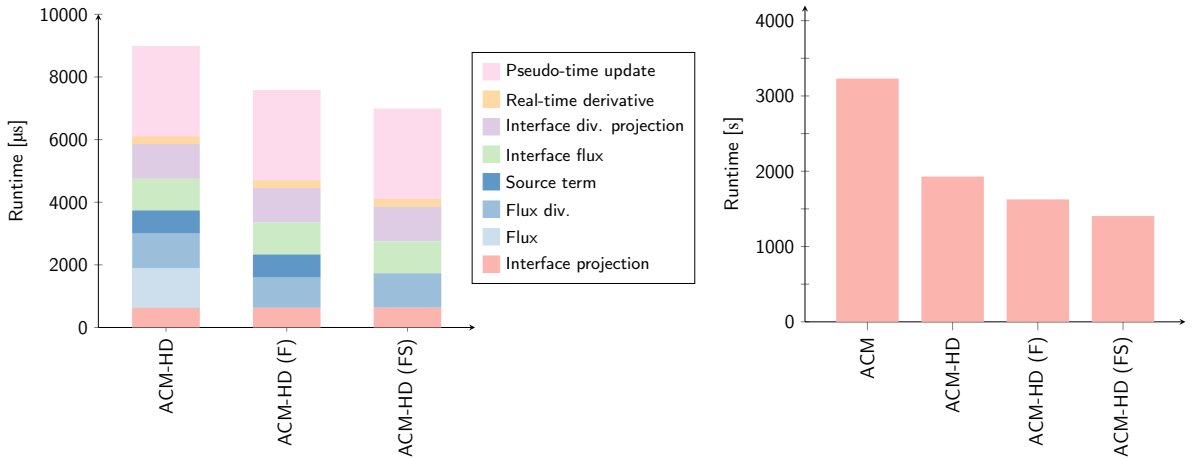

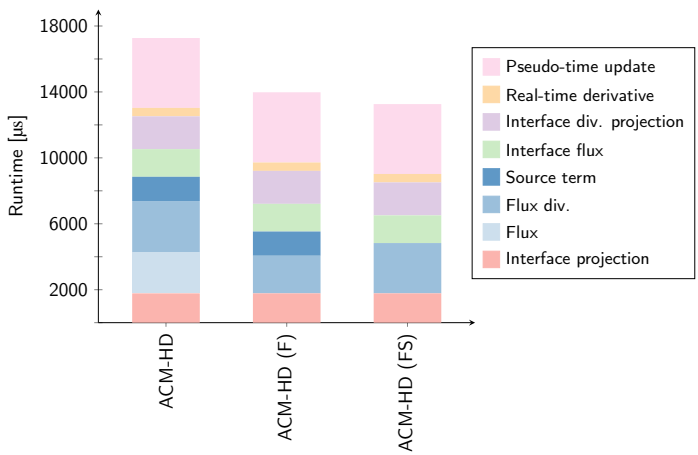
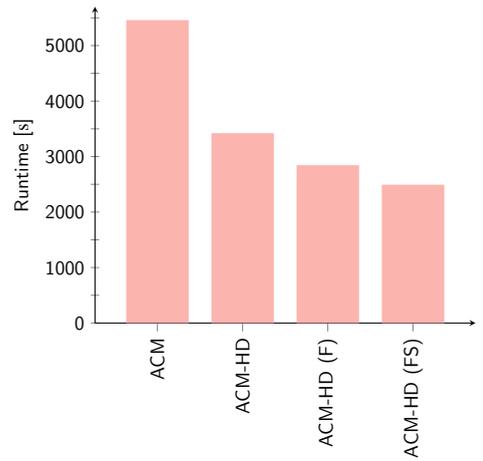
\begin{figure}[tbhp]
        \centering
        \subfloat[Single step runtime comparison of ACM-HD fusions, $p=4$.]{\label{fig:tgv_p4_fp64}\adjustbox{width=0.56\linewidth,valign=b}{\begin{tikzpicture}[spy using outlines={rectangle, height=3cm,width=2.5cm, magnification=3, connect spies}]
    \pgfplotsset{scaled y ticks=false}
    \begin{axis}
    [
        axis line style={latex-latex},
        axis y line=left,
        axis x line=left,
        ybar stacked,
        bar width=35pt,
        xmode=linear, % not log
        ymode=linear, % not log
        ylabel = {Runtime [\SI{}{\micro\second}]},
        ymin = 0, ymax = 19000,
        symbolic x coords={1,2,3},
        xtick={1,2,3},
        xticklabels={ACM-HD,ACM-HD (F),ACM-HD (FS)},
        ytick={0,2000,4000,6000,8000,10000,12000,14000,16000,18000},
        yticklabels={ ,2000, ,6000, ,10000, ,14000, ,18000},
        enlarge x limits={abs=1*\pgfplotbarwidth},
        legend cell align={left},
        legend style={font=\footnotesize, at={(0.97, 0.97)},anchor=north east},
        %nodes near coords,
        %nodes near coords align={vertical},
        x tick label style={/pgf/number format/.cd, fixed, fixed zerofill, precision=0, /tikz/.cd,rotate=90},
        y tick label style={/pgf/number format/.cd, fixed, fixed zerofill, precision=0, /tikz/.cd},
        label style={font=\normalsize},
        tick label style={font=\normalsize},
    ]
        
        \addplot+[ybar,fill={Pastel1-A}, draw=none] coordinates {(1, 1780) (2, 1780) (3, 1780)};
        
        \addplot+[ybar,fill={Set1-B!25},  draw=none] coordinates {(1, 2470) (2, 0) (3, 0)};
        \addplot+[ybar,fill={Set1-B!50}, draw=none] coordinates {(1, 3130) (2, 2280) (3, 3050)};
        \addplot+[ybar,fill={Set1-B!80}, draw=none] coordinates {(1, 1470) (2, 1470) (3, 0)};
        
        \addplot+[ybar,fill={Pastel1-C}, draw=none] coordinates {(1, 1680) (2, 1680) (3, 1680)};
        \addplot+[ybar,fill={Pastel1-D}, draw=none] coordinates {(1, 1990) (2, 2000) (3, 2000)};
        
        \addplot+[ybar,fill={Pastel1-E}, draw=none] coordinates {(1, 510) (2, 508) (3, 512)};
        \addplot+[ybar,fill={Pastel1-H}, draw=none] coordinates {(1, 4240) (2, 4260) (3, 4240)};

    \end{axis}
    
    \begin{scope}[on above layer]
        \draw[fill=Pastel1-A,Pastel1-A] (7,1.0) rectangle (7.3,1.3) node[black,right,above=-0.15,anchor=west]{Interface projection};
        \draw[fill=Set1-B!25,Set1-B!25] (7,1.5) rectangle (7.3,1.8) node[black,right,above=-0.15,anchor=west]{Flux};
        \draw[fill=Set1-B!50,Set1-B!50] (7,2.0) rectangle (7.3,2.3) node[black,right,above=-0.15,anchor=west]{Flux div.};
        \draw[fill=Set1-B!50,Set1-B!80] (7,2.5) rectangle (7.3,2.8) node[black,right,above=-0.15,anchor=west]{Source term};
        \draw[fill=Pastel1-C,Pastel1-C] (7,3.0) rectangle (7.3,3.3) node[black,right,above=-0.15,anchor=west]{Interface flux};
        \draw[fill=Pastel1-D,Pastel1-D] (7,3.5) rectangle (7.3,3.8) node[black,right,above=-0.15,anchor=west]{Interface div. projection};
        \draw[fill=Pastel1-E,Pastel1-E] (7,4.0) rectangle (7.3,4.3) node[black,right,above=-0.15,anchor=west]{Real-time derivative};
        \draw[fill=Pastel1-H,Pastel1-H] (7,4.5) rectangle (7.3,4.8) node[black,right,above=-0.15,anchor=west]{Pseudo-time update};
    \end{scope}
    \draw[black] (6.8,0.8) rectangle (10.85,5.0);
    
\end{tikzpicture}}}
        ~
        \subfloat[Runtime comparison with $P$-multigrid, $p=4$ for $t_\text{end}=10$.]{\label{fig:tgv_pmg_fp64}\adjustbox{width=0.38\linewidth,valign=b}{\begin{tikzpicture}[spy using outlines={rectangle, height=3cm,width=2.5cm, magnification=3, connect spies}]
    \pgfplotsset{scaled y ticks=false}
    \begin{axis}
    [
        axis line style={latex-latex},
        axis y line=left,
        axis x line=left,
        ybar,
        bar width=32pt,
        xmode=linear, % not log
        ymode=linear, % not log
        ylabel = {Runtime [\SI{}{\second}]},
        ymin = 0, ymax = 5700,
        symbolic x coords={1,2,3,4},
        xtick={1,2,3,4},
        xticklabels={ACM,ACM-HD,ACM-HD (F),ACM-HD (FS)},
        ytick={0,500,1000,1500,2000,2500,3000,3500,4000,4500,5000,5500},
        yticklabels={0, ,1000, ,2000, ,3000, ,4000, ,5000, },
        enlarge x limits={abs=1*\pgfplotbarwidth},
        legend cell align={left},
        legend style={font=\footnotesize, at={(0.97, 0.97)},anchor=north east},
        %nodes near coords,
        %nodes near coords align={vertical},
        x tick label style={/pgf/number format/.cd, fixed, fixed zerofill, precision=0, /tikz/.cd,rotate=90},
        y tick label style={/pgf/number format/.cd, fixed, fixed zerofill, precision=0, /tikz/.cd},
        label style={font=\normalsize},
        tick label style={font=\normalsize},
    ]
        
        \addplot[ybar,fill={Pastel1-A}, draw=none] coordinates {(1, 5458.29927) (2,3421.23252) (3,2843.58298) (4,2492.18659)};
        
    \end{axis}
\end{tikzpicture}}}
        \caption{\label{fig:tgv_runtime_fp64}TGV runtime comparison, FP-64, on a NVIDIA V100.}
    \end{figure}

    Initially, we consider the case of $p=4$, FP-32, for a grid of $32^3$ elements and a single pseudo-time step run on a single NVIDIA V100 GPU. The results of this are presented in \cref{fig:tgv_p4_fp32,fig:tgv_p4_fp64} for single and double precision floating-point respectively, in which each shows three bars. These bars represent the time taken for a time step by: ACM-HD without any fusion, \emph{ACM-HD}; ACM-HD with flux divergence fusion, \emph{ACM-HD (F)}; and ACM-HD with flux divergence and source term fusion, \emph{ACM-HD (FS)}. These results show that when applied with PyFR the fused flux divergence kernel (in blue) achieves a $2.46\times$ speedup for both precisions, which is slightly lower than expected given the results from the tests in \cref{sec:method}, and is attributable to the differences between the Titan V and V100 architectures --- namely, the different FLOPS to bandwidth ratio. When the source term is also fused into the flux divergence kernel, the speedup was measured to be $2.83\times$ and $2.32\times$ for single and double precision, respectively. The flux and source fusion is lower than optimal primarily due to the performance prior to the source term fusion being less than the expected optimal. Furthermore, fusing the source term has a larger than expected cost, increasing the kernel runtime by ${\sim}14.3\%$ and ${\sim}33.8\%$, and is a result of not storing $q_x$, $r_y$, and $s_z$ directly in shared memory; which is why the cost doubles when moving from single to double precision. Although fusing the source term does give a net reduction in run time, one of the other lines strategies may be better suited in this case. A final point on the comparison in \cref{fig:tgv_p4_fp32,fig:tgv_p4_fp64} is that several optimisation routes are known to the authors which are able to reduce the runtime of pseudo-time update stage, causing the impact of the fused kernel to be proportionally greater in the future.
    We intend to improve the performance of this kernel through a combination of fusing what are currently individual AXPY type kernels together, and by increasing their granularity such that multiple threads are assigned to each element (whereas currently one thread loops through all solution points --- something which can limit the ability of the kernel to fully saturate global memory bandwidth).  Beyond this, there are medium term goals to further fuse this revised kernel into the final update kernel involved in an RHS evaluation, further saving memory bandwidth and reducing overhead.
    
\begin{figure}[tbhp]
        \centering
        \adjustbox{width=0.7\linewidth,valign=b}{\begin{tikzpicture}[spy using outlines={rectangle, height=3cm,width=2.5cm, magnification=3, connect spies}]
    \begin{scope}[on behind layer]
        \draw[black!00] (-0.5,-0.5) rectangle (16.0,5.5);
    \end{scope}

    \tikzstyle{pmg}=[draw, circle, color={black}, text=black, minimum width=0.05cm, minimum height=0.05cm]
    \tikzstyle{connect}=[-,black,ultra thick]

    \draw[black, -latex, ultra thick] (0,-0.2) -- (0,5) node[midway,rotate=90, above, yshift=5ex]{\LARGE$p$};
    \draw[black, -latex, ultra thick] (-0.2,0) -- (15.5,0) node[midway, below, yshift=-1ex]{\Large{Stage}};
    
    \draw[black] (-0.2,4) -- (0.1,4) node[left,xshift=-2ex]{\Large $4$};
    \draw[black] (-0.1,3) -- (0.1,3) node[left,xshift=-2ex]{\Large $3$};
    \draw[black] (-0.1,2) -- (0.1,2) node[left,xshift=-2ex]{\Large $2$};
    \draw[black] (-0.1,1) -- (0.1,1) node[left,xshift=-2ex]{\Large $1$};

    \node[pmg, fill={OrRd-J}] at (0,4) (p4r) {};
    \node[pmg, fill={OrRd-G}] at (1,3) (p3r) {};
    \node[pmg, fill={OrRd-E}] at (2,2) (p2r) {};
    \node[pmg, fill={OrRd-B}] at (3,1) (p1r) {};
    \node[pmg, fill={OrRd-B}] at (4,1) (p1s1) {};
    \node[pmg, fill={OrRd-B}] at (5,1) (p1s2) {};
    \node[pmg, fill={OrRd-B}] at (6,1) (p1s3) {};
    \node[pmg, fill={OrRd-E}] at (7,2) (p2p) {};
    \node[pmg, fill={OrRd-G}] at (8,3) (p3p) {};
    \node[pmg, fill={OrRd-J}] at (9,4) (p4p) {};
    \node[pmg, fill={OrRd-J}] at (10,4) (p4s1) {};
    \node[pmg, fill={OrRd-J}] at (11,4) (p4s2) {};
    \node[pmg, fill={OrRd-J}] at (12,4) (p4s3) {};
    \node[pmg, fill={OrRd-J}] at (13,4) (p4s4) {};
    \node[pmg, fill={OrRd-J}] at (14,4) (p4s5) {};
    \node[pmg, fill={OrRd-J}] at (15,4) (p4s6) {};

    \draw[connect] (p4r) -- (p3r);
    \draw[connect] (p3r) -- (p2r);
    \draw[connect] (p2r) -- (p1r);
    \draw[connect] (p1r) -- (p1s1);
    \draw[connect] (p1s1) -- (p1s2);
    \draw[connect] (p1s2) -- (p1s3);    
    \draw[connect] (p1s3) -- (p2p);
    \draw[connect] (p2p) -- (p3p);
    \draw[connect] (p3p) -- (p4p);
    \draw[connect] (p4p) -- (p4s1);
    \draw[connect] (p4s1) -- (p4s2);
    \draw[connect] (p4s2) -- (p4s3);
    \draw[connect] (p4s3) -- (p4s4);
    \draw[connect] (p4s4) -- (p4s5);
    \draw[connect] (p4s5) -- (p4s6);
    
\end{tikzpicture}}
        \caption{\label{fig:pmg}$P$-multigrid cycle: [(4, 1),\; (3, 1),\; (2, 1),\; (1, 4),\; (2, 1),\; (3, 1),\; (4, 7)].}
    \end{figure}
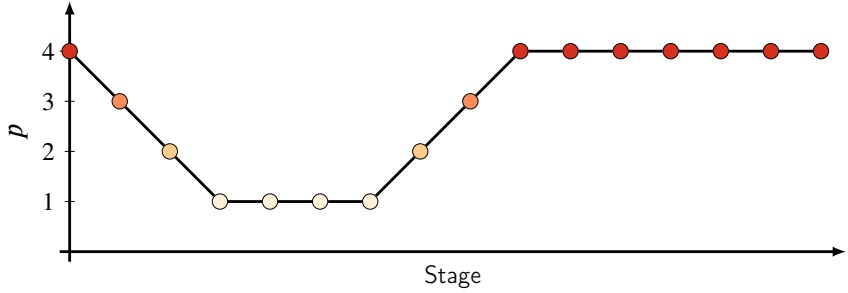

    We now consider running from $t=0$ to $t=10$ for both single and double precision. In doing so, we will use several established convergence acceleration methods, namely $P$-multigrid~\citep{Loppi2018} and spatially varying pseudo-time stepping~\citep{Loppi2019}. For the double-precision run the same number of pseudo-iterations  were used as single-precision for each physical step. Based on the work of \citet{Loppi2019} and \citet{Trojak2020}, we made use of the asymmetric $P$-multigrid cycle shown diagrammatically in \cref{fig:pmg}.
    
    The physical time-integration was performed with a BDF2 scheme with $\Delta t=0.01$. The pseudo iterations were performed using a RK34 embedded pair method with locally adaptive pseudo-stepping with the initial $\Delta t/\Delta \tau = 10$. The single-precision convergence was monitored via the $l_2$ residual in velocity field with a tolerance of $2\times10^{-4}$. As the stability of the pseudo-stepping increases when spatial order decreases, the $\Delta\tau$ may be increased within the $P$-multigrid cycle. For this, a constant factor of $1.75$ was found to be appropriate. The exact input files used can be found in the ESM. For the common interface flux, a Rusanov type approximate Riemann solver was used for the hyperbolic portion of the Riemann solve. For the standard ACM formulation, the viscous common interface flux was approximated using LDG with $\beta=0.5$ and $\tau=0.1$. For ACM-HD, it was found that taking the mean of left and right flux at the interfaces for the additional equations was sufficient. 
    
    To ensure the validity of the numerical results, the global integral property --- enstrophy --- was calculated. This functional has the following form:
    \begin{equation}
        \epsilon(t) = \frac{\nu}{|\Omega|}\int_\Omega \pmb{\omega}(t,\mathbf{x})^T\pmb{\omega}(t,\mathbf{x})\;\mathrm{d}\mathbf{x}
    \end{equation}
     where $|\Omega|$ is the volume of the domain, and $\pmb{\omega}=\nabla\times\mathbf{V}$. Here, the slightly modified definition of enstrophy is used, as, in the incompressible limit, this form should be identical to the kinetic energy dissipation rate. A comparison between the ACM and ACM-HD results is made in \cref{fig:tgv_functional}, which shows that only a small difference is visible when the kinetic energy dissipation rate is at its peak. From the work of \citet{Mengaldo2018}, the observed difference is within the range of the effect of the approximate Riemann solver, which changes between the two runs due to one system being purely hyperbolic and the other mixed. 
    
    \begin{figure}[tbhp]
        \centering
        \adjustbox{width=0.4\linewidth,valign=b}{\begin{tikzpicture}[spy using outlines={rectangle, height=3cm,width=2.5cm, magnification=3, connect spies}]
		\begin{axis}[
            axis line style={latex-latex},
            axis y line=left,
            axis x line=left,
            xmode=linear, % not log
            ymode=linear, % not log
		    xlabel={$t$},
		    xtick={0,2,...,10},
    		xmin=0,xmax=10.1,
    		ylabel={$\epsilon$},
    		ylabel style={rotate=-90},
    		ytick={0,0.002,...,0.014},
    		ymin=0,ymax=0.014,
            %x tick label style={/pgf/number format/.cd, fixed, fixed zerofill, precision=1, /tikz/.cd},
            y tick label style={/pgf/number format/.cd, fixed, fixed zerofill, precision=1, /tikz/.cd},
    		legend style={at={(0.03,0.97)},anchor=north west,font=\small},
    		legend cell align={left},
    		]
    	
        \addplot[line width=1pt,color={Set1-A}] table[x=t, y=dkep, col sep=comma]{./figs/data/tgv/van_rees_2011_dkep_PSP512.csv};
                                                      
        \addplot[line width=1pt,color={Set1-B}] table[x expr=(\thisrowno{0}), y expr=(\thisrowno{2}*2.5197e-06), 
                                                      col sep=comma]{./figs/data/tgv/integral_acm.csv};
                                                      
        \addplot[line width=1pt,color={Set1-C}] table[x expr=(\thisrowno{0}), y expr=(\thisrowno{2}*2.5197e-06), 
                                                      col sep=comma]{./figs/data/tgv/integral_hype_fs.csv};
        \addlegendentry{Reference}
        \addlegendentry{ACM}
        \addlegendentry{ACM-HD}
    \end{axis}
\end{tikzpicture}}
        \caption{\label{fig:tgv_functional}TGV enstrophy comparison at FP-32 with reference DNS data from \citet{vanRees2011} with ACM and ACM-HD, $p=4$ and $160^3$ grid points. The reference DNS was calculated with a pseudo-spectral method for $512^3$ grid points.}
    \end{figure}
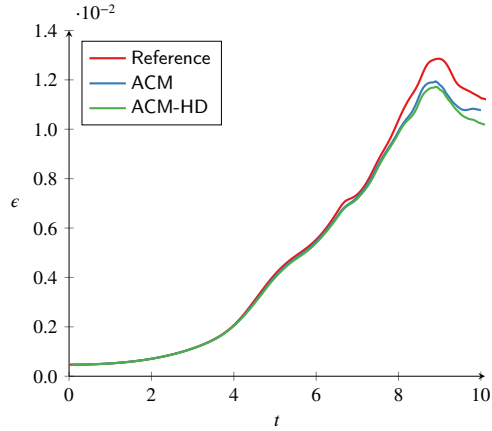    
    
    \cref{fig:tgv_pmg_fp32,fig:tgv_pmg_fp64} displays a comparison of the conventional mixed hyperbolic-parabolic ACM approach with the various ACM-HD approaches. A single conventional ACM step is faster than an ACM-HD (FS) step; for example, by ${\sim}9.8\%$ at $p=4$ FP-32. However, ACM-HD has an increased rate of convergence, and, as the system is purely hyperbolic, the stability limit scales with $h^{-1}$ rather than $h^{-2}$~\citep{Watson2018}. In combination, kernel fusion and hyperbolic diffusion in this case is found to give a significant speedup of ${\sim}2.3\times$ and ${\sim}2.2\times$, for FP-32 and FP-64 respectively.

\section{Conclusions}\label{sec:conclusions}
In this work, the hyperbolic diffusion technique has been applied to the artificial compressibility method used to solve the incompressible Navier--Stokes equations. The resulting system of equations was solved with the high-order flux reconstruction method, where, due to the removal of the algorithmic steps needed to construct the viscous terms, optimisation could be performed. In particular, it was found that the evaluation of the fluxes could be fused with the calculation of the flux divergence to give a theoretical speedup of $4\times$ over the un-fused operation. 

This work has focused on the treatment of tensor product elements with constant Jacobians and NVIDIA GPU architectures. To this end, two approaches were taken to fuse the kernels, which offered differing degrees of parallelism. These were a planar method and a lines method, in which each thread formed the calculation for a plane and line, respectively. For the reduced parallelism of the planar method, it was found that a fully unrolled approach could produce optimal kernels at low orders, but, due to high register usage, high orders were challenging. To maximise the use of GPU resources, a novel generation time memory manager was developed which was able to increase the performance of the planar method in some cases. The lines method was able to successfully increase the performance of fused kernels at higher orders, largely due to increased parallelism. Furthermore, the lines method did not benefit from unrolling, and instead varying the shared memory used per point was successful in increasing performance by diversifying memory usage. 

Finally, the optimal kernels were applied to the Taylor--Green vortex test case via the PyFR solver framework. It was observed that the kernel fusion could reduce the runtime of the hyperbolised ACM system by ${\sim}25\%$. A comparison was then made between the standard ACM formulation and the optimised hyperbolic form, and it was found that the latter was ${\sim}2.3\times$ faster when converged to the same accuracy. This not only facilitates the faster computation of incompressible flows, but also demonstrates the more general utility of these kernel fusion approaches.

Future work will be required to generalise these approaches to non-constant spatial Jacobians, although it is expected that in the case of linear elements this will not pose a significant challenge. This is because only the locations of the element corners need to be loaded --- a strategy that is already applied in the PyFR flux kernels. However, for fully non-linear elements more complex fusion strategies will be needed that incorporate the Jacobian into the shared memory layout. Furthermore, AMD has indicated that the quantity of shared memory on their upcoming future architectures will increase, which will broaden the scope of the techniques developed here.

\section*{Acknowledgements}\label{sec:ack}
The authors would like to gratefully acknowledge the support of the NVIDIA Corporation with the donation of one NVIDIA TITAN V GPU through their seeding grant program.

\bibliographystyle{cas-model2-names}
\bibliography{reference}


\clearpage
\begin{appendices}
\section{\label{app:lines}Lines Appendix}

    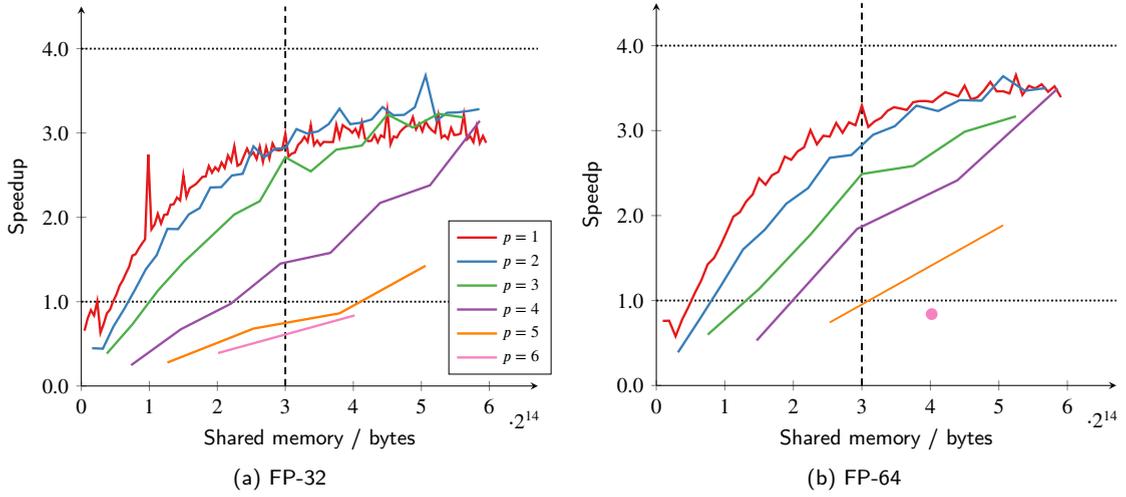
\begin{figure}[tbhp]
        \centering
        \subfloat[FP-32]{\label{fig:l24_smem_rt_32}\adjustbox{width=0.447\linewidth, valign=b}{\begin{tikzpicture}[spy using outlines={rectangle, height=3cm,width=2.5cm, magnification=3, connect spies}]
\definecolor{agrey}{rgb}{0.6,0.6,0.6}
    \pgfplotsset{scaled x ticks=false}
    \begin{axis}
    [
        axis line style={latex-latex},
        axis y line=left,
        axis x line=left,
        xmode=linear, % not log
        ymode=linear, % not log
        xlabel = {Shared memory / bytes},
        ylabel = {Speedup},
        xmin = 0, xmax = 110000,
        ymin = 0, ymax = 4.5,
        xtick = {0, 2^14, 2^15, 3*2^14, 4*2^14, 5*2^14, 6*2^14},
        xticklabels = {$0$, $1$, $2$, $3$, $4$, $5$, $6$},
        legend cell align={left},
        legend style={font=\scriptsize, at={(1.03, 0.03)},anchor=south east},
        %axis line style={draw=none},
        %tick style={draw=none},
        x tick label style={/pgf/number format/.cd, fixed, fixed zerofill, precision=1, /tikz/.cd},
        y tick label style={/pgf/number format/.cd, fixed, fixed zerofill, precision=1, /tikz/.cd},
    ]
        \addplot[thick, black, densely dotted, forget plot] coordinates{(0,4) (110000,4)};
        \addplot[thick, black, densely dotted, forget plot] coordinates{(0,1) (110000,1)};
        \addplot[thick, black, densely dashed, forget plot] coordinates{(3*2^14,0) (3*2^14,4.5)};
    
        \addplot[line width=1pt,color={Set1-A}] table[x expr=(\thisrowno{6}), y expr=(192224/\thisrowno{7}), 
                                                      col sep=comma]{./figs/data/lines/24/gimmik_data_lines_p1_fp32.csv};
        \addlegendentry{$p=1$}
                                                      
        \addplot[line width=1pt,color={Set1-B}] table[x expr=(\thisrowno{6}), y expr=(675392/\thisrowno{7}), 
                                                      col sep=comma]{./figs/data/lines/24/gimmik_data_lines_p2_fp32.csv};
        \addlegendentry{$p=2$}
                                                      
        \addplot[line width=1pt,color={Set1-C}] table[x expr=(\thisrowno{6}), y expr=(1624578/\thisrowno{7}), 
                                                      col sep=comma]{./figs/data/lines/24/gimmik_data_lines_p3_fp32.csv};
        \addlegendentry{$p=3$}
                                                      
        \addplot[line width=1pt,color={Set1-D}] table[x expr=(\thisrowno{6}), y expr=(3213824/\thisrowno{7}), 
                                                      col sep=comma]{./figs/data/lines/24/gimmik_data_lines_p4_fp32.csv};
        \addlegendentry{$p=4$}
                                                      
        \addplot[line width=1pt,color={Set1-E}] table[x expr=(\thisrowno{6}), y expr=(5928512/\thisrowno{7}), 
                                                      col sep=comma]{./figs/data/lines/24/gimmik_data_lines_p5_fp32.csv};
        \addlegendentry{$p=5$}
                                                      
        \addplot[line width=1pt,color={Set1-H}] table[x expr=(\thisrowno{6}), y expr=(13513248/\thisrowno{7}), 
                                                      col sep=comma]{./figs/data/lines/24/gimmik_data_lines_p6_fp32.csv};
        \addlegendentry{$p=6$}
    \end{axis}
    \node[black] at (6.65,-0.5) {$\cdot2^{14}$};
\end{tikzpicture}}}
        ~
        \subfloat[FP-64]{\label{fig:l24_smem_rt_64}\adjustbox{width=0.447\linewidth, valign=b}{\begin{tikzpicture}[spy using outlines={rectangle, height=3cm,width=2.5cm, magnification=3, connect spies}]
\definecolor{agrey}{rgb}{0.6,0.6,0.6}
    \pgfplotsset{scaled x ticks=false}
    \begin{axis}
    [
        axis line style={latex-latex},
        axis y line=left,
        axis x line=left,
        xmode=linear, % not log
        ymode=linear, % not log
        xlabel = {Shared memory / bytes},
        ylabel = {Speedp},
        xmin = 0, xmax = 110000,
        ymin = 0, ymax = 4.5,
        xtick = {0, 2^14, 2^15, 3*2^14, 4*2^14, 5*2^14, 6*2^14},
        xticklabels = {$0$, $1$, $2$, $3$, $4$, $5$, $6$},
        legend cell align={left},
        legend style={font=\scriptsize, at={(1.1, 0.03)},anchor=south east},
        %axis line style={draw=none},
        %tick style={draw=none},
        x tick label style={/pgf/number format/.cd, fixed, fixed zerofill, precision=1, /tikz/.cd},
        y tick label style={/pgf/number format/.cd, fixed, fixed zerofill, precision=1, /tikz/.cd},
    ]
        \addplot[thick, black, densely dotted, forget plot] coordinates{(0,4) (110000,4)};
        \addplot[thick, black, densely dotted, forget plot] coordinates{(0,1) (110000,1)};
        \addplot[thick, black, densely dashed, forget plot] coordinates{(3*2^14,0) (3*2^14,4.5)};
    
        \addplot[line width=1pt,color={Set1-A}] table[x expr=(\thisrowno{6}), y expr=(381536/\thisrowno{7}), col sep=comma]{./figs/data/lines/24/gimmik_data_lines_p1_fp64.csv};
                                                      
        \addplot[line width=1pt,color={Set1-B}] table[x expr=(\thisrowno{6}), y expr=(1312640/\thisrowno{7}), col sep=comma]{./figs/data/lines/24/gimmik_data_lines_p2_fp64.csv};
                                                      
        \addplot[line width=1pt,color={Set1-C}] table[x expr=(\thisrowno{6}), y expr=(3124064/\thisrowno{7}), col sep=comma]{./figs/data/lines/24/gimmik_data_lines_p3_fp64.csv};
                                                      
        \addplot[line width=1pt,color={Set1-D}] table[x expr=(\thisrowno{6}), y expr=(7707616/\thisrowno{7}), col sep=comma]{./figs/data/lines/24/gimmik_data_lines_p4_fp64.csv};
                                                      
        \addplot[color={Set1-E}, thick] table[x expr=(\thisrowno{6}), y expr=(17476064/\thisrowno{7}), col sep=comma]{./figs/data/lines/24/gimmik_data_lines_p5_fp64.csv};
                                                      
        \addplot[line width=1pt,color={Set1-H}, mark=*, mark options={solid}, smooth] table[x expr=(\thisrowno{6}), y expr=(31571264/\thisrowno{7}), col sep=comma]{./figs/data/lines/24/gimmik_data_lines_p6_fp64.csv};
        
    \end{axis}
    \node[black] at (6.65,-0.5) {$\cdot2^{14}$};
\end{tikzpicture}}}
        \caption{\label{fig:lines24_smem_rt}Method of lines performance with 24 variables and varying shared memory usage.}
    \end{figure}
    
    \begin{figure}[tbhp]
        \centering
        \subfloat[FP-32]{\label{fig:l18_smem_rt_32}\adjustbox{width=0.447\linewidth, valign=b}{\begin{tikzpicture}[spy using outlines={rectangle, height=3cm,width=2.5cm, magnification=3, connect spies}]
\definecolor{agrey}{rgb}{0.6,0.6,0.6}
    \pgfplotsset{scaled x ticks=false}
    \begin{axis}
    [
        axis line style={latex-latex},
        axis y line=left,
        axis x line=left,
        xmode=linear, % not log
        ymode=linear, % not log
        xlabel = {Shared memory / bytes},
        ylabel = {Speedup},
        xmin = 0, xmax = 110000,
        ymin = 0, ymax = 4.5,
        xtick = {0, 2^14, 2^15, 3*2^14, 4*2^14, 5*2^14, 6*2^14},
        xticklabels = {$0$, $1$, $2$, $3$, $4$, $5$, $6$},
        legend cell align={left},
        legend style={font=\scriptsize, at={(1.03, 0.03)},anchor=south east},
        %axis line style={draw=none},
        %tick style={draw=none},
        x tick label style={/pgf/number format/.cd, fixed, fixed zerofill, precision=1, /tikz/.cd},
        y tick label style={/pgf/number format/.cd, fixed, fixed zerofill, precision=1, /tikz/.cd},
    ]
        \addplot[thick, black, densely dotted, forget plot] coordinates{(0,4) (110000,4)};
        \addplot[thick, black, densely dotted, forget plot] coordinates{(0,1) (110000,1)};
        \addplot[thick, black, densely dashed, forget plot] coordinates{(3*2^14,0) (3*2^14,4.5)};
    
        \addplot[line width=1pt,color={Set1-A}] table[x expr=(\thisrowno{6}), y expr=(192224/\thisrowno{7}), 
                                                      col sep=comma]{./figs/data/lines/18/gimmik_data_lines_p1_fp32.csv};
        \addlegendentry{$p=1$}
                                                      
        \addplot[line width=1pt,color={Set1-B}] table[x expr=(\thisrowno{6}), y expr=(675392/\thisrowno{7}), 
                                                      col sep=comma]{./figs/data/lines/18/gimmik_data_lines_p2_fp32.csv};
        \addlegendentry{$p=2$}
                                                      
        \addplot[line width=1pt,color={Set1-C}] table[x expr=(\thisrowno{6}), y expr=(1624578/\thisrowno{7}), 
                                                      col sep=comma]{./figs/data/lines/18/gimmik_data_lines_p3_fp32.csv};
        \addlegendentry{$p=3$}
                                                      
        \addplot[line width=1pt,color={Set1-D}] table[x expr=(\thisrowno{6}), y expr=(3213824/\thisrowno{7}), 
                                                      col sep=comma]{./figs/data/lines/18/gimmik_data_lines_p4_fp32.csv};
        \addlegendentry{$p=4$}
                                                      
        \addplot[line width=1pt,color={Set1-E}] table[x expr=(\thisrowno{6}), y expr=(5928512/\thisrowno{7}), 
                                                      col sep=comma]{./figs/data/lines/18/gimmik_data_lines_p5_fp32.csv};
        \addlegendentry{$p=5$}
                                                      
        \addplot[line width=1pt,color={Set1-H}] table[x expr=(\thisrowno{6}), y expr=(13513248/\thisrowno{7}), 
                                                      col sep=comma]{./figs/data/lines/18/gimmik_data_lines_p6_fp32.csv};
        \addlegendentry{$p=6$}
    \end{axis}
    \node[black] at (6.65,-0.5) {$\cdot2^{14}$};
\end{tikzpicture}}}
        ~
        \subfloat[FP-64]{\label{fig:l18_smem_rt_64}\adjustbox{width=0.447\linewidth, valign=b}{\begin{tikzpicture}[spy using outlines={rectangle, height=3cm,width=2.5cm, magnification=3, connect spies}]
\definecolor{agrey}{rgb}{0.6,0.6,0.6}
    \pgfplotsset{scaled x ticks=false}
    \begin{axis}
    [
        axis line style={latex-latex},
        axis y line=left,
        axis x line=left,
        xmode=linear, % not log
        ymode=linear, % not log
        xlabel = {Shared memory / bytes},
        ylabel = {Speedup},
        xmin = 0, xmax = 110000,
        ymin = 0, ymax = 4.5,
        xtick = {0, 2^14, 2^15, 3*2^14, 4*2^14, 5*2^14, 6*2^14},
        xticklabels = {$0$, $1$, $2$, $3$, $4$, $5$, $6$},
        legend cell align={left},
        legend style={font=\scriptsize, at={(1.1, 0.03)},anchor=south east},
        %axis line style={draw=none},
        %tick style={draw=none},
        x tick label style={/pgf/number format/.cd, fixed, fixed zerofill, precision=1, /tikz/.cd},
        y tick label style={/pgf/number format/.cd, fixed, fixed zerofill, precision=1, /tikz/.cd},
    ]
        \addplot[thick, black, densely dotted, forget plot] coordinates{(0,4) (110000,4)};
        \addplot[thick, black, densely dotted, forget plot] coordinates{(0,1) (110000,1)};
        \addplot[thick, black, densely dashed, forget plot] coordinates{(3*2^14,0) (3*2^14,4.5)};
    
        \addplot[line width=1pt,color={Set1-A}] table[x expr=(\thisrowno{6}), y expr=(381536/\thisrowno{7}), col sep=comma]{./figs/data/lines/18/gimmik_data_lines_p1_fp64.csv};
                                                      
        \addplot[line width=1pt,color={Set1-B}] table[x expr=(\thisrowno{6}), y expr=(1312640/\thisrowno{7}), col sep=comma]{./figs/data/lines/18/gimmik_data_lines_p2_fp64.csv};
                                                      
        \addplot[line width=1pt,color={Set1-C}] table[x expr=(\thisrowno{6}), y expr=(3124064/\thisrowno{7}), col sep=comma]{./figs/data/lines/18/gimmik_data_lines_p3_fp64.csv};
                                                      
        \addplot[line width=1pt,color={Set1-D}] table[x expr=(\thisrowno{6}), y expr=(7707616/\thisrowno{7}), col sep=comma]{./figs/data/lines/18/gimmik_data_lines_p4_fp64.csv};
                                                      
        \addplot[color={Set1-E}, thick] table[x expr=(\thisrowno{6}), y expr=(17476064/\thisrowno{7}), col sep=comma]{./figs/data/lines/18/gimmik_data_lines_p5_fp64.csv};
                                                      
        \addplot[line width=1pt,color={Set1-H}, mark=*, mark options={solid}, smooth] table[x expr=(\thisrowno{6}), y expr=(31571264/\thisrowno{7}), col sep=comma]{./figs/data/lines/18/gimmik_data_lines_p6_fp64.csv};
        
    \end{axis}
    \node[black] at (6.65,-0.5) {$\cdot2^{14}$};
\end{tikzpicture}}}
        \caption{\label{fig:lines18_smem_rt}Method of lines performance with 18 variables and varying shared memory usage.}
    \end{figure}
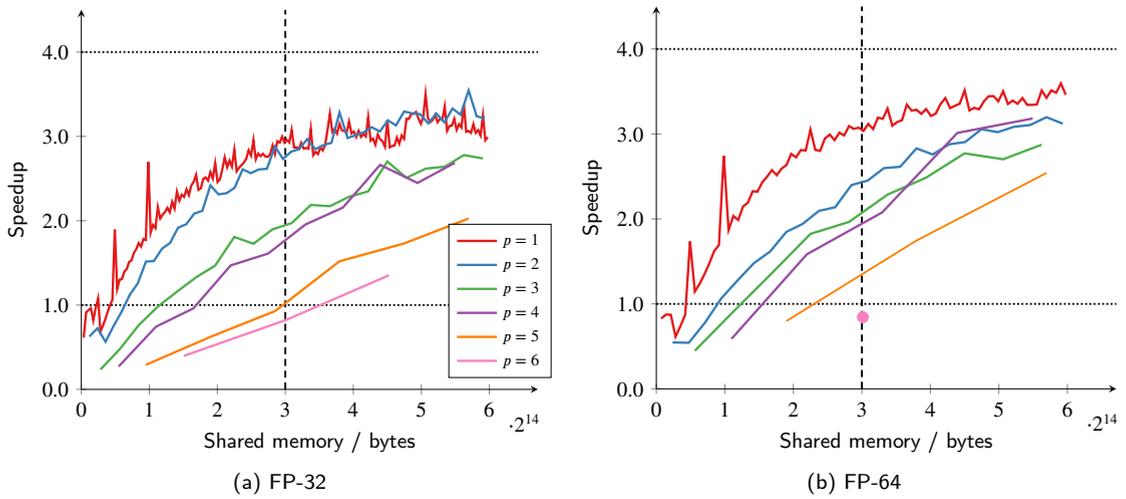
    
    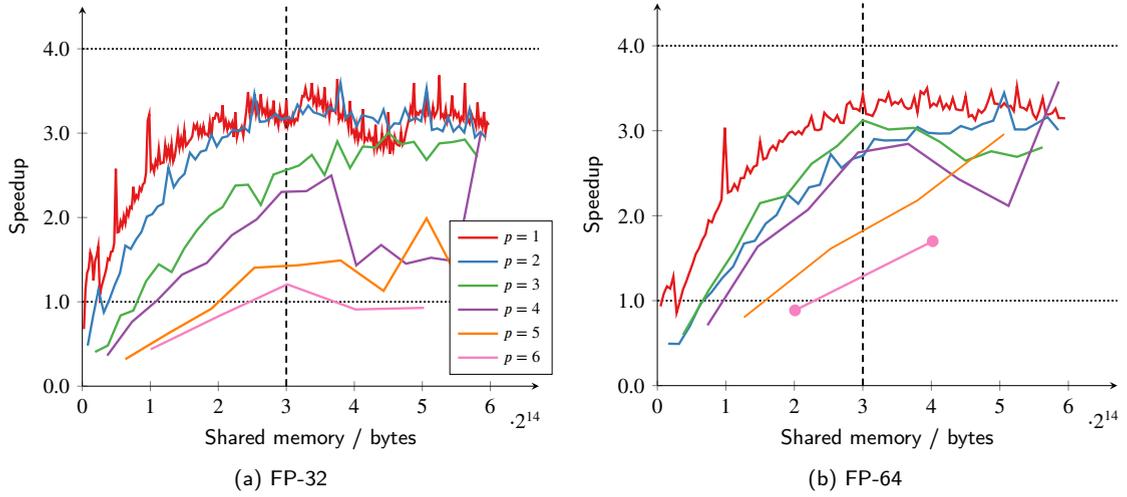
\begin{figure}[tbhp]
        \centering
        \subfloat[FP-32]{\label{fig:l15_smem_rt_32}\adjustbox{width=0.447\linewidth, valign=b}{\begin{tikzpicture}[spy using outlines={rectangle, height=3cm,width=2.5cm, magnification=3, connect spies}]
\definecolor{agrey}{rgb}{0.6,0.6,0.6}
    \pgfplotsset{scaled x ticks=false}
    \begin{axis}
    [
        axis line style={latex-latex},
        axis y line=left,
        axis x line=left,
        xmode=linear, % not log
        ymode=linear, % not log
        xlabel = {Shared memory / bytes},
        ylabel = {Speedup},
        xmin = 0, xmax = 110000,
        ymin = 0, ymax = 4.5,
        xtick = {0, 2^14, 2^15, 3*2^14, 4*2^14, 5*2^14, 6*2^14},
        xticklabels = {$0$, $1$, $2$, $3$, $4$, $5$, $6$},
        legend cell align={left},
        legend style={font=\scriptsize, at={(1.03, 0.03)},anchor=south east},
        %axis line style={draw=none},
        %tick style={draw=none},
        x tick label style={/pgf/number format/.cd, fixed, fixed zerofill, precision=1, /tikz/.cd},
        y tick label style={/pgf/number format/.cd, fixed, fixed zerofill, precision=1, /tikz/.cd},
    ]
        \addplot[thick, black, densely dotted, forget plot] coordinates{(0,4) (110000,4)};
        \addplot[thick, black, densely dotted, forget plot] coordinates{(0,1) (110000,1)};
        \addplot[thick, black, densely dashed, forget plot] coordinates{(3*2^14,0) (3*2^14,4.5)};
    
        \addplot[line width=1pt,color={Set1-A}] table[x expr=(\thisrowno{6}), y expr=(192224/\thisrowno{7}), 
                                                      col sep=comma]{./figs/data/lines/15/gimmik_data_lines_p1_fp32.csv};
        \addlegendentry{$p=1$}
                                                      
        \addplot[line width=1pt,color={Set1-B}] table[x expr=(\thisrowno{6}), y expr=(675392/\thisrowno{7}), 
                                                      col sep=comma]{./figs/data/lines/15/gimmik_data_lines_p2_fp32.csv};
        \addlegendentry{$p=2$}
                                                      
        \addplot[line width=1pt,color={Set1-C}] table[x expr=(\thisrowno{6}), y expr=(1624578/\thisrowno{7}), 
                                                      col sep=comma]{./figs/data/lines/15/gimmik_data_lines_p3_fp32.csv};
        \addlegendentry{$p=3$}
                                                      
        \addplot[line width=1pt,color={Set1-D}] table[x expr=(\thisrowno{6}), y expr=(3213824/\thisrowno{7}), 
                                                      col sep=comma]{./figs/data/lines/15/gimmik_data_lines_p4_fp32.csv};
        \addlegendentry{$p=4$}
                                                      
        \addplot[line width=1pt,color={Set1-E}] table[x expr=(\thisrowno{6}), y expr=(5928512/\thisrowno{7}), 
                                                      col sep=comma]{./figs/data/lines/15/gimmik_data_lines_p5_fp32.csv};
        \addlegendentry{$p=5$}
                                                      
        \addplot[line width=1pt,color={Set1-H}] table[x expr=(\thisrowno{6}), y expr=(13513248/\thisrowno{7}), 
                                                      col sep=comma]{./figs/data/lines/15/gimmik_data_lines_p6_fp32.csv};
        \addlegendentry{$p=6$}
    \end{axis}
    \node[black] at (6.65,-0.5) {$\cdot2^{14}$};
\end{tikzpicture}}}
        ~
        \subfloat[FP-64]{\label{fig:l15_smem_rt_64}\adjustbox{width=0.447\linewidth, valign=b}{\begin{tikzpicture}[spy using outlines={rectangle, height=3cm,width=2.5cm, magnification=3, connect spies}]
\definecolor{agrey}{rgb}{0.6,0.6,0.6}
    \pgfplotsset{scaled x ticks=false}
    \begin{axis}
    [
        axis line style={latex-latex},
        axis y line=left,
        axis x line=left,
        xmode=linear, % not log
        ymode=linear, % not log
        xlabel = {Shared memory / bytes},
        ylabel = {Speedup},
        xmin = 0, xmax = 110000,
        ymin = 0, ymax = 4.5,
        xtick = {0, 2^14, 2^15, 3*2^14, 4*2^14, 5*2^14, 6*2^14},
        xticklabels = {$0$, $1$, $2$, $3$, $4$, $5$, $6$},
        legend cell align={left},
        legend style={font=\scriptsize, at={(1.1, 0.03)},anchor=south east},
        %axis line style={draw=none},
        %tick style={draw=none},
        x tick label style={/pgf/number format/.cd, fixed, fixed zerofill, precision=1, /tikz/.cd},
        y tick label style={/pgf/number format/.cd, fixed, fixed zerofill, precision=1, /tikz/.cd},
    ]
        \addplot[thick, black, densely dotted, forget plot] coordinates{(0,4) (110000,4)};
        \addplot[thick, black, densely dotted, forget plot] coordinates{(0,1) (110000,1)};
        \addplot[thick, black, densely dashed, forget plot] coordinates{(3*2^14,0) (3*2^14,4.5)};
    
        \addplot[line width=1pt,color={Set1-A}] table[x expr=(\thisrowno{6}), y expr=(381536/\thisrowno{7}), 
                                                      col sep=comma]{./figs/data/lines/15/gimmik_data_lines_p1_fp64.csv};
                                                      
        \addplot[line width=1pt,color={Set1-B}] table[x expr=(\thisrowno{6}), y expr=(1312640/\thisrowno{7}), 
                                                      col sep=comma]{./figs/data/lines/15/gimmik_data_lines_p2_fp64.csv};
                                                      
        \addplot[line width=1pt,color={Set1-C}] table[x expr=(\thisrowno{6}), y expr=(3124064/\thisrowno{7}), 
                                                      col sep=comma]{./figs/data/lines/15/gimmik_data_lines_p3_fp64.csv};
                                                      
        \addplot[line width=1pt,color={Set1-D}] table[x expr=(\thisrowno{6}), y expr=(7707616/\thisrowno{7}), 
                                                      col sep=comma]{./figs/data/lines/15/gimmik_data_lines_p4_fp64.csv};
                                                      
        \addplot[color={Set1-E}, thick] table[x expr=(\thisrowno{6}), y expr=(17476064/\thisrowno{7}), 
                                                      col sep=comma]{./figs/data/lines/15/gimmik_data_lines_p5_fp64.csv};
                                                      
        \addplot[line width=1pt,color={Set1-H}, mark=*, mark options={solid}, smooth] table[x expr=(\thisrowno{6}), y expr=(31571264/\thisrowno{7}), col sep=comma]{./figs/data/lines/15/gimmik_data_lines_p6_fp64.csv};
        
    \end{axis}
    \node[black] at (6.65,-0.5) {$\cdot2^{14}$};
\end{tikzpicture}}}
        \caption{\label{fig:lines15_smem_rt}Method of lines performance with 15 variables and varying shared memory usage.}
    \end{figure}
\section{Example Function Listings}
\noindent
Source code not included on ArXiV.
\end{appendices}

% Show the list of todo's in the document.  Needed to avoid stupid warnings/errors when using the todo package
%\todos

\end{document}